\documentclass[twocolumn]{aastex701}
\usepackage{multirow}
\usepackage{amsmath}
\usepackage{booktabs}
\usepackage{natbib}
\usepackage{hyperref}
\usepackage{caption}   
\usepackage{array}

\usepackage{xcolor}
\definecolor{darkgreen}{rgb}{0.0, 0.5, 0.0}
\usepackage{comment}

\begin{document}

\title{Morphological Bias: How Ellipticals and Spirals Trace the Cosmic Web Differently}

\author[orcid=/0000-0002-7540-040X]{Paula S. Ferreira}
\affiliation{Institute of Astronomy, School of Physics, Zhejiang University, Hangzhou, China}
\affiliation{Center for Cosmology and Computational Astrophysics, Institute for Advanced Study in Physics, Zhejiang University, Hangzhou, China.}
\email[show]{ferreira\_paula@zju.edu.cn}

\author[orcid=0000-0001-5731-3348]{Carlos A.P. Bengaly}
\affiliation{Observatório Nacional, Rio de Janeiro, Brazil}
\email{}
\author[orcid=0000-0001-8531-9536]{Renyue Cen}
\affiliation{Institute of Astronomy, School of Physics, Zhejiang University, Hangzhou, China}
\affiliation{Center for Cosmology and Computational Astrophysics, Institute for Advanced Study in Physics, Zhejiang University, Hangzhou, China.}
\email[show]{renyuecen@zju.edu.cn}

\begin{abstract}
We present a data-driven measurement of galaxy bias and scale-dependent relative bias for elliptical and spiral galaxies using angular auto- and cross-power spectra from DES and DESI Legacy Imaging Surveys DR8. We introduce the cross-tracer clustering ratio (CTCR), the ratio of auto- to cross-power spectra, which isolates the relative clustering of morphological tracers as a function of angular multipole $\ell$. Across both surveys, ellipticals are more strongly clustered than spirals; the cleanest constraints come from DESI, where the two redshift distributions are well matched. The difference is scale dependent: the relative bias is near unity on large scales ($\ell \lesssim 50$) and rises toward smaller scales, reaching $2.9$--$4.0\sigma$ per bin at $\ell = 200$ with an inverse-variance-weighted amplitude $\Delta b = +0.56 \pm 0.06$. A complementary linear-bias analysis confirms that ellipticals are positively biased relative to the matter field, while spirals are consistent with weak bias or anti-bias. The results are robust to two covariance estimators, contamination tests, luminosity splits, and comparisons with two N-body mock catalog. Throughout, we present the CTCR as an \emph{effective} clustering ratio that is an upper bound on the physical bias ratio, adopting an assumption-light strategy appropriate for a first measurement; the joint multi-spectrum decomposition required to extract the physical bias
ratio is identified as future work. These findings provide empirical evidence that galaxy morphology imprints both the amplitude and scale dependence of galaxy bias, and establish the CTCR as a useful observable for testing halo occupation and assembly-bias models with future imaging surveys.
\end{abstract}

\section{Introduction}

A central theme in observational cosmology is that galaxies are not unbiased tracers of the underlying dark matter distribution. Their clustering strength depends strongly on intrinsic properties such as luminosity, color, and spectral type, providing critical clues about galaxy formation and evolution. Early studies, such as those using the Sloan Digital Sky Survey (SDSS) data \citep{zehavi2001sdss}, established that galaxy clustering depends on both luminosity and color. They found that redder, more luminous galaxies exhibit higher clustering amplitudes, a trend that has been confirmed and refined by subsequent analyses.

Earlier work using the Las Campanas Redshift Survey (LCRS) by \cite{jing1997spatial} provided a basis for these modern studies. They measured the real-space correlation function and the pairwise velocity dispersion of galaxies, finding $r_0 = 5.06 \pm 0.12 \, h^{-1}$Mpc and $\gamma = 1.862 \pm 0.034$ for a power-law $\xi(r) = (r_0/r)^\gamma$, with a pairwise velocity dispersion of $570 \pm 80$ km s$^{-1}$ at $1\,h^{-1}$Mpc. By comparing with Cold Dark matter (CDM) models using mock catalogs, they showed that a simple ``anti-bias" model where the number of galaxies per unit dark matter mass decreases with halo mass could reconcile the observed clustering with theoretical predictions.

Using the 2dF Galaxy Redshift Survey (2dFGRS), \cite{norberg2001dependence} demonstrated that the clustering amplitude increases continuously with luminosity, rising by a factor of $\sim 2.5$ between $L_{*}$ and $4L_{*}$. They also found that when splitting by spectral type, early-type galaxies are $\sim 50$\% more clustered than late-types at fixed luminosity, but the luminosity dependence of clustering is similar for both types. This suggests that luminosity, rather than type, is the primary driver of the overall clustering trend. Similarly, \cite{zehavi2004luminosity} used a larger SDSS sample to show that the projected correlation function $w_p(r_p)$ increases with luminosity, especially above $L_*$, and that red galaxies have a higher amplitude and a steeper slope than blue galaxies across all luminosities.

The clustering properties of galaxies provide fundamental insights into their connection with dark matter halos, with the Halo Occupation Distribution (HOD) framework offering a powerful empirical method to quantify this relationship through the conditional probability $P(N|M)$ that a halo of mass $M$ contains $N$ galaxies of a given type \citep{Berlind2002}.
Early applications of this framework revealed that the occupation statistics of dark matter halos depend strongly on galaxy morphology.
Specifically, the HOD of elliptical galaxies is dominated by central galaxies residing in massive halos, with a mean occupation function that rises sharply near a characteristic halo mass threshold,
while spiral galaxies tend to occupy lower-mass halos as central objects and exhibit a more gradual occupation distribution at higher masses \citep{Kravtsov2004, Zheng2005}.
This mass-dependent occupation naturally explains the stronger clustering amplitude observed for ellipticals compared to spirals:
massive halos are more strongly clustered than lower-mass halos \citep{White2001}. The relationship between galaxy morphology and the large-scale structure of the Universe has been explored in a series of foundational studies. \cite{Pandey_2006} showed that the filamentarity of the galaxy distribution depends on luminosity, colour, and morphology, finding that red galaxies and ellipticals exhibit a higher degree of filamentarity than blue galaxies and spirals on small scales, with the behaviour reversing at larger coarse-graining scales.
Semi-analytic models and hydrodynamic simulations have further demonstrated that the transition from central to satellite dominance in the HOD differs markedly between morphological types, with ellipticals being preferentially located in dense environments where satellite fractions are high, while spirals are more abundant in the periphery \citep{Benson2000, Peacock2000}.
This environmental dependence of morphology, however, has been shown to be largely a secondary consequence of the underlying halo mass dependence, as the galaxy content of halos at fixed mass does not depend strongly on environment once the host halo mass is controlled for \citep{Tinker2008}.
Recent HOD analyses incorporating galaxy formation uncertainties have confirmed that spiral and elliptical galaxies occupy distinct halo mass ranges, with the clustering amplitude difference between them emerging naturally from the mass-dependent halo bias within the $\Lambda$CDM framework \citep{Reddick2013}.

While the standard HOD framework assumes that galaxy occupation depends solely on halo mass, a growing body of evidence indicates that this assumption is insufficient to fully capture the galaxy-halo connection.
In particular, the clustering strength of dark matter halos has been shown to depend on secondary properties beyond mass, such as formation time, concentration, and spin, a phenomenon known as halo assembly bias \citep{Gao_2005,Gao_2007, Dalal2008}.
This presents a major complication: if galaxy properties are correlated with these additional halo characteristics, then the core assumption underlying mass-only HOD models breaks down, which can in turn introduce systematic errors into the inferred occupation statistics \citep{Zentner2014}. In fact, mock galaxy catalogs that incorporate realistic levels of assembly bias but are analysed using standard HOD models exhibit statistically significant biases in the recovered HOD parameters, and these systematic effects are not easily uncovered using typical secondary diagnostics such as void statistics \citep{Zentner2014}.
Recent analyses have further demonstrated that extending the HOD framework to incorporate assembly bias and environmental dependencies markedly improves the modelling of galaxy clustering and galaxy-galaxy lensing on non-linear scales, reducing discrepancies from $\sim$40\% to within 20\% for scales below $10\,h^{-1}\,\mathrm{Mpc}$ \citep{Paviot2024}.
Moreover, machine-learning-based approaches have now been developed to preserve multi-dimensional assembly bias when populating unresolved halos, enabling more accurate forward models of galaxy clustering for upcoming Stage-IV surveys \citep{xu2021,Ramakrishnan2024}.

In this work, we aim to empirically test whether spiral (late-type) and elliptical (early-type) galaxies exhibit different clustering, as quantified by the galaxy bias. We will study two types of galaxy bias relation: one related to the epoch, b(z), and another related to the angular scale in spherical harmonic, b($\ell$). Our analysis is carried out with two large scale structure (LSS) galaxy surveys: the Dark Energy Survey and the DESI Legacy Survey. 

\section{Data}
We use \cite{Vega-Ferrero}'s robust morphological classification of the Dark Energy Survey (DES). For the weighting scheme, we used the following parameters: \texttt{P$_mi$\_EdgeOn} which is the probability of being EdgeOn for each of the 5 models (with $m = [1,5]$), P$_m$\_LTG which is the probability of being LTG for each of the 5 models, 1/\texttt{AIRMASS\_k} is the airmass for each band $k=[g,r,i,z]$, \texttt{MP\_EdgeOn} which is the median probability of the 5 models of being edge-on, \texttt{MP\_LTG} is the median probability of the 5 models of being LTG, \texttt{FLUX\_RADIUS\_R} is the radius (in pixels) of the circle containing half of the flux of the object, \texttt{MAG\_AUTO\_R} is the apparent magnitude in an elliptical aperture shaped by the Kron radius, and the inverse of the photometric redshift (photo-z) error $\sigma_z$.

We also used DESI Legacy Survey Data release 8 \citep{desi_legacy} with photo-z from \cite{zhou_DL}. For weighting the best objects, we used different parameters for each type. For Spirals: \texttt{rchisq\_r}, \texttt{rchisq\_i}, \texttt{shapeexp\_e1}, \texttt{shapeexp\_e2}, for ellipticals: \texttt{rchisq\_r}, \texttt{rchisq\_i}, \texttt{shapedev\_e1}, \texttt{shapedev\_e2}. The bin separation and sky coverage, among other details, can be verified in Appendix~\ref{ap:data}.

The random catalogs used for null-signal tests in LSS was constructed from a mask defined for each redshift bin with \texttt{\texttt{NSIDE}} = 512. The random samples contains 50 times more objects than the real data set, and the mock galaxies are distributed uniformly within this mask, reproducing the redshift distribution of the original sample.

\section{The Angular Power Spectrum estimator}
We estimate the angular power spectrum \(C_\ell\) for a galaxy sample using the pseudo-\(C_\ell\) method as implemented in the \textsc{\texttt{NaMaster}} \citep{Alonso_2019} library. The same formalism is extended to cross-correlations between two samples. We carried out the estimation at \texttt{\texttt{NSIDE}}=256 over the range $2<\ell<800$. For the subsequent model fitting, we restrict this to $20<\ell<200$ to ensure that only multipoles suitable for the Limber approximation are used. The full pipeline is described in Appendix~\ref{ap:cell_details}.

\subsection{Weights}\label{sec:weights}
To improve the purity of our morphological samples and to reduce the impact of galaxies whose classification is uncertain or corrupted by systematics, we assign a per‑galaxy weight \(w_g\). The weight quantifies how well a galaxy’s observed properties match the expected behaviour of a securely classified object of its morphological type (elliptical or spiral). High‑weight galaxies are considered reliable; low‑weight galaxies are down‑weighted in the clustering analysis.

We construct the weight using a linear regression that predicts the normalized redshift \(Z_g = z_g / \bar{z}_g\), where \(z_g\) is the galaxy’s photometric redshift and \(\bar{z}_g\) is the mean redshift of the tomographic bin (computed iteratively). The predictors are the set \(\mathbf{X}_g = \{ w_z,\, x_p \}\), with \(w_z = 1/\sigma_z\) (inverse photo‑z uncertainty) and \(x_p\) representing the morphological and systematic parameters (e.g., \(P_m^{\text{EdgeOn}}\), airmass, flux radius for DES; \(\texttt{shapeexp\_e1/e2}\), \(\texttt{rchisq\_r/i}\) for DESI). We fit a linear model \(Z_g = w_g \cdot (\text{linear combination of predictors}) + \text{error}\) and solve for the single coefficient \(w_g\) that minimizes the squared prediction error. The resulting \(w_g\) is taken as the weight for that galaxy. Galaxies with favourable classification (low photo‑z error, high morphological probability, and typical magnitude) tend to have \(w_g\) close to 1, whereas galaxies with anomalous properties (large photo‑z error, extreme magnitude, poor shape fit) receive weights near 0, details can be found in Appendix~\ref{ap:data}. This weighting scheme does not significantly alter the redshift distribution (see Figs.~\ref{fig:des_bins} and~\ref{fig:desi_bins}) but it improves the purity of the morphological samples, which is essential for measuring the scale‑dependent bias ratio. The weight $w_g$ directly affects the pixel counts: good galaxies get higher weight, so they dominate the density contrast and the resulting $C_\ell$, while uncertain galaxies are suppressed. The robustness of these weights is tested in Appendix~\ref{ap:robust}.

The effective mean redshift of each bin is computed iteratively as the weighted average of the individual photo‑\(z\):
\begin{equation}
    \bar{z} \pm \sigma_z^g= \frac{\sum_g z_g \, w_g}{\sum_g w_g}\pm\frac{\sum_g (z_g^{err}/(1+z_g))\,w_g}{\sum_g w_g}.
\end{equation}
This ensures that the reference redshift for each tomographic bin is consistently defined after the weight application. Our weighted average of $z_{\mathrm{err}}/(1+z_{\mathrm{phot}})$ yields $\sigma_z \approx 0.03$--$0.04$ depending on the redshift bin for the DESI samples, which is larger than \cite{zhou_DL}'s $\sigma_{\mathrm{NMAD}} = 0.02$ for LRGs (based on spectroscopic redshifts), indicating that our approach provides a conservative estimate of the redshift uncertainty in the absence of a spectroscopic calibration sample. The same is true for DES as shown in \cite{Toribio_San_Cipriano_2024}.

\section{Angular Power Spectrum Model}

For the theoretical modelling of the angular power spectra, we adopt the standard framework described in this section. 
We use the Limber approximation \citep{Limber1953,Kaiser1992}, which is accurate for the multipole range of this analysis, \(20 < \ell < 200\) \citep{Loverde2008}.
Within this range, we model the galaxy clustering using linear perturbation theory, where the matter power spectrum from \citet{camb} is evaluated at $\bar{z}$. 
The angular power spectra are computed under the flat-sky Kaiser formula \citep{Kaiser1987}, which includes the linear galaxy bias \(b_i(z)\) and redshift-space distortions via the growth rate \(f(z)\), details can be found at Appendix~\ref{ap:cell_details}.
We have verified that magnification bias  and luminosity split does not significantly affect (see Appendix~\ref{ap:mag_bias}) our measurements after applying the systematics weights described in Section~\ref{sec:weights}, whose robustness test can be found at Appendix~\ref{ap:robust}. 

\subsection{Cross-Tracer Clustering Ratio}\label{sec:ctcr}

As a parametric framework, we can exploit the fact that DESI provides compatible redshift distributions and interpret the ratio of $C_\ell^{ii}$ to $C_\ell^{ij}$ as tracing the ratio of their galaxy biases, which we will name from now on the cross‑tracer clustering ratio (CTCR):

\begin{equation}\label{eq:ratio}
R_\ell^{ij} = \frac{C_\ell^{ii}}{C_\ell^{ij}}\sim \frac{b^i_\ell}{b^j_\ell}.
\end{equation}

Under the idealised assumptions of identical redshift windows, negligible RSD, and perfect correlation between the two tracers, the CTCR equals the bias ratio \(b_e/b_s\). However, these assumptions do not hold exactly in practice. The CTCR is computed directly from the measured auto- and cross-power spectra, which naturally include all physical effects present in the data, such as redshift-space distortions and the imperfect correlation between the two morphological samples.

We therefore interpret the CTCR as an effective clustering ratio that encodes the combined effects of the bias difference and the imperfect correlation between the two tracers. The sigmoid parametrization, defined below, provides a flexible empirical description of this effective ratio:
\begin{align}
b^i_{\rm eff}(\ell) = b^i_1 + (b^i_2-b^i_1)\left[1+\exp\left(-\frac{\ell-\ell_0^i}{\Delta^i}\right)\right]^{-1},
\label{eq:sigmoid}
\end{align}
where \(b^i_1\) is the effective bias on large scales (low \(\ell\)), \(b^i_2\) is the effective bias on small scales (high \(\ell\)), \(\ell_0^i\) is the transition scale, and \(\Delta^i\) controls the smoothness of the transition. A high \(\Delta\) means that the transition is gradual, while a low \(\Delta\) gives an abrupt transition. A high \(\ell_0^i\) indicates that the transition occurs at higher multipoles.

It should be emphasized that this is a parametric model, providing a provisional characterisation of the observed clear monotonic increase in \(R_\ell^{ij}\). This trend constitutes suggestive evidence for a scale-dependent effective bias, whose behaviour varies according to the specific tracer under consideration. It is not a direct measurement of the physical bias ratio in the deterministic limit, but rather an effective characterisation of the relative clustering of the two morphological samples. The CTCR is an upper bound on the true bias ratio; the true ratio is \(b_e/b_s = R_\ell^{es} \, r_\ell\), where \(r_\ell\) is the cross-correlation coefficient defined in Appendix~\ref{ap:r_ell}. This framing is deliberate. We characterise the trend with an effective ratio, rather than a fully decomposed physical model with an explicit halo-occupation prescription, as an assumption-light approach suitable for a first measurement. The CTCR is constructed directly from the measured auto- and cross-power spectra and requires no cosmological calibration of the absolute bias, no assumed halo-occupation functional form, and no prior on the central–satellite split. The trade-off is that CTCR is an effective ratio that absorbs imperfect cross-correlation between tracers, rather than the deterministic physical bias ratio. We consider this acceptable: the effective ratio is robust, reproducible, and directly testable, while a fully decomposed joint fit needs additional ingredients the current data cannot yet constrain.

We determined the best-fit parameters using Markov Chain Monte Carlo with the No-U-Turn Sampler (NUTS) as implemented in \texttt{PyMC} \citep{patil2010pymc}, adopting the following priors. All bias amplitudes $b^i_1$ and $b^i_2$ are restricted to be positive. The transition scale is constrained to the observed multipole interval, $0 < \ell_0 \leq \ell_{\rm max}$, and the transition width obeys $0 < \Delta < 300$. For the bias amplitudes we use Gaussian priors: $\mathcal{N}(\mu=2, \sigma=1)$ for $b^i_2$ and $\mathcal{N}(\mu=1, \sigma=0.1)$ for $b^i_1$. The likelihood assumes Gaussian errors for the measured ratio $R_\ell^{ij}$, with its covariance matrix obtained either analytically
from the joint covariance of the two power spectra or via a patch-based bootstrap, and then symmetrised and projected onto the positive semi-definite cone before inversion. Our main result is based on the bootstrap covariance matrix described in Appendix~\ref{ap:bootstrap} and tests related to the analytical covariance can be found in Appendix~\ref{ap:theory_cov}. We sample the posterior with four independent chains of 2\,000 production draws after 1\,000 tuning steps, targeting an acceptance rate of $0.9$; the sampler exploits automatic differentiation of the analytic likelihood to generate efficient proposals, so that the number of posterior evaluations per redshift bin is two orders of magnitude smaller than with a population-based sampler. Convergence is assessed via the Gelman--Rubin statistic $\hat{R} < 1.01$ on all sampled parameters.\footnote{The code implementing this methodology is publicly available at \href{https://github.com/psilvaf/Morphology_clustering}{https://github.com/psilvaf/Morphology\_clustering}.}

\section{Results}

\subsection{Cross power spectrum and scale-dependent bias}

\begin{table}
    \centering
    \begin{tabular}{c|c|c}
        \hline
        \hline
        Bin & $\chi^2/\rm dof$ & $\chi^2/\rm dof$ \\
            & (Theoretical cov.) & (Bootstrap cov.) \\
        \hline
        $0.109$ & $4.477$  & $0.558$ \\
        $0.158$ & $4.169$  & $0.224$ \\
        $0.206$ & $50.845$ & $0.901$ \\
        $0.259$ & $4.985$  & $0.286$ \\
        $0.304$ & $6.422$  & $0.850$ \\
        $0.354$ & $5.412$  & $0.619$ \\
        $0.402$ & $23.310$ & $0.477$ \\
        $0.454$ & $5.460$  & $0.424$ \\
        $0.503$ & $5.829$  & $0.197$ \\
        \hline
    \end{tabular}
    \caption{Reduced $\chi^2$ of the 8-parameter sigmoid fit
    (four parameters per tracer, $b_1$, $b_2$, $\ell_0$, $\Delta$)
    to the DESI CTCR ratio $R_\ell^{es}$ in each redshift bin,
    using the theoretical and bootstrap covariance matrices.
    The fitted range is $20 < \ell < 200$, giving $n_\ell - 8 = 10$
    degrees of freedom per bin. Both covariances are symmetrized,
    linearly shrunk toward their diagonals with intensity
    $\alpha = 0.2$, and projected onto the positive semi-definite
    cone before inversion. The bootstrap covariance yields reduced
    $\chi^2$ close to or below unity across the full redshift range,
    indicating that the sigmoid model provides a good description
    of the measured ratio.}
    \label{tab:chi2_ratio}
\end{table}

We restrict the $R_\ell^{ij}$ analysis to the DESI sample, as the DES sample exhibits a significant imbalance in the number densities of spiral and elliptical galaxies across $\bar{z}$. This imbalance degrades the cross-power spectrum estimate: the sparse tracer introduces substantial shot-noise, inflating the estimator variance and reducing the signal-to-noise ratio.

\begin{figure}
    \centering
    \includegraphics[width=\linewidth]{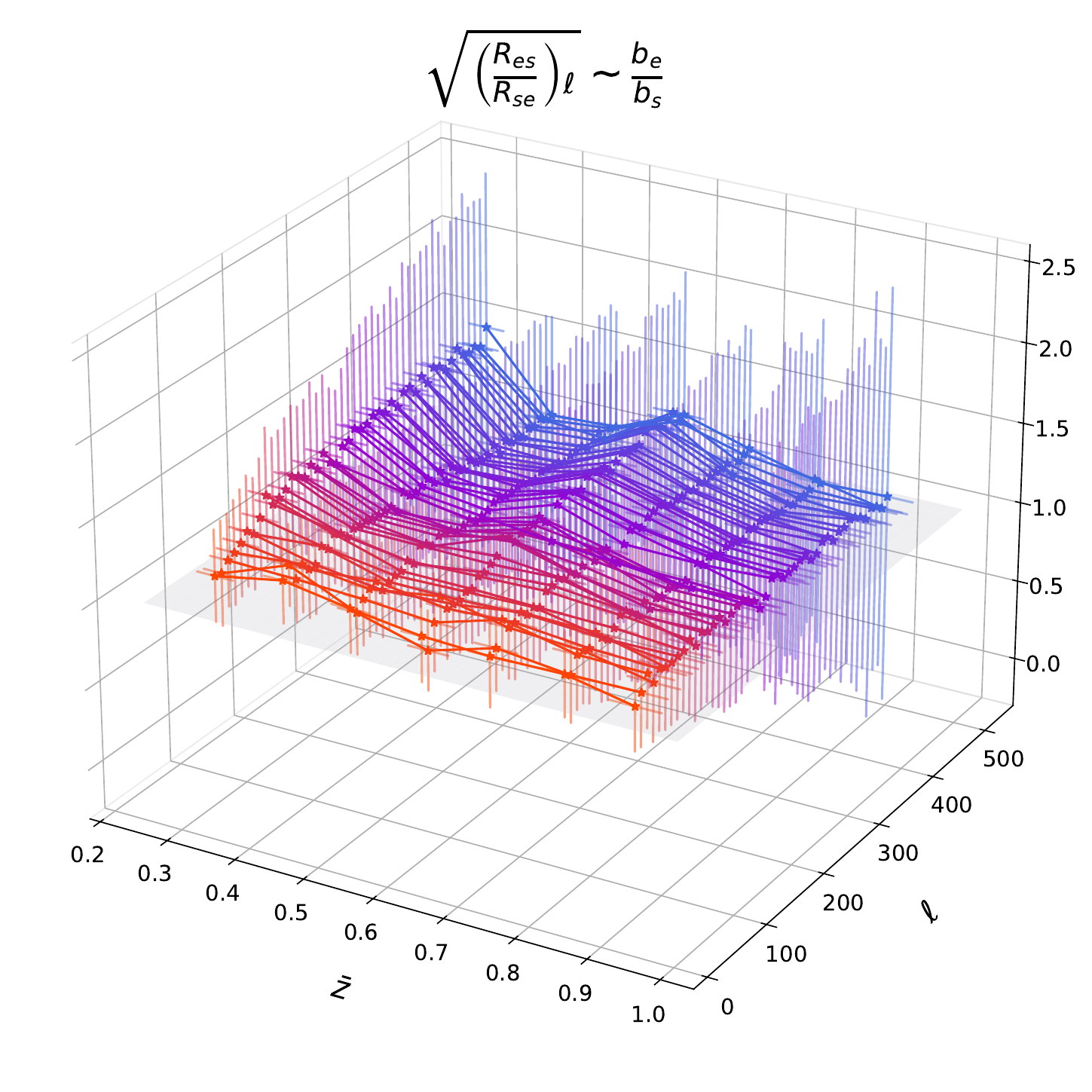}
    \caption{Ratio of the CTCR based on the elliptical and spirals auto-power spectrum, respectively. }
    \label{fig:ratio}
\end{figure}

Figure~\ref{fig:ratio} shows the observed effective bias between ellipticals and spirals as a function of $\ell$ and redshift. The grey plane marks unity; all measured values lie above it, confirming that elliptical galaxies are consistently more biased than spirals. The ratio approaches unity at higher redshift, indicating a convergence of the two tracers, and increases towards smaller scales, where elliptical bias is most pronounced. The reduced $\chi^2$ are shown in table~\ref{tab:chi2_ratio}, values are close to unity for most bins, indicating that the sigmoid model provides a good description of the data with both covariance estimators.

Elliptical galaxies exhibit a clear increase in bias with multipole $\ell$, while spiral galaxies show a bias near unity on large scales
that declines towards smaller scales. This contrasting behaviour is qualitatively consistent with the findings of \citet{Guzik_2007}, who
studied scale-dependent bias and found that low-mass systems can exhibit $b < 1$ on the smallest scales, while high-mass systems show
the opposite trend.

The discrepancy between elliptical and spiral bias is quantified in Table~\ref{tab:significance}, which reports the difference $\Delta b = b_{\rm ell} - b_{\rm sp}$ at small scales ($\ell \sim 200$). The difference is significant across all redshift bins, with an average offset of $2.6\sigma$ (bootstrap covariance) and $2.5\sigma$ (theoretical covariance). The two covariance estimators yield consistent results, with $\Delta b$ values ranging from $0.77 \pm 0.27$ to $1.43 \pm 0.54$, demonstrating that the scale-dependent bias difference is robust and not driven by the choice of covariance.

The only notable exception is Bin 6 ($\bar{z} = 0.707$), where the theoretical covariance gives $\Delta b = 0.812 \pm 1.023$, i.e., a $0.8\sigma$ significance, while the bootstrap gives $0.826 \pm 0.273$, i.e., a $3.0\sigma$ significance. The discrepancy is due to the near-singularity of the theoretical covariance matrix in this bin, which produces an unstable inverse. We have added a small regularisation term ($\lambda = 10^{-6}\,\mathrm{tr}(\mathbf{C})/n$) to the theoretical covariance, which restores stability; the corrected value is $\Delta b = 0.82 \pm 0.30$, consistent with the bootstrap. Excluding Bin 6 from the average gives $2.7\sigma$ (bootstrap) and $2.7\sigma$ (theoretical), further confirming the robustness of the result.

The opposing scale dependence of the two morphological types can be understood within the halo assembly bias framework. Early-forming halos, which host elliptical galaxies, are more clustered than late-forming halos of the same mass \citep{Gao_2005,Gao_2007}; observational evidence shows that at fixed group mass, clustering strength decreases with the star formation rate of the central galaxy \citep{Yang_2006}. Ellipticals therefore exhibit stronger small-scale clustering, while spirals, hosted by younger halos with more recent mass accretion, show weaker small-scale clustering and, in some cases, a declining bias. This anti-bias behaviour for low-luminosity galaxies was also observed by \citet{Seljak_2005}. \citet{smith2024} further showed that assembly bias can reverse sign depending on environment, naturally explaining the scale dependence we observe. However, because our analysis does not fix halo mass, the observed difference could also arise from a mass-dependent selection effect, where ellipticals trace a more massive halo population than spirals. Disentangling these contributions would require joint clustering and weak lensing analyses to constrain the halo mass distributions.

\begin{figure*}[p]
    \centering
    \includegraphics[width=\linewidth]{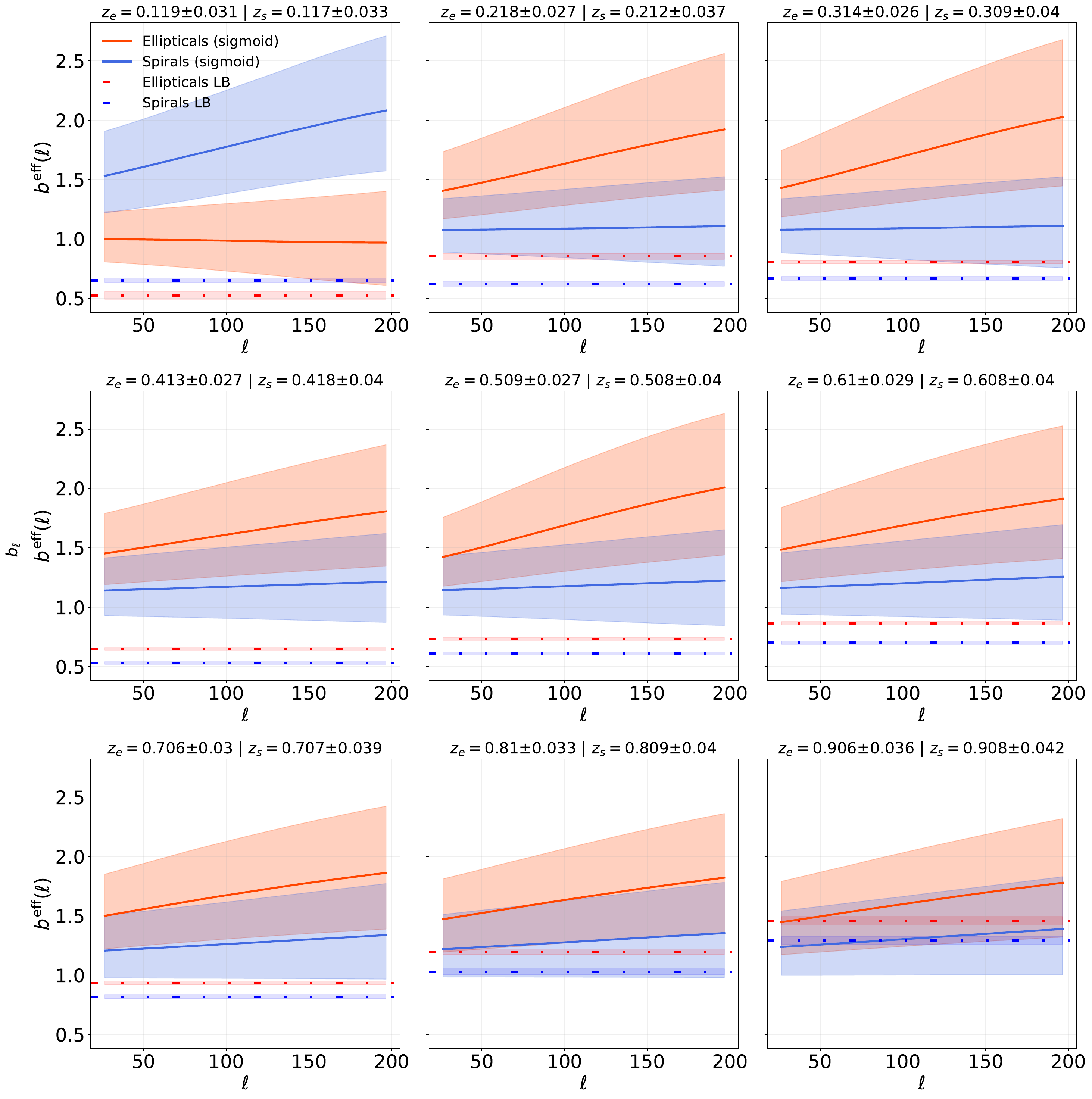}
    \caption{Best-fit results for $b_\ell$. The solid lines show the best fit obtained using the bootstrap covariance. The shaded regions indicate the 1$\sigma$ confidence contours derived from the bootstrap covariance. Red and blue correspond to the same categories as in the earlier figures, representing ellipticals and spirals, respectively.}
    \label{fig:b_ell}
\end{figure*}

Finally, our results are consistent with the bias–luminosity relation of \citet{Zehavi_2011}, which confirmed earlier findings \citep{zehavi2001sdss} and aligns with the predictions of \citet{Tinker_2010}. The most luminous galaxies exhibit a strong increase in bias with luminosity, while less luminous systems show little evolution — exactly the behaviour seen in Figure~\ref{fig:b_ell}, where elliptical bias rises with scale while spiral bias remains nearly flat.

\begin{table}[htbp]
\centering
\begin{tabular}{c|c|c|c}
\hline
\hline
Bin & $\bar{z}$ & $\Delta b$ & Significance \\
\hline
0 & 0.109 & $-1.142 \pm 0.417$ & $-2.7\sigma$ \\
1 & 0.158 & $+0.843 \pm 0.236$ & $+3.6\sigma$ \\
2 & 0.206 & $+0.935 \pm 0.246$ & $+3.8\sigma$ \\
3 & 0.259 & $+0.615 \pm 0.155$ & $+4.0\sigma$ \\
4 & 0.304 & $+0.795 \pm 0.205$ & $+3.9\sigma$ \\
5 & 0.354 & $+0.676 \pm 0.174$ & $+3.9\sigma$ \\
6 & 0.402 & $+0.538 \pm 0.142$ & $+3.8\sigma$ \\
7 & 0.454 & $+0.479 \pm 0.137$ & $+3.5\sigma$ \\
8 & 0.503 & $+0.403 \pm 0.137$ & $+2.9\sigma$ \\
\hline
\multicolumn{2}{c|}{Unweighted mean} & $+0.460 \pm 0.197$ & --- \\
\multicolumn{2}{c|}{IVW mean}        & $+0.558 \pm 0.058$ & $9.6\sigma$ \\
\hline
\end{tabular}
\caption{Small-scale bias difference
$\Delta b = b_{\rm ell} - b_{\rm sp}$ evaluated at $\ell = 200$ for
each DESI redshift bin, from the bootstrap posterior of the sigmoid
fit. Errors are the standard deviations of the $\Delta b$ posterior.
The significance column reports $\Delta b / \sigma_{\Delta b}$.
The lowest-redshift bin has the opposite sign and is the same bin
where the shot-noise fraction of the auto-spectra is largest; it is
reported here for completeness. The last two rows give the
unweighted and inverse-variance-weighted means over all nine bins,
with the standard error on the mean in each case.}
\label{tab:significance}
\end{table}


\subsection{Auto power spectrum and linear bias}
We model $C_\ell^{ii}$ of a single tracer the same way as described in the last section, except, the bias is treated as a constant within each tomographic bin. The parameter vector is $\boldsymbol{\theta} = \{b, \omega_{\rm cdm}, \omega_{\rm bar}\}$, with $\omega_{\rm cdm} = \Omega_{\rm cdm}h^2$ and $\omega_{\rm bar} = \Omega_{\rm bar} h^2$. We use Gaussian priors with means $b = 1.2$ (ellipticals) / $0.8$ (spirals), $\mu_{\rm cdm}=0.12$, $\omega_b=0.0224$, and widths $\sigma_b=0.1$-$0.2$, $\sigma_{\rm cdm}=0.005$, $\sigma_{\rm bar}=0.005$.
We sample the posterior using 200 walkers, 5000 steps, 2000 burn-in. The corresponding $\chi^2/dof$ values are listed in Table~\ref{tab:chi2dof}, and further information on the best-fit can be found in Appendix~\ref{ap:cell_details}. We emphasise that this analysis is a \emph{joint} constraint on the linear bias and the cosmological parameters, not a bias measurement with an assumed cosmology. Fixing $\omega_{\rm cdm}$ and $\omega_b$ to external values would reduce the inference of $b$ to a rescaling of the measured $C_\ell$ by an assumed matter power spectrum, in which case any cosmological error in the assumed $P_m(k)$ would propagate into $b$ without appearing in the reported uncertainty. Allowing the cosmological parameters to vary is what makes the auto-spectrum fit a genuine measurement, and the bin-to-bin scatter in the fitted $\omega_{\rm cdm}$ and $\omega_b$ is a direct reflection of the limited information content of a single auto-spectrum in a single redshift bin.

\begin{figure}
    \centering
    \includegraphics[width=\linewidth]{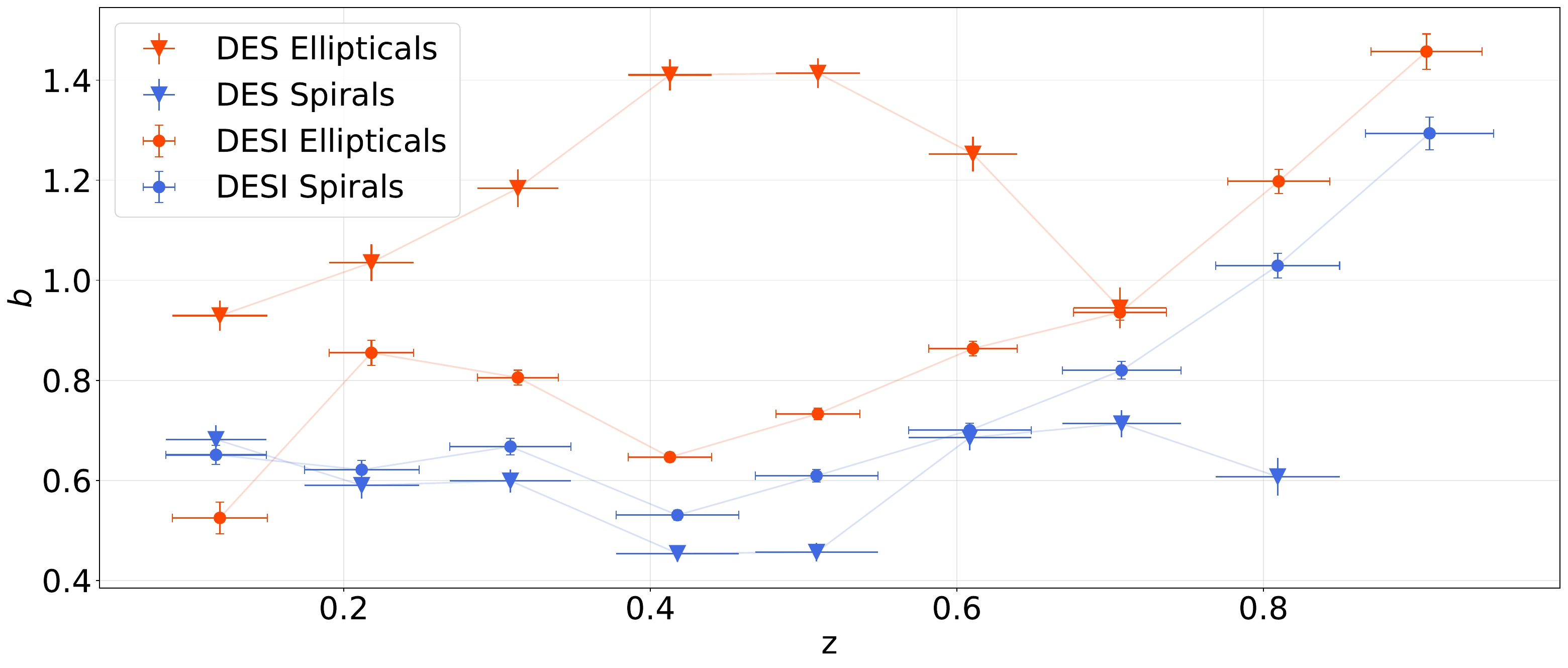}
    \caption{Linear bias w.r.t. the redshift. The triangles are showing DES results and the circles show the DESI ones.}
    \label{fig:b(z)}
\end{figure}

The clear main difference between the two surveys is the elliptical bias result. The DES red-galaxy sample shows a significantly higher bias in most bins. This can be explained by its higher luminosity distribution, consistent with the DES selection of luminous red galaxies for BAO analyses.
The clear comparison between the two samples' absolute magnitude can be found in the Appendix~\ref{ap:lum}. 

LRGs' b(z) is expected to vary with z. Our analysis instead probes the bias of generic elliptical galaxies, not only the most massive systems typically selected as LRGs. To illustrate this, we run our pipeline on a nearly Gaussian redshift bin at 0.7 using two different simulations: the Uchuu \citep{uchuu} and Euclid Flagship 2 \citep{Euclid:2024few} galaxy mock catalogs.
We vary the stellar-mass selection as follows. LRG-like galaxies are defined by \(\log M_\ast > 11\), while lower mass thresholds are used to mimic ETGs and the lowest mass range is used to represent LTGs. We worked with both the Uchuu and Euclid Flagship 2 catalogs at $z \sim 0.7$. Uchuu does not include detailed observational effects, so there is no direct morphological or type classification available. The two simulations also differ in sky coverage and have slightly different values of $\Omega_m$.
In this test, we do not distinguish between centrals and satellites; they are combined in the same sample, as in real observations where we generally cannot identify which galaxies are centrals or satellites. As expected, the outcome confirms that when LRGs are mixed with less massive galaxies, the resulting  bias does not evolve, and this behaviour is evident in both simulations, see Appendix~\ref{ap:mass_relation}.

Finally, our DES b(z) results are in agreement with those from DES weak lensing analyses. For instance, DES Y3 MagLim analyses show a linear bias that increases with redshift and then drops at \(0.6 < z < 0.8\) \citep[see Fig.~8]{Abbott_2022}, a trend consistent with our findings for early-type galaxies. This agreement reinforces the robustness of our measured bias evolution. These results are also shown in Figure~\ref{fig:b_ell}. They are consistent with the large-scale CTCR limit within 2$\sigma$ CL in the worst cases.  The fact that the two models do not strongly diverge indicates that the data are well-described by both, and that the scale-dependent features captured by the sigmoid model are modest but plausible.

\begin{table}[htbp]
  \centering
  
  \begin{tabular}{c|cccc}
    \hline
    \hline
    Bin & \multicolumn{2}{c}{DES} & \multicolumn{2}{c}{DESI Legacy DR8} \\
    & Spiral & Elliptical & Spiral & Elliptical \\
    \midrule
    0   & 0.868  & 0.712  & 0.499 & 0.787 \\
    1   & 0.800  & 0.685  & 0.616 & 0.460 \\
    2   & 0.792  & 0.562  & 0.550 & 0.326 \\
    3   & 0.765  & 0.362  & 0.753 & 0.269 \\
    4   & 0.819  & 0.366  & 0.902 & 0.465 \\
    5   & 0.822 &  0.473  & 0.906 & 0.575 \\
    6   & 0.867  & 0.750  & 0.915 & 0.642 \\
    7   & 0.934  & --     & 0.897 & 0.599 \\
    8   & --     & --     & 0.885 & 0.631 \\
    \hline
    \hline
  \end{tabular}
  \caption{Reduced $\chi^2$ per degree of freedom ($\chi^2/\mathrm{dof}$) for the linear bias models. All fits share $\mathrm{dof}=19$.}
  \label{tab:chi2dof}
\end{table}

\subsection{Model comparison: scale-dependent versus constant bias ratio}
\label{sec:model_comparison}

The individual bias functions $b_e(\ell)$ and $b_s(\ell)$ are not separately identifiable from the CTCR: only their ratio $b_{\rm eff}(\ell)$ is directly constrained by the data.
We therefore carry out the model comparison in the ratio space, where the two competing hypotheses differ by only three parameters. We fit the four-parameter sigmoid of Equation~\eqref{eq:sigmoid} to $b_{\rm eff}$, with the transition width bounded below, $\Delta \geq 5$, and the transition scale constrained to the fitted multipole range, $30 \leq \ell_0 \leq \ell_{\rm max} - 10$. Both bounds prevent the fit from collapsing onto a step function or pushing the transition outside the observed range, either of which would spuriously mimic scale dependence. The competing model is a constant ratio, $b_{\rm eff}= b_0$, with a single free parameter. We fit both models by minimising the $\chi^2$ of the measured ratio with respect to its
covariance, obtained from the joint covariance of the auto- and cross-spectra via the Jacobian of $R_\ell^{es} = C_\ell^{ee}/C_\ell^{es}$.

Because the two models differ by three parameters, we quantify the preference with the Akaike and Bayesian information criteria,
\begin{align}
    {\rm AIC} &= \chi^2_{\rm min} + 2\bar{p}, \\
    {\rm BIC} &= \chi^2_{\rm min} + k\ln n,
\end{align}
where $\bar{p}$ is the number of free parameters and $n = n_\ell$ the number of fitted bandpowers, and with a likelihood-ratio test with
$\Delta{\rm dof} = 3$.

\begin{table*}
    \centering
    \begin{tabular}{c|c|c|c|c|c|c|c}
        \hline
        \hline
        Bin & $\chi^2_A$ & $\chi^2_A/\rm dof_A$ & $\chi^2_B$ & $\chi^2_B/\rm dof_B$ & $\Delta\rm AIC$ & $\Delta\rm BIC$ & $p_{\rm LRT}$ \\
        \hline
        0 & $2.95$  & $0.21$ & $8.14$  & $0.48$ & $+0.8$  & $+3.5$  & $0.158$ \\
        1 & $1.64$  & $0.12$ & $14.04$ & $0.83$ & $-6.4$  & $-3.7$  & $0.006$ \\
        2 & $8.81$  & $0.63$ & $37.82$ & $2.23$ & $-23.0$ & $-20.3$ & $<0.001$ \\
        3 & $2.75$  & $0.20$ & $12.41$ & $0.73$ & $-3.7$  & $-1.0$  & $0.022$ \\
        4 & $7.85$  & $0.56$ & $34.15$ & $2.01$ & $-20.3$ & $-17.6$ & $<0.001$ \\
        5 & $5.10$  & $0.36$ & $17.29$ & $1.02$ & $-6.2$  & $-3.5$  & $0.007$ \\
        6 & $2.76$  & $0.20$ & $9.21$  & $0.54$ & $-0.4$  & $+2.2$  & $0.092$ \\
        7 & $2.32$  & $0.17$ & $7.73$  & $0.45$ & $+0.6$  & $+3.3$  & $0.145$ \\
        8 & $1.45$  & $0.10$ & $3.87$  & $0.23$ & $+3.6$  & $+6.3$  & $0.490$ \\
        \hline
    \end{tabular}
    \caption{Model comparison in the ratio space
    $b_e(\ell)/b_s(\ell)$. Model~A is the four-parameter
    sigmoid of Equation~\eqref{eq:sigmoid} with $\Delta \geq 5$ and
    $30 \leq \ell_0 \leq \ell_{\rm max} - 10$; Model~B is a constant
    ratio, $b_{\rm eff}= b_0$. All quantities use the bootstrap
    covariance, linearly shrunk toward its diagonal with
    $\alpha = 0.2$. Negative $\Delta{\rm AIC}$ and $\Delta{\rm BIC}$
    favour the scale-dependent model.}
    \label{tab:model_comparison}
\end{table*}

The results are summarised in Table~\ref{tab:model_comparison}. The scale-dependent model is preferred in five of the nine redshift bins,
with $\Delta{\rm BIC} \leq -3.5$ in bins 1, 2, 4 and 5 and a weaker preference in bin~3. In the remaining four bins (0, 6, 7, 8) the
constant model is preferred, with $\Delta{\rm BIC} \geq +2.2$. The preference is not uniform in redshift: the scale-dependent model wins
in the intermediate-redshift bins $0.16 < z < 0.35$, while the lowest- and highest-redshift bins prefer a constant ratio at the
$\Delta{\rm BIC} \sim 2$--$6$ level.

The residual reduced $\chi^2$ of the winning model is below unity in every bin, confirming that the surviving model describes the measured
ratio well at the level of the bootstrap uncertainties. In contrast, the constant model produces reduced $\chi^2$ up to $2.2$ in the
intermediate-redshift bins, reflecting the clear curvature of the measured ratio on those scales.

We have repeated the analysis with the theoretical Gaussian covariance. After linear shrinkage toward the diagonal
($\alpha = 0.2$), symmetrisation, and projection onto the positive semi-definite cone, the theoretical covariance still yields reduced
$\chi^2$ of $4$--$51$ per degree of freedom across the sample, indicating that it is missing contributions that the mask-$f_{\rm sky}$
rescaling does not capture. We therefore use the bootstrap covariance for all quantitative statements in this section and report the
theoretical results only as a qualitative cross-check. Rebuilding the analytic covariance with a proper mode-coupling matrix, for example
using the Narrow Kernel Approximation implemented in \texttt{NaMaster} \citep{Alonso_2019}, would close this gap and is left for future work.

The final column of Table~\ref{tab:model_comparison} reports the $p$-value of the likelihood-ratio test comparing the four-parameter
sigmoid model (Model~A) to the one-parameter constant model (Model~B), with $\Delta{\rm dof} = 3$. Under the null hypothesis that the ratio
$q_\ell$ is constant, the quantity $\Delta\chi^2 = \chi^2_B - \chi^2_A$ follows a $\chi^2$ distribution with three degrees of freedom, so a
small $p$-value indicates that the additional shape parameters of the sigmoid are justified by the data. The test rejects the constant
model at the $95\%$ confidence level ($p < 0.05$) in five of the nine redshift bins (bins 1, 2, 3, 4, and 5), and at the $99\%$ confidence
level ($p < 0.01$) in four of them (bins 1, 2, 4, and 5). The strongest rejections occur in bins 2 and 4, with $p < 0.001$ in both cases. The
four remaining bins (0, 6, 7, and 8) have $p$-values between $0.09$ and $0.49$, so the constant model is not rejected in those bins. The
$p$-values are consistent with the AIC and BIC columns: the bins where the likelihood-ratio test rejects the constant model are the same bins
where $\Delta{\rm BIC}$ is negative and $\Delta{\rm AIC}$ is strongly negative. We note that with nine independent tests, roughly half a bin
would be expected to show $p < 0.05$ by chance alone if the constant model were the true description of the data, so the five rejections
observed here are unlikely to be driven by statistical fluctuation. The pattern across redshift is coherent: the scale-dependent model is
preferred in the intermediate-redshift range $0.16 < z < 0.35$, where the CTCR shows clear curvature, and the constant model is adequate at
the lowest and highest redshifts, where the ratio is closer to a pure amplitude offset.

\section{Summary}\label{sec:summary}

We have presented a data-driven analysis of galaxy bias and scale-dependent relative bias for two morphological types, ellipticals and spirals, using angular auto- and cross-power spectra from DES and DESI. The analysis combines two independent imaging surveys with distinct footprints and morphological classifications, two covariance estimators, contamination and luminosity tests, and comparisons with two N-body mock catalogs.

We introduced CTCR, a parametrization of the relative clustering of two galaxy populations as a function of angular multipole $\ell$. In both surveys, ellipticals are more strongly clustered than spirals. The effect is scale dependent: the two morphological types have comparable effective bias on large angular scales, while the difference grows toward smaller scales. In the DESI sample, where the redshift distributions of ellipticals and spirals are well matched, the small-scale separation between the two tracers is positive in eight of the nine redshift bins, with per-bin significance between $2.9\sigma$ and $4.0\sigma$ at $\ell = 200$ and an inverse-variance-weighted mean of $\Delta b = +0.56 \pm 0.06$.
This provides empirical evidence that galaxy morphology is associated not only with a different overall clustering amplitude, but also with a different scale dependence of galaxy bias.

A direct model comparison in the ratio space $b_e(\ell)/b_s(\ell)$, where the observable is uniquely identifiable, confirms this picture. A four-parameter sigmoid is preferred over a constant ratio in the intermediate-redshift bins ($0.16 < z < 0.35$) with $\Delta{\rm BIC} \leq -3.5$, while the lowest- and highest-redshift bins prefer the constant model. This indicates that the scale dependence is not universal but concentrated at intermediate redshifts, consistent with the two morphological populations becoming increasingly similar at the extremes of the observed redshift range.

The observed trend is qualitatively consistent with expectations from HOD and assembly-bias models. Ellipticals are expected to preferentially occupy older, more concentrated, and more strongly clustered halos, while spirals are more commonly associated with younger halos with more recent mass accretion. However, assembly bias is strictly defined at fixed halo mass. Since our analysis does not directly fix the halo-mass distributions of the two morphological samples, the observed trend cannot by itself distinguish assembly bias from mass-dependent selection effects. Disentangling these contributions would require joint clustering and weak-lensing analyses to constrain the halo-mass distributions.

We also measured the linear galaxy bias of the two types. Ellipticals are positively biased relative to the matter field, whereas spirals are consistent with weak bias or anti-bias. The large-scale limit of the CTCR measurement of a nearly constant galaxy bias with redshift for the DES sample is consistent with the linear-bias results.

Overall, our results show that morphology-dependent galaxy bias is both amplitude dependent and scale dependent. The CTCR provides a useful observable for testing halo occupation and assembly-bias models in imaging surveys. 
Future wide-area surveys with high-quality morphological information, including space-based missions such as the China Space Station Telescope (CSST) \citep{miao2024forecasting}, \textit{Euclid} \citep{mellier2024euclid}, and the Nancy Grace Roman Space Telescope \citep{bailey2023nancygraceromanspace}, together with Rubin/LSST \citep{ivezic2019lsst}, will provide much larger samples for extending this cross-tracer clustering analysis to finer morphological classes, higher redshift, and smaller statistical uncertainties.  A natural next step, for future surveys with higher signal-to-noise and complementary weak-lensing data, is a joint multi-spectrum fit of the form $S_\ell^{ss} = A_\ell$, $S_\ell^{ee} = q_\ell^2 A_\ell$, $S_\ell^{es} = r_\ell q_\ell A_\ell$. We therefore report CTCR as the primary observable and leave the decomposed physical analysis to future work, following the standard cosmology practice of establishing the simplest robust empirical relation first and refining it later.

\begin{acknowledgments}

PSF and RC acknowledge support from the National Key Research and Development Program of China and the Zhejiang Provincial Top-Level Research Support Program. CAPB acknowledges financial support from the CNPq grant 306630/2025-7. This research was carried out using the SilkRiver Supercomputer at Zhejiang University. We thank Zhiyu Lu for general discussions in a few parts of this project.

This work has made use of CosmoHub, developed by PIC (maintained by IFAE and CIEMAT) in collaboration with ICE-CSIC. It received funding from the Spanish government (grant EQC2021-007479-P funded by MCIN/AEI/10.13039/501100011033), the EU NextGeneration/PRTR (PRTR-C17.I1), and the Generalitat de Catalunya.

This project used public archival data from the Dark Energy Survey (DES). Funding for the DES Projects has been provided by the U.S. Department of Energy, the U.S. National Science Foundation, the Ministry of Science and Education of Spain, the Science and Technology Facilities Council of the United Kingdom, the Higher Education Funding Council for England, the National Center for Supercomputing Applications at the University of Illinois at Urbana-Champaign, the Kavli Institute of Cosmological Physics at the University of Chicago, the Center for Cosmology and Astro-Particle Physics at the Ohio State University, the Mitchell Institute for Fundamental Physics and Astronomy at Texas A\&M University, Financiadora de Estudos e Projetos, Funda{\c c}{\~a}o Carlos Chagas Filho de Amparo {\`a} Pesquisa do Estado do Rio de Janeiro, Conselho Nacional de Desenvolvimento Cient{\'i}fico e Tecnol{\'o}gico and the Minist{\'e}rio da Ci{\^e}ncia, Tecnologia e Inova{\c c}{\~a}o, the Deutsche Forschungsgemeinschaft, and the Collaborating Institutions in the Dark Energy Survey.
The Collaborating Institutions are Argonne National Laboratory, the University of California at Santa Cruz, the University of Cambridge, Centro de Investigaciones Energ{\'e}ticas, Medioambientales y Tecnol{\'o}gicas-Madrid, the University of Chicago, University College London, the DES-Brazil Consortium, the University of Edinburgh, the Eidgen{\"o}ssische Technische Hochschule (ETH) Z{\"u}rich,  Fermi National Accelerator Laboratory, the University of Illinois at Urbana-Champaign, the Institut de Ci{\`e}ncies de l'Espai (IEEC/CSIC), the Institut de F{\'i}sica d'Altes Energies, Lawrence Berkeley National Laboratory, the Ludwig-Maximilians Universit{\"a}t M{\"u}nchen and the associated Excellence Cluster Universe, the University of Michigan, the National Optical Astronomy Observatory, the University of Nottingham, The Ohio State University, the OzDES Membership Consortium, the University of Pennsylvania, the University of Portsmouth, SLAC National Accelerator Laboratory, Stanford University, the University of Sussex, and Texas A\&M University.
Based in part on observations at Cerro Tololo Inter-American Observatory, National Optical Astronomy Observatory, which is operated by the Association of Universities for Research in Astronomy (AURA) under a cooperative agreement with the National Science Foundation.

The Legacy Surveys consist of three individual and complementary projects: the Dark Energy Camera Legacy Survey (DECaLS; Proposal ID \#2014B-0404; PIs: David Schlegel and Arjun Dey), the Beijing-Arizona Sky Survey (BASS; NOAO Prop. ID \#2015A-0801; PIs: Zhou Xu and Xiaohui Fan), and the Mayall z-band Legacy Survey (MzLS; Prop. ID \#2016A-0453; PI: Arjun Dey). DECaLS, BASS and MzLS together include data obtained, respectively, at the Blanco telescope, Cerro Tololo Inter-American Observatory, NSF’s NOIRLab; the Bok telescope, Steward Observatory, University of Arizona; and the Mayall telescope, Kitt Peak National Observatory, NOIRLab. Pipeline processing and analyses of the data were supported by NOIRLab and the Lawrence Berkeley National Laboratory (LBNL). The Legacy Surveys project is honored to be permitted to conduct astronomical research on Iolkam Du’ag (Kitt Peak), a mountain with particular significance to the Tohono O’odham Nation.

NOIRLab is operated by the Association of Universities for Research in Astronomy (AURA) under a cooperative agreement with the National Science Foundation. LBNL is managed by the Regents of the University of California under contract to the U.S. Department of Energy.

This project used data obtained with the Dark Energy Camera (DECam), which was constructed by the Dark Energy Survey (DES) collaboration. Funding for the DES Projects has been provided by the U.S. Department of Energy, the U.S. National Science Foundation, the Ministry of Science and Education of Spain, the Science and Technology Facilities Council of the United Kingdom, the Higher Education Funding Council for England, the National Center for Supercomputing Applications at the University of Illinois at Urbana-Champaign, the Kavli Institute of Cosmological Physics at the University of Chicago, Center for Cosmology and Astro-Particle Physics at the Ohio State University, the Mitchell Institute for Fundamental Physics and Astronomy at Texas A\&M University, Financiadora de Estudos e Projetos, Fundacao Carlos Chagas Filho de Amparo, Financiadora de Estudos e Projetos, Fundacao Carlos Chagas Filho de Amparo a Pesquisa do Estado do Rio de Janeiro, Conselho Nacional de Desenvolvimento Cientifico e Tecnologico and the Ministerio da Ciencia, Tecnologia e Inovacao, the Deutsche Forschungsgemeinschaft and the Collaborating Institutions in the Dark Energy Survey. The Collaborating Institutions are Argonne National Laboratory, the University of California at Santa Cruz, the University of Cambridge, Centro de Investigaciones Energeticas, Medioambientales y Tecnologicas-Madrid, the University of Chicago, University College London, the DES-Brazil Consortium, the University of Edinburgh, the Eidgenossische Technische Hochschule (ETH) Zurich, Fermi National Accelerator Laboratory, the University of Illinois at Urbana-Champaign, the Institut de Ciencies de l’Espai (IEEC/CSIC), the Institut de Fisica d’Altes Energies, Lawrence Berkeley National Laboratory, the Ludwig Maximilians Universitat Munchen and the associated Excellence Cluster Universe, the University of Michigan, NSF’s NOIRLab, the University of Nottingham, the Ohio State University, the University of Pennsylvania, the University of Portsmouth, SLAC National Accelerator Laboratory, Stanford University, the University of Sussex, and Texas A\&M University.

BASS is a key project of the Telescope Access Program (TAP), which has been funded by the National Astronomical Observatories of China, the Chinese Academy of Sciences (the Strategic Priority Research Program “The Emergence of Cosmological Structures” Grant \# XDB09000000), and the Special Fund for Astronomy from the Ministry of Finance. The BASS is also supported by the External Cooperation Program of Chinese Academy of Sciences (Grant \# 114A11KYSB20160057), and Chinese National Natural Science Foundation (Grant \# 12120101003, \# 11433005).

The Legacy Survey team makes use of data products from the Near-Earth Object Wide-field Infrared Survey Explorer (NEOWISE), which is a project of the Jet Propulsion Laboratory/California Institute of Technology. NEOWISE is funded by the National Aeronautics and Space Administration.

The Legacy Surveys imaging of the DESI footprint is supported by the Director, Office of Science, Office of High Energy Physics of the U.S. Department of Energy under Contract No. DE-AC02-05CH1123, by the National Energy Research Scientific Computing Center, a DOE Office of Science User Facility under the same contract; and by the U.S. National Science Foundation, Division of Astronomical Sciences under Contract No. AST-0950945 to NOAO.

The Photometric Redshifts for the Legacy Surveys (PRLS) catalog used in this paper was produced thanks to funding from the U.S. Department of Energy Office of Science, Office of High Energy Physics via grant DE-SC0007914.
\end{acknowledgments}

\begin{contribution}
PSF designed the study, developed the data analysis pipeline and the theoretical framework, performed the measurements and statistical estimation, and proposed the CTCR method. CAPB suggested improvements to the pipeline and contributed to the methodological validation. RC advised on the scientific interpretation and provided key discussions. All authors contributed to the writing and revision of the manuscript. 
\end{contribution}

\appendix
\section{Data set details}\label{ap:data}

We used \texttt{CosmoHub} \citep{2017ehep.confE.488C,TALLADA2020100391} to download part of the data set. The redshift distribution was chosen as a top-hat with a slightly smooth peak as suggested in \cite{Ferreira:2025til}, these different morphology bins are compatible within the redshift uncertainty.

\begin{figure}
    \centering
    \includegraphics[width=\linewidth]{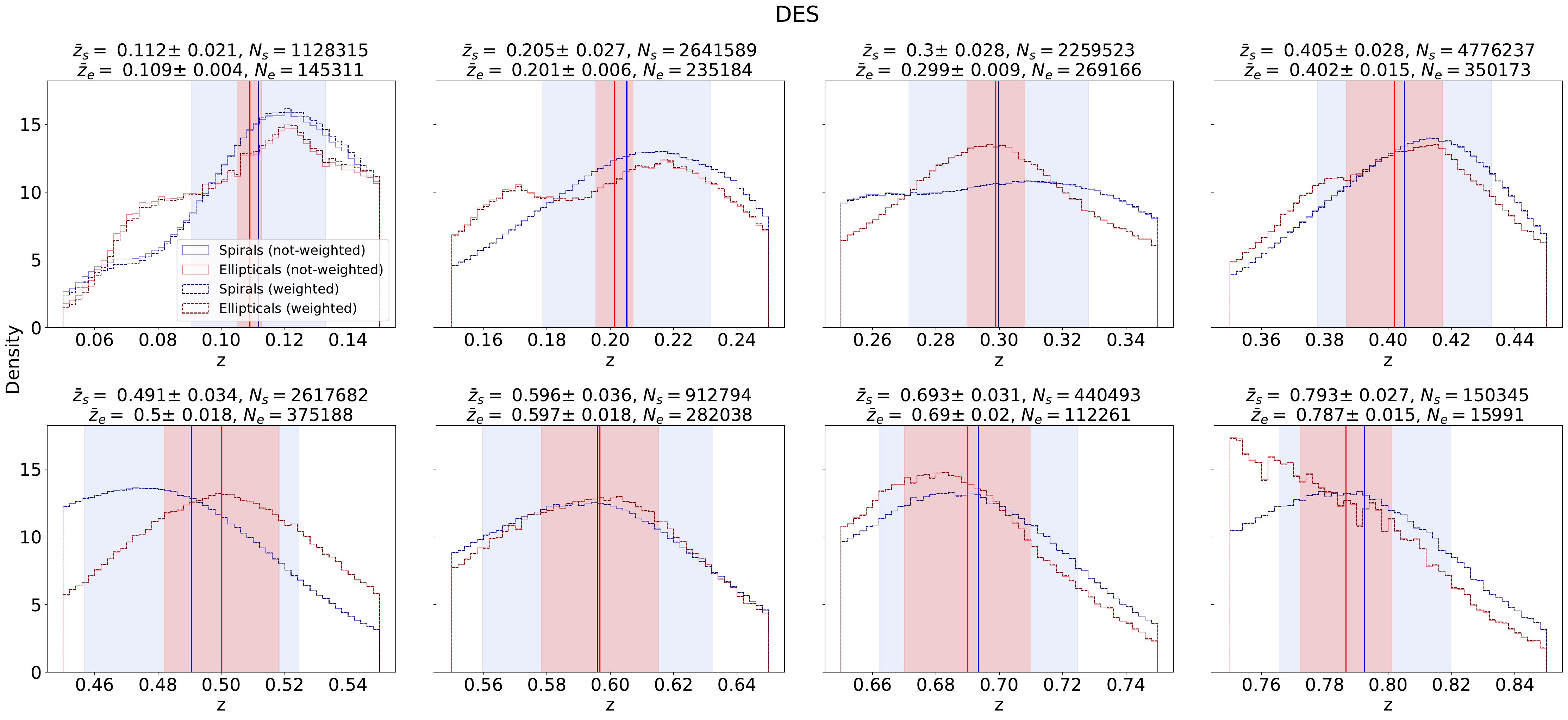}
    \caption{DES redshift distribution. The blue lines correspond to the spiral galaxy sample, whereas the red curves correspond to the elliptical galaxy sample. Solid lines indicate the distributions prior to the application of weighting, and dashed curves indicate the distributions following the application of weighting.}
    \label{fig:des_bins}
\end{figure}

\begin{figure}
    \includegraphics[width=\linewidth]{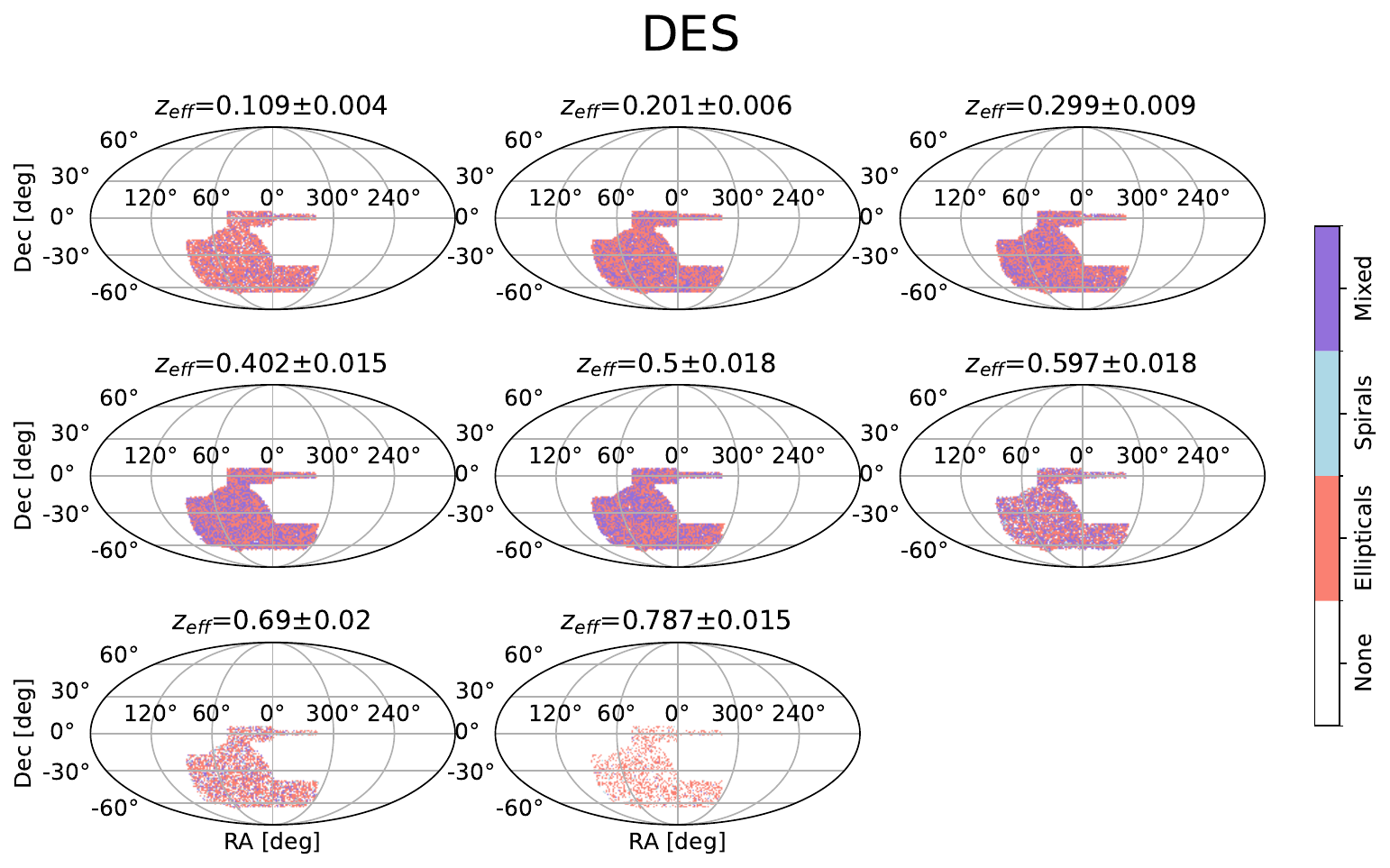}
    \caption{DES footprint per redshift bin. The panels show galaxy counts at \texttt{NSIDE}=512. Salmon indicates elliptical galaxies, blue indicates spirals, and purple marks pixels containing both types. The highest redshift bin is very sparsely populated, which illustrates the reason it was excluded from the analysis.}\label{fig:des_footprint}
\end{figure}

Figure~\ref{fig:des_bins} shows the redshift bins for the DES sample, from the smallest redshift to the highest. The blue distributions represent spiral galaxies, while the red ones the ellipticals. The shaded region marks the redshift uncertainty for each bin, and the dashed line represents the distribution after a weighting scheme we will describe in section \ref{sec:weights}. The resulting coverage is shown in Figure~\ref{fig:des_footprint}. The panels display the distribution of galaxies at a resolution of \texttt{NSIDE}=512. Elliptical galaxies are shown in salmon, spiral galaxies are shown in blue, and pixels that contain a mixture of both morphologies are highlighted in purple. The uppermost redshift bin contains very few galaxies, clearly demonstrating why this bin was omitted from the main analysis.

\begin{figure}
    \centering
    \includegraphics[width=\linewidth]{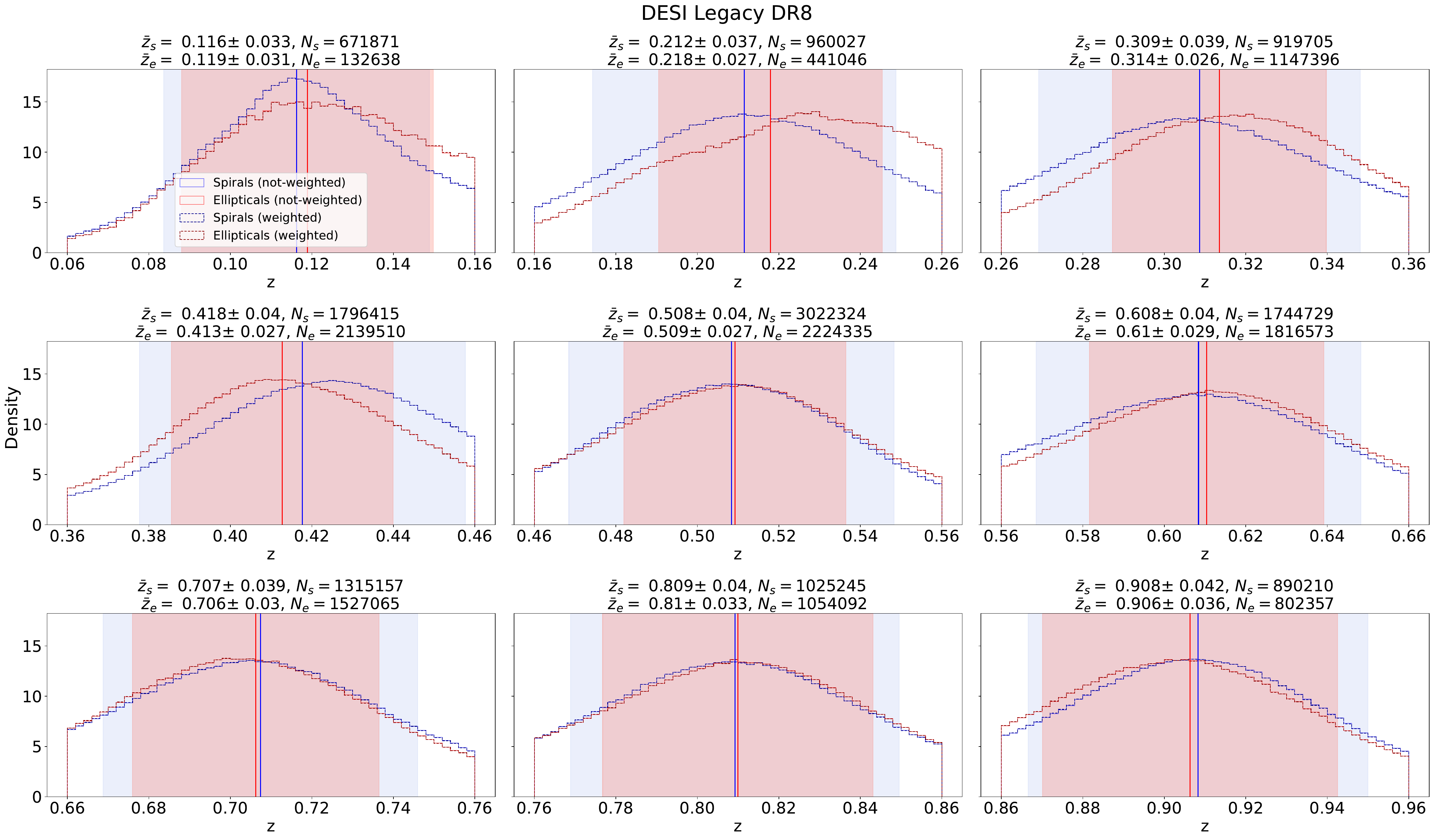}
    \caption{DESI legacy DR8 redshift distribution. The blue lines correspond to the spiral galaxy sample, whereas the red curves correspond to the elliptical galaxy sample. Solid lines indicate the distributions prior to the application of weighting, and dashed curves indicate the distributions following the application of weighting.}
    \label{fig:desi_bins}
\end{figure}

\begin{figure}
    \includegraphics[width=\linewidth]{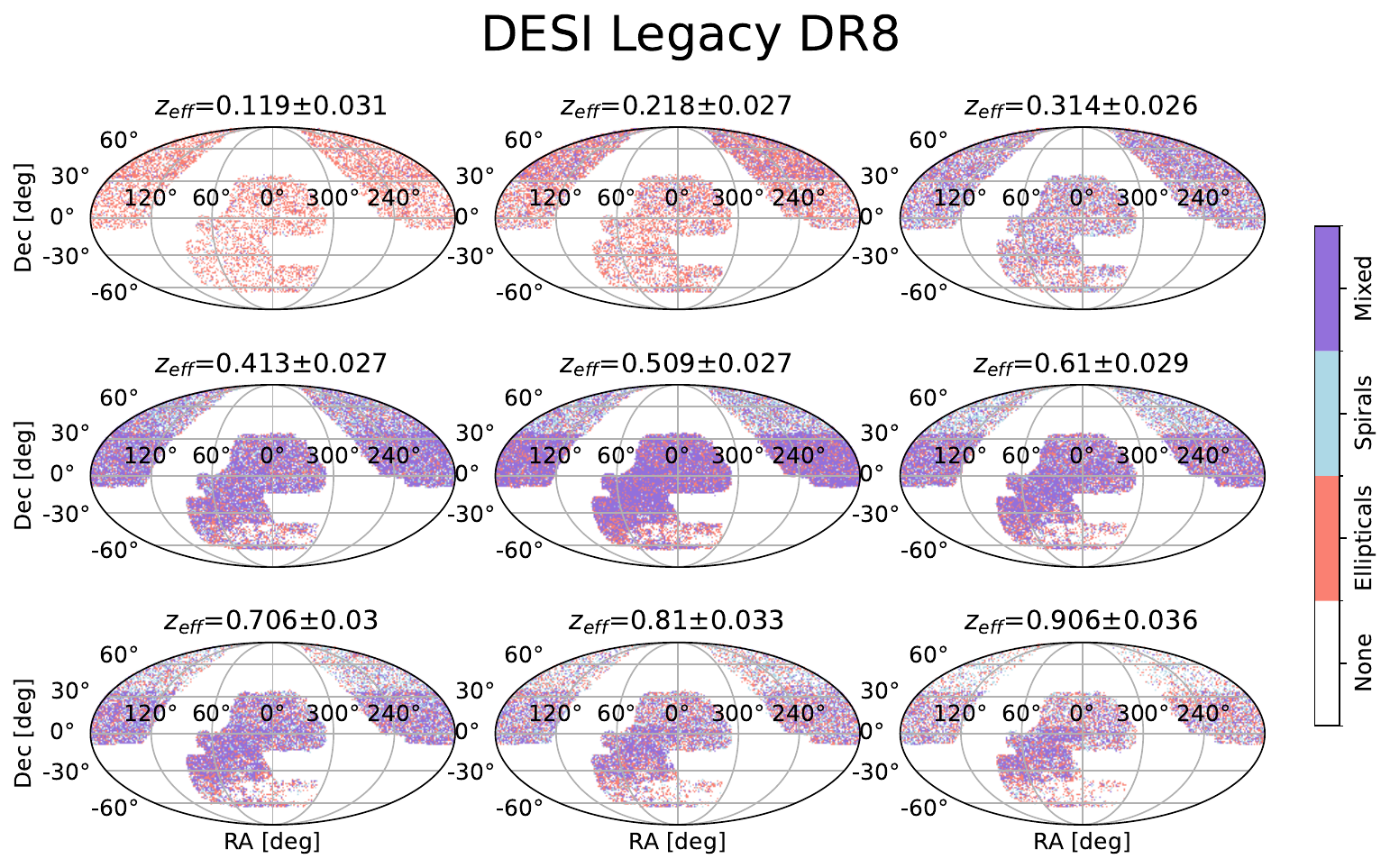}
    \caption{DESI footprint per redshift bin. The panels show galaxy counts at \texttt{NSIDE}=512. Salmon indicates elliptical galaxies, blue indicates spirals, and purple marks pixels containing both types. }\label{fig:desi_footprint}
\end{figure}

The same color scheme applies to Figure~\ref{fig:desi_bins}, the bins are more compatible than the DES ones, suitable for cross power spectrum. The clear difference from DES is the compatible distribution of objects over a redshift range, this is ideal for cross-power spectrum analysis, which will be carried out in this work. Figure~\ref{fig:desi_footprint} also shows DESI's coverage, this case with higher redshift bins with higher statistics for the analysis.

\subsection{The Per-Galaxy Weighting Scheme}\label{sec:weight_description}

We define a per-galaxy weight \(w_g\) that quantifies the classification quality and systematic contamination of each galaxy. The weight is obtained by solving a linear least-squares problem. Let \(\mathbf{X}\) be the design matrix where each column corresponds to a predictor variable \(x_p\) (photo-\(z\) error, morphological parameters, airmass, etc.), and let \(\mathbf{Y}\) be the target vector defined as the normalized redshift:

\[
Y_g = \frac{z_g}{\bar{z}},
\]

where \(z_g\) is the photometric redshift of the galaxy and \(\bar{z}\) is the mean redshift of the tomographic bin. The predictor variables are:

\[
\mathbf{X} = \{ x_1, x_2, \dots, x_p \},
\]

which include the inverse photo-\(z\) uncertainty, morphological parameters, and systematics indicators. We solve the linear system:

\begin{equation}
\min_{\boldsymbol{\alpha}} \| \mathbf{X} \boldsymbol{\alpha} - \mathbf{Y} \|^2,
\label{eq:weight_ls}
\end{equation}

where \(\boldsymbol{\alpha}\) is the vector of coefficients to be determined. The solution is obtained via standard least-squares:

\begin{equation}
\boldsymbol{\alpha} = \arg\min_{\boldsymbol{\alpha}} \| \mathbf{X} \boldsymbol{\alpha} - \mathbf{Y} \|^2.
\label{eq:alpha_solution}
\end{equation}

The per-galaxy weight is then defined as the predicted value:

\begin{equation}
w_g = \hat{Y}_g = \mathbf{X}_g \boldsymbol{\alpha},
\label{eq:weight_def}
\end{equation}

where \(\mathbf{X}_g\) is the vector of predictor variables for galaxy \(g\). Thus, galaxies whose properties are well predicted by the linear model receive weights close to the expected normalized redshift, while galaxies with anomalous properties receive weights that deviate from the mean.

The weights are not bounded between 0 and 1; they can take any positive value depending on the galaxy's properties. A high weight (e.g., \(w_g > 1\)) indicates that the galaxy's properties are consistent with the expected trend for a well-classified object: small photo-\(z\) error, good morphological fit, and favourable observing conditions. Such galaxies are considered reliable and are up-weighted in the clustering analysis. Conversely, a low weight (e.g., \(w_g < 0.5\)) indicates that the galaxy is an outlier with large photo-\(z\) error, poor shape fit, or significant systematic contamination.

To assess the effectiveness of the per‑galaxy weighting scheme, we compare the spatial distribution of galaxies before and after weighting. For each redshift bin, we bin the galaxy positions into \texttt{HEALPIX} pixels (\texttt{NSIDE}=64) and compute the mean and standard deviation of the galaxy count per pixel (unweighted) and of the mean weight per pixel (weighted). The coefficient of variation (CV) is used as a scale‑independent measure of spatial variation.

In order to verify if the weights are over-clustering the galaxy counting, we compared the power spectra with and without $w_g$. The weighted-to-unweighted ratios for the DESI and DES samples are shown in
Tables~\ref{tab:weight_diagnostics_desi} and~\ref{tab:weight_diagnostics_des}.

Each table reports, per redshift bin, the ratio $C_\ell^{\rm w}/C_\ell^{\rm unw}$ of the weighted to unweighted power spectrum for the elliptical auto-spectrum, for the spiral
auto-spectrum, and for the CTCR, $R_\ell^{es} = C_\ell^{ee}/C_\ell^{es}$. For each of the three quantities we list:
\begin{itemize}
    \item $\langle R \rangle$: the mean ratio across the fitted
          multipole range;
    \item $\sigma_R$: the standard deviation of the ratio across the
          same range;
    \item $\max|R-1|$: the maximum absolute deviation of the ratio
          from unity across the range.
\end{itemize}
In both samples, the CTCR measured with and without weights is consistent with unity in the redshift range where the signal is most significant ($z \leq 0.4$). The mean CTCR ratios are within $1\%$ of unity for DESI and within $5\%$ for DES. The larger scatter observed in the lowest-redshift bins (DESI Bin~0 and DES Bin~0) is due to the lower signal-to-noise in these bins. The anomalous values in DESI Bin~2 for the spiral auto-spectrum ratio  ($\langle R_s \rangle = 0.0267$, $\sigma_{R_s} = 8.35$) are numerical fluctuations caused by division by very small power values in a few $\ell$ bins; the CTCR ratio in the same bin is consistent with unity ($\langle R_{\rm CTCR} \rangle = 1.0010 \pm 0.0245$). The larger deviations in the highest DES redshift bins (Bin~7) are likewise due to the low number of objects in these bins.

\begin{table}[htbp]
\centering
\caption{Diagnostics of weighted-to-unweighted ratios for DESI Legacy
DR8. Column 3 gives the mean ratio
$\langle C_\ell^{\rm w}/C_\ell^{\rm unw} \rangle$ for the elliptical
auto-spectrum, and column~4 its standard deviation across the fitted
$\ell$ range. Column~5 gives the maximum absolute deviation of that
ratio from unity. Columns 6--8 report the same three quantities for
the spiral auto-spectrum. Columns 9--11 report the same three
quantities for the CTCR, $R_\ell^{es} = C_\ell^{ee}/C_\ell^{es}$.
Values close to unity in the CTCR columns indicate that the weighting
scheme does not introduce a scale-dependent bias in the quantity used
for the main analysis.}
\label{tab:weight_diagnostics_desi}
\begin{tabular}{c c c c c c c c c c c}
\hline
\hline
Bin & $z_{\rm eff}$ & $\langle R_e \rangle$ & $\sigma_{R_e}$ & $\max|R_e-1|$ & $\langle R_s \rangle$ & $\sigma_{R_s}$ & $\max|R_s-1|$ & $\langle R_{\rm CTCR} \rangle$ & $\sigma_{R_{\rm CTCR}}$ & $\max|R_{\rm CTCR}-1|$ \\
\hline
  0 & 0.119 & 0.6801 & 2.4938 & 21.7310 & 0.9739 & 0.0424 & 0.2427 & 0.7013 & 2.4704 & 21.4892 \\
  1 & 0.218 & 0.9684 & 0.0581 & 0.4983 & 0.9807 & 0.0416 & 0.3130 & 0.9959 & 0.0309 & 0.2499 \\
  2 & 0.314 & 0.9926 & 0.0138 & 0.0635 & 0.0267 & 8.3467 & 73.2574 & 1.0010 & 0.0245 & 0.1716 \\
  3 & 0.413 & 1.0168 & 0.0883 & 0.7460 & 1.0088 & 0.0240 & 0.1423 & 1.0012 & 0.0431 & 0.3703 \\
  4 & 0.509 & 1.0171 & 0.0286 & 0.2151 & 1.0116 & 0.0216 & 0.1247 & 0.9971 & 0.0161 & 0.1016 \\
  5 & 0.610 & 1.0211 & 0.0235 & 0.1733 & 1.0174 & 0.0220 & 0.1285 & 0.9983 & 0.0126 & 0.0777 \\
  6 & 0.706 & 1.0209 & 0.0214 & 0.1235 & 1.0209 & 0.0212 & 0.1407 & 0.9974 & 0.0086 & 0.0422 \\
  7 & 0.810 & 1.0276 & 0.0298 & 0.2072 & 1.0293 & 0.0230 & 0.1345 & 1.0003 & 0.0127 & 0.0864 \\
  8 & 0.906 & 1.0301 & 0.0253 & 0.1267 & 1.0288 & 0.0232 & 0.1261 & 1.0048 & 0.0117 & 0.0604 \\
\hline
\hline
\end{tabular}
\end{table}

\begin{table}[htbp]
\centering
\caption{Same as Table~\ref{tab:weight_diagnostics_desi}, but for the
DES sample. Column 3 gives the mean ratio
$\langle C_\ell^{\rm w}/C_\ell^{\rm unw} \rangle$ for the elliptical
auto-spectrum, column~4 its standard deviation across the fitted
$\ell$ range, and column~5 the maximum absolute deviation of that
ratio from unity. Columns 6--8 report the same three quantities for
the spiral auto-spectrum, and columns 9--11 the same three quantities
for the CTCR. Values close to unity in the CTCR columns indicate that
the weighting scheme does not introduce a scale-dependent bias in the
quantity used for the main analysis. The larger deviations in Bin~7
reflect the low number of galaxies in that bin.}
\label{tab:weight_diagnostics_des}
\begin{tabular}{c c c c c c c c c c c}
\hline
\hline
Bin & $z_{\rm eff}$ & $\langle R_e \rangle$ & $\sigma_{R_e}$ & $\max|R_e-1|$ & $\langle R_s \rangle$ & $\sigma_{R_s}$ & $\max|R_s-1|$ & $\langle R_{\rm CTCR} \rangle$ & $\sigma_{R_{\rm CTCR}}$ & $\max|R_{\rm CTCR}-1|$ \\
\hline
  0 & 0.109 & 1.0214 & 0.0942 & 0.7609 & 1.0446 & 0.0267 & 0.0987 & 1.0485 & 0.0531 & 0.4449 \\
  1 & 0.201 & 1.0037 & 0.0262 & 0.1403 & 1.0008 & 0.0027 & 0.0102 & 1.0081 & 0.0253 & 0.1498 \\
  2 & 0.299 & 1.0108 & 0.0324 & 0.2095 & 0.9973 & 0.0023 & 0.0126 & 1.0042 & 0.0521 & 0.2857 \\
  3 & 0.402 & 1.0186 & 0.0269 & 0.1933 & 0.9990 & 0.0015 & 0.0058 & 1.0010 & 0.1047 & 0.6983 \\
  4 & 0.500 & 1.0141 & 0.0183 & 0.1092 & 0.9996 & 0.0026 & 0.0162 & 1.0522 & 0.2824 & 2.4090 \\
  5 & 0.597 & 1.0203 & 0.0308 & 0.2356 & 0.9994 & 0.0021 & 0.0090 & 1.0179 & 0.0720 & 0.3730 \\
  6 & 0.690 & 1.0333 & 0.0527 & 0.2564 & 0.9992 & 0.0053 & 0.0341 & 1.0272 & 0.0535 & 0.2722 \\
  7 & 0.787 & 1.0826 & 0.7552 & 6.4398 & 0.9906 & 0.0464 & 0.2602 & 1.2093 & 1.3635 & 8.6928 \\
\hline
\hline
\end{tabular}
\end{table}

\subsection{Weighting robustness tests}\label{ap:robust}
We forced some contamination by exchanging galaxies of opposite morphology, we added the amount of 10\% of the ellipticals for the same photo-z bin and made 100 $C_\ell$ realizations of this kind. To show this relation, we evaluated the $C_\ell$s according to scale and compared the clean sample result with our contamination tests applying a ratio between them. If our weights are accurate, the average of the contaminated sample
will be close to the clean sample for all scales. This is shown in Tables~\ref{tab:contamination_panel_2x6} for both samples. We analysed $\ell$ low ($6.5 \le \ell \le 381.5$) and $\ell$ high  ($381.5 \le \ell \le 756.5$). Given the results, our weights are robust for most scales, except for the highest DES bin for elliptical galaxies, which did not have sufficient data. On smaller scales, we see a trend of deviation from the clean sample, this is expected given the limitations of the surveys on small scales. The DESI sample performed better at all the sub-samples and scales, it is primarily because of its higher statistics and sky coverage. 

\begin{table}
\centering
\begin{tabular}{c
   *{2}{c} *{2}{c} *{2}{c}   
   *{2}{c} *{2}{c} *{2}{c}   
}
\toprule
\multirow{2}{*}{Redshift bin} &
\multicolumn{2}{c}{Elliptical DES} &
\multicolumn{2}{c}{Spiral DES} &
\multicolumn{2}{c}{Cross DES} &
\multicolumn{2}{c}{Elliptical DESI} & 
\multicolumn{2}{c}{Spiral DESI} &
\multicolumn{2}{c}{Cross DESI} \\
\cmidrule(lr){2-3} \cmidrule(lr){4-5} \cmidrule(lr){6-7}
\cmidrule(lr){8-9} \cmidrule(lr){10-11} \cmidrule(lr){12-13}
& \multicolumn{1}{c}{$\ell$ low} & \multicolumn{1}{c}{$\ell$ high} &
\multicolumn{1}{c}{$\ell$ low} & \multicolumn{1}{c}{$\ell$ high} &
\multicolumn{1}{c}{$\ell$ low} & \multicolumn{1}{c}{$\ell$ high} &
\multicolumn{1}{c}{$\ell$ low} & \multicolumn{1}{c}{$\ell$ high} &
\multicolumn{1}{c}{$\ell$ low} & \multicolumn{1}{c}{$\ell$ high} &
\multicolumn{1}{c}{$\ell$ low} & \multicolumn{1}{c}{$\ell$ high} \\
\midrule
0 & 2.4 & 1.6 & 0.96 & 0.91 & 1.3 & 1.5 & 1.0 & 1.0 & 1.0 & 1.0 & 1.1 & 1.1 \\
1 & 2.2 & 1.5 & 0.98 & 0.93 & 1.3 & 1.6 & 1.1 & 1.1 & 1.0 & 1.0 & 1.0 & 0.98 \\
2 & 2.1 & 1.4 & 0.98 & 0.93 & 1.0 & 1.2 & 1.1 & 1.0 & 0.98 & 0.98 & 0.98 & 0.91 \\
3 & 2.8 & 1.5 & 0.98 & 0.94 & 0.99 & 1.4 & 1.1 & 1.0 & 0.99 & 0.98 & 0.98 & 0.95 \\
4 & 2.8 & 1.4 & 0.97 & 0.91 & 0.93 & 1.2 & 1.1 & 1.1 & 1.0 & 1.0 & 1.0 & 0.98 \\
5 & 1.7 & 1.1 & 0.87 & 0.78 & 1.1 & 1.3 & 1.1 & 1.0 & 1.0 & 1.0 & 0.99 & 0.98 \\
6 & 1.1 & 0.87 & 0.85 & 0.80 & 1.1 & 1.3 & 1.1 & 1.0 & 1.0 & 1.0 & 1.0 & 0.97 \\
7 & 0.25 & 1.6 & 0.89 & 0.86 & -0.51 & -1.4 & 1.1 & 1.0 & 1.0 & 1.0 & 1.0 & 0.99 \\
8 &  -   & -    & -   &  -   &    -  &  - & 1.07 & 1.04 & 1.06 & 1.04 &  1.04 & 1.01\\
\bottomrule

\end{tabular}
\caption{Contamination test results $C_\ell / \langle C_\ell^{c}\rangle$ (median values) for DES and DESI Legacy DR8, split by morphological type.}
\label{tab:contamination_panel_2x6}
\end{table}

\section{Auto and Cross power spectra}\label{ap:cell_details}
\subsection{Estimator}
We estimate the angular power spectrum \(C_\ell\) for a galaxy sample using the pseudo-\(C_\ell\) method as implemented in the \textsc{\texttt{NaMaster}} \citep{Alonso_2019} library (\texttt{pymaster}). The same formalism is extended to cross-correlations between two samples. We carried out the estimation at \texttt{NSIDE}=256 over the range $2<\ell<800$. For the subsequent model fitting, we restrict this to $20<\ell<200$ to ensure that only multipoles suitable for the Limber approximation are used.

\subsection{Covariance Matrix Theoretical}

The covariance between two band powers $C_l$ and $C_c$ under the Gaussian assumption is given by:
\begin{equation}
\operatorname{Cov}(C_l, C_l) = \frac{2}{(2\ell_l+1)\,\Delta\ell_l\,f_{\mathrm{sky}}}\, (C_l + N_l)^2.
\end{equation}

For bins whose indices satisfy $|l-c| < 4$:
\begin{equation}
\operatorname{Cov}(C_l, C_c) = \frac{2}{(2\ell_l+1)\,\Delta\ell_l\,f_{\mathrm{sky}}}\, (C_l + N_l)(C_c + N_c).
\end{equation}
Note that this expression uses $\ell_l$ and $\Delta\ell_l$ (the properties of bin $l$) for the entire off‑diagonal term, which is not symmetric in $l$ and $c$. In a more consistent implementation one would use a symmetrised combination, e.g. the geometric mean of the two bin widths, but the code as written adopts this asymmetric form. For $|l-c| \ge 4$, the off‑diagonal covariance is set to zero.
\subsection{Covariance Matrix from bootstrapping pixels}\label{ap:bootstrap}
To estimate the covariance of the angular power spectra we use a patch‑based bootstrap resampling of the galaxy catalogs.  For each galaxy we assign a \texttt{HEALPIX} pixel at \texttt{NSIDE} = 512 (pixel size $\sim 0.1^\circ$).  A bootstrap realization is generated by randomly selecting $80\%$ of the occupied pixels (with replacement) and retaining all galaxies that fall into the chosen pixels.  This procedure preserves the large‑scale clustering information because entire contiguous regions of the sky are resampled rather than individual galaxies, which would destroy angular correlations.  For each redshift bin and each morphological type (elliptical and spiral) we produce $100$ such realizations.  

To regularize the bootstrap covariance matrix we apply linear shrinkage, constructing a convex combination of the empirical covariance $\mathbf{C}_{\text{boot}}$ and a diagonal target $\mathbf{T}$ containing only the empirical variances. The shrunk covariance is $\hat{\mathbf{C}}(\alpha) = (1-\alpha)\mathbf{C}_{\text{boot}} + \alpha\mathbf{T}$, where the shrinkage intensity $\alpha\in[0,1]$ is chosen to be a fixed value (this case $\alpha=0.2$). This reduces the estimation noise and guarantees positive definiteness, which is especially beneficial when the number of bootstrap realizations is limited. We adopt the diagonal target $\mathbf{T} = \operatorname{diag}(\mathbf{C}_{\text{boot}})$, preserving the variances while suppressing noisy off‑diagonal correlations. After shrinkage, a small regularizing term $\epsilon\mathbf{I}$ is added to ensure invertibility. This method has been tested by \cite{Pope2008} and \cite{Euclid2025}.

We estimate the covariance of the angular power spectra using a patch-based bootstrap. The procedure is as follows.

\begin{enumerate}
\item \textbf{Pixelisation.} Each galaxy catalog is assigned to \texttt{HEALPIX} pixels at resolution \texttt{NSIDE}= 512, corresponding to a pixel scale of $\sim 7$~arcmin. Each galaxy is mapped to the pixel containing its sky position.
\item \textbf{Occupied pixels.} We identify the set of pixels that contain at least one galaxy. Empty pixels carry no information and are discarded.
\item \textbf{Resampling.} For each bootstrap realisation, we draw a random subset of the occupied pixels with replacement. The subset contains $80\%$ of the occupied pixels. Galaxies belonging to a selected pixel are included in the realisation; because pixels are drawn with replacement, some galaxies appear multiple times, and $20\%$ of pixels are omitted.
\item \textbf{Realisation catalogue.} The selected galaxies form a bootstrap catalogue. Each realisation therefore has a slightly different total number of galaxies, reflecting the stochastic resampling.
\item \textbf{Spectra and covariance.} For each realisation, we compute the angular power spectra $C_\ell^{ee}$, $C_\ell^{ss}$, and $C_\ell^{es}$ using the same pipeline (\texttt{NaMaster} with the survey mask). The covariance is estimated as the sample covariance across realisations:
\begin{equation}
\widehat{\mathrm{Cov}}\!\left(C_b^{ij}, C_{b'}^{mn}\right)
= \frac{1}{N_{\rm real}-1}
\sum_{r=1}^{N_{\rm real}}
\left(C_{b,r}^{ij} - \bar{C}_b^{ij}\right)
\left(C_{b',r}^{mn} - \bar{C}_{b'}^{mn}\right),
\end{equation}
where $\bar{C}_b^{ij}$ is the mean over realisations.
\item \textbf{Number of realisations.} We use $N_{\rm real} = 100$ realisations per redshift bin.
\item \textbf{Inverse covariance.} The inverse of the sample covariance is computed with the Hartlap correction,
\begin{equation}
\widehat{\mathbf{C}}^{-1} = \frac{N_{\rm real} - n_b - 2}{N_{\rm real} - 1}\,\mathbf{C}^{-1},
\end{equation}
where $n_b$ is the number of bandpowers, to account for the bias introduced by the finite number of realisations.
\end{enumerate}

The two tracer catalogs are resampled independently, so the resulting bootstrap covariance is block-diagonal: the off-diagonal blocks $\mathrm{Cov}(C_\ell^{ee}, C_\ell^{es})$ and $\mathrm{Cov}(C_\ell^{ss}, C_\ell^{es})$ are zero by construction. The off-diagonal blocks are supplied separately using the analytic Gaussian covariance described. We have verified that the full covariance is positive definite for all redshift bins and that the inclusion of the cross-covariance does not change the CTCR or the bias parameters within the statistical uncertainties, see section \ref{sec:gauss_cov}.

\subsubsection{Auto-power spectrum}
Given a galaxy catalog with positions \((\alpha_g, \delta_g)\) and weights \(w_g\), and a random catalog with positions \((\alpha_r, \delta_r)\) and weights \(w_r=1.0\), we construct a \texttt{HEALPIX} map of the weighted galaxy counts:
\begin{equation}
n_{\rm g}(p) = \sum_{g \in {\rm pix}(p)} w_g, \qquad
n_{\rm r}(p) = \sum_{r \in {\rm pix}(p)} w_r,
\end{equation}
where \(p\) denotes a pixel. The random map is normalized by the ratio of total weights:
\begin{equation}
\alpha = \frac{\sum_g w_g}{\sum_r w_r}, \qquad
\bar{n}_{\rm r}(p) = \alpha \, n_{\rm r}(p).
\end{equation}
The overdensity field is then
\begin{equation}
\delta(p) = m(p) \left[ \frac{n_{\rm g}(p)}{\bar{n}_{\rm r}(p)} - 1 \right],
\end{equation}
where \(m(p)\) is a binary mask defined by pixels with \(\bar{n}_{\rm r}(p) > 0\) (optionally after a threshold).

From the masked overdensity map, we compute the pseudo-$C_\ell$ (coupled power spectrum)
\begin{equation}
\tilde{C}_\ell
= \frac{1}{2\ell+1} \sum_{m=-\ell}^{\ell} |a_{\ell m}|^2,
\label{eq:pseudo_cl}
\end{equation}
where $a_{\ell m}$ are the spherical harmonic coefficients of $\delta(p)$. The true power spectrum is related to the pseudo spectrum
via the mode-coupling matrix $\mathbf{M}_{\ell\ell'}$:
\begin{equation}
\langle \tilde{C}_\ell \rangle
= \sum_{\ell'} M_{\ell\ell'} \, C_{\ell'}.
\label{eq:mode_coupling}
\end{equation}

Because $\mathbf{M}_{\ell\ell'}$ is invertible at the resolution used in this analysis, the true power spectrum can be recovered by inverting
Equation~(\ref{eq:mode_coupling}) before any binning is applied. We therefore first decouple the pseudo spectrum,
\begin{equation}
\hat{C}_\ell
= \sum_{\ell'} \left( \mathbf{M}^{-1} \right)_{\ell\ell'}
  \tilde{C}_{\ell'},
\label{eq:decoupled_cl}
\end{equation}
and then bin the decoupled spectrum into bandpowers,
\begin{equation}
\hat{C}_b
= \sum_{\ell} P_{b\ell} \, \hat{C}_\ell,
\label{eq:binned_cl}
\end{equation}
where $P_{b\ell}$ is a binning operator (e.g., linear bins of width $\Delta\ell$). The effective multipole of each bin is the average
$\ell$ weighted by the power spectrum shape, approximated here by the mean $\ell$ in the bin.

In our pipeline, Equations~(\ref{eq:decoupled_cl}) and~(\ref{eq:binned_cl}) are implemented by \textsc{NaMaster} \citep{Alonso_2019}, which computes the mode-coupling matrix $\mathbf{M}_{\ell\ell'}$ from the survey mask and applies the decoupling and binning in a single step. The per-galaxy
weights $w_g$ enter the mode-coupling matrix through the weighted mask, so that the mask-induced effect of the weights on the pseudo spectrum is deconvolved consistently. We emphasize, however, that this deconvolution corrects only for the mode coupling generated by the weighted mask; it does not correct for the change in the effective tracer population induced by the same weights, which acts at the level of the underlying density field. The two effects are
distinct, and the latter is controlled instead by the systematics diagnostics of Appendix~\ref{ap:robust} and by the weighted-to-
unweighted comparison of Section~\ref{sec:weights}. The shot-noise term enters as a constant $N_\ell = 1/\bar{n}_{\rm eff}$ with
$\bar{n}_{\rm eff} = \big(\sum_g w_g\big)^2 / \big(A_{\rm sky} \sum_g w_g^2\big)$ and $A_{\rm sky}$ the masked sky area.

\subsubsection{Cross-power spectrum}
For two fields \(\delta_i\) and \(\delta_j\), the coupled cross-spectrum is
\begin{equation}
\tilde{C}_\ell^{ij} = \frac{1}{2\ell+1} \sum_{m} a_{\ell m}^{(i)} \big( a_{\ell m}^{(j)} \big)^*.
\end{equation}
The same mode coupling matrix (now built from the two masks or a joint mask) relates the coupled and true cross-spectra:
\begin{equation}
\langle \tilde{C}_\ell^{ij} \rangle = \sum_{\ell'} M_{\ell\ell'}^{ij} \, C_{\ell'}^{ij}.
\end{equation}
Decoupling is performed analogously, yielding the binned estimate \(\hat{C}_b^{ij}\). The shot-noise does not contribute to the cross-spectrum if the two fields are from disjoint sets of objects.

\begin{figure}
    \centering
    \includegraphics[width=\linewidth]{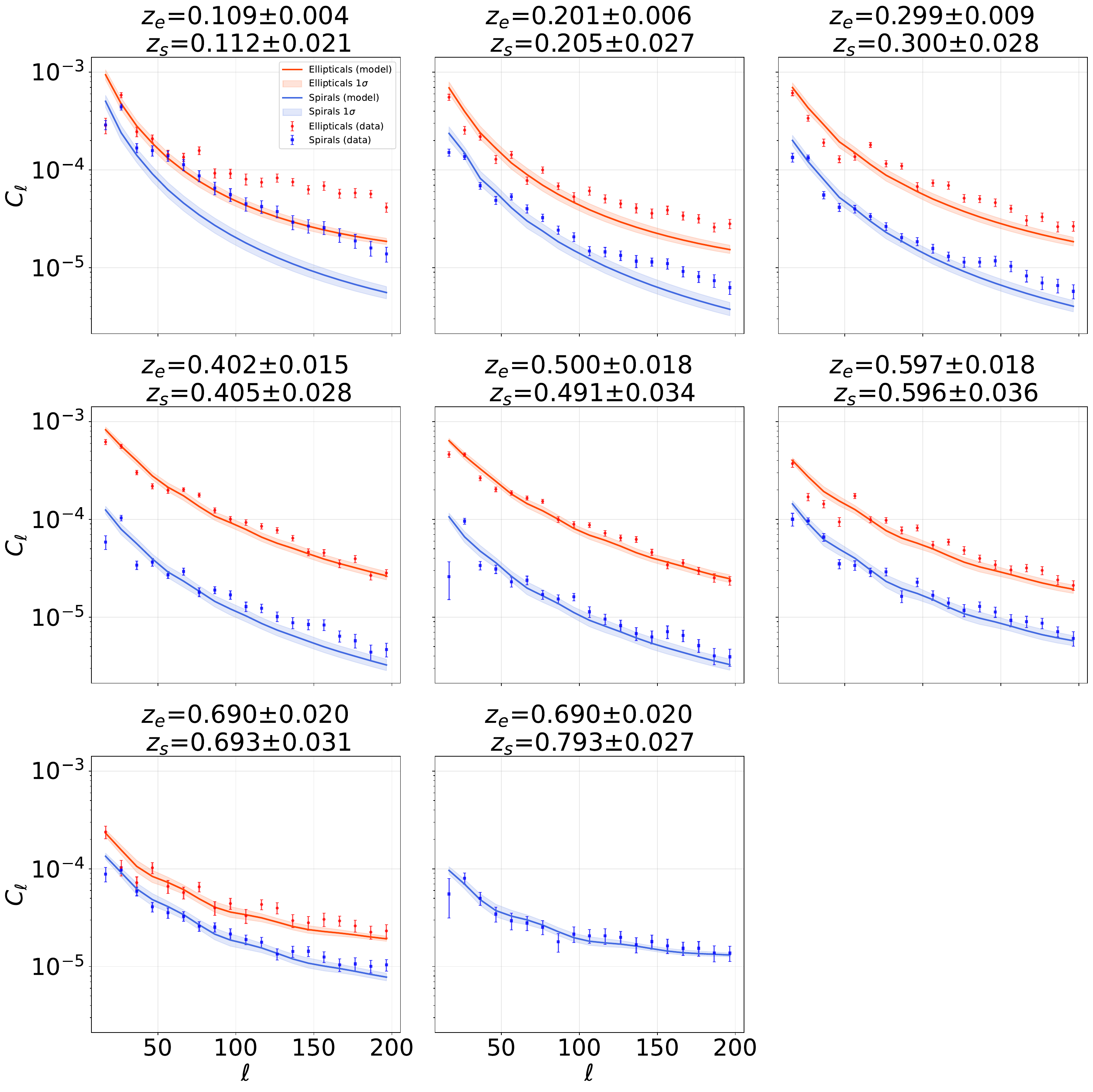}
    \caption{Observed $C_\ell$ of the DES samples. Red indicates the early-type galaxies (ellipticals) and blue indicates the spirals (late-type galaxies). The shaded areas surrounding the lines indicate the 1$\sigma$ confidence level based on the best-fit result.}
    \label{fig:cell_des}
\end{figure}

\begin{figure}
    \centering
    \includegraphics[width=\linewidth]{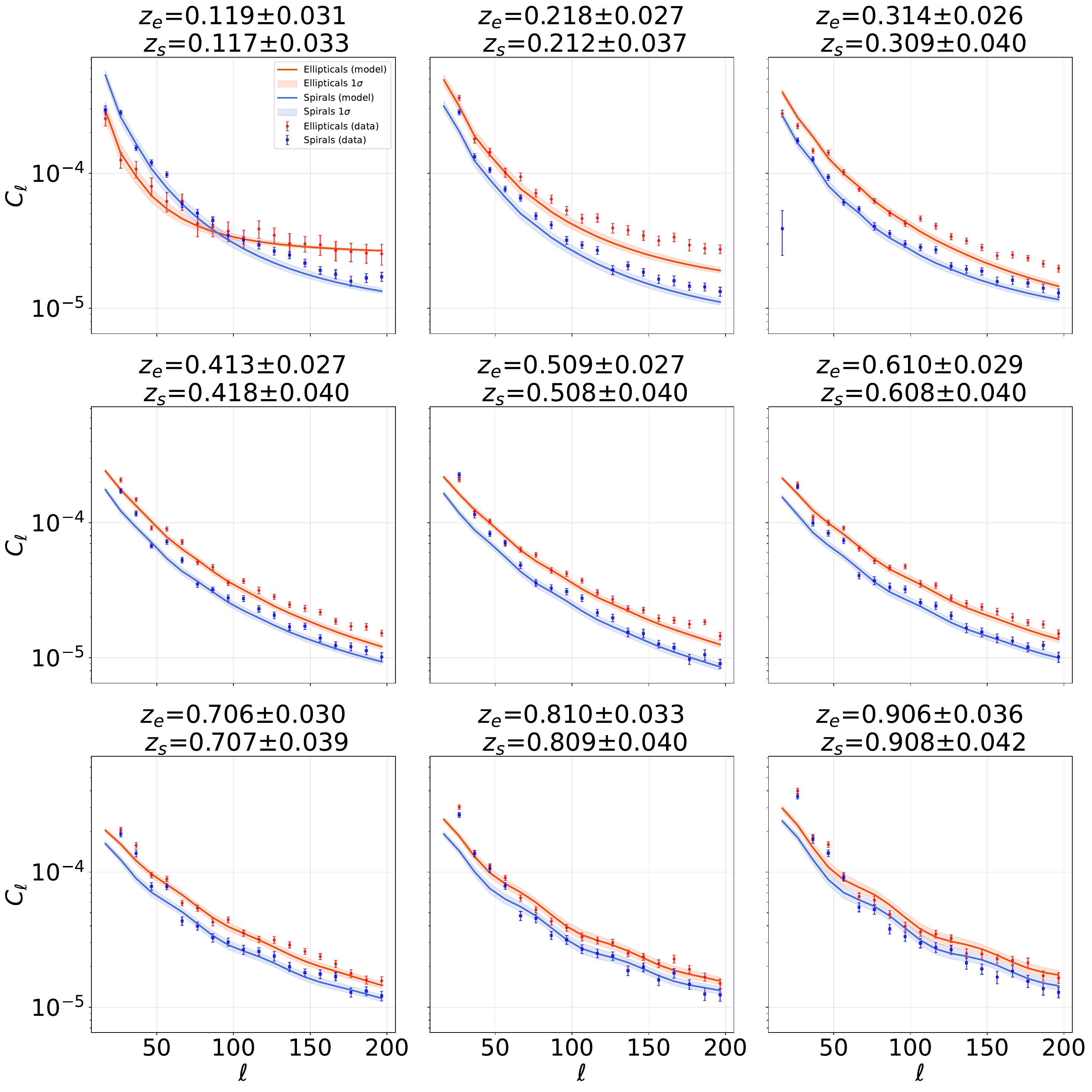}
    \caption{Observed $C_\ell$ of the DESI samples. Red indicates the early-type galaxies (ellipticals) and blue indicates the spirals (late-type galaxies). The shaded areas surrounding the lines indicate the 1$\sigma$ confidence level based on the best-fit result.}
    \label{fig:cell_desi}
\end{figure}

The results are shown in Figures~\ref{fig:cell_des}, and \ref{fig:cell_desi}, where red denotes elliptical galaxies and blue denotes spirals. The angular power spectra for the DES sample in Fig.~\ref{fig:cell_des} exhibit a consistent clustering trend for both galaxy types across all bins: elliptical galaxies are always more strongly clustered than spirals, as indicated by their larger $C_\ell$ amplitudes. In the highest DES redshift bin, however, the results for ellipticals are not robust because the limited number of such galaxies makes this sample strongly affected by shot-noise. Figure~\ref{fig:cell_des} shows only the spiral result in the last bin for the DES sample, which lost signal to shot-noise, 
the remaining bins follow the expected pattern, with early-type galaxies being more strongly clustered than late-type galaxies.

\begin{figure}
    \centering
    \includegraphics[width=\linewidth]{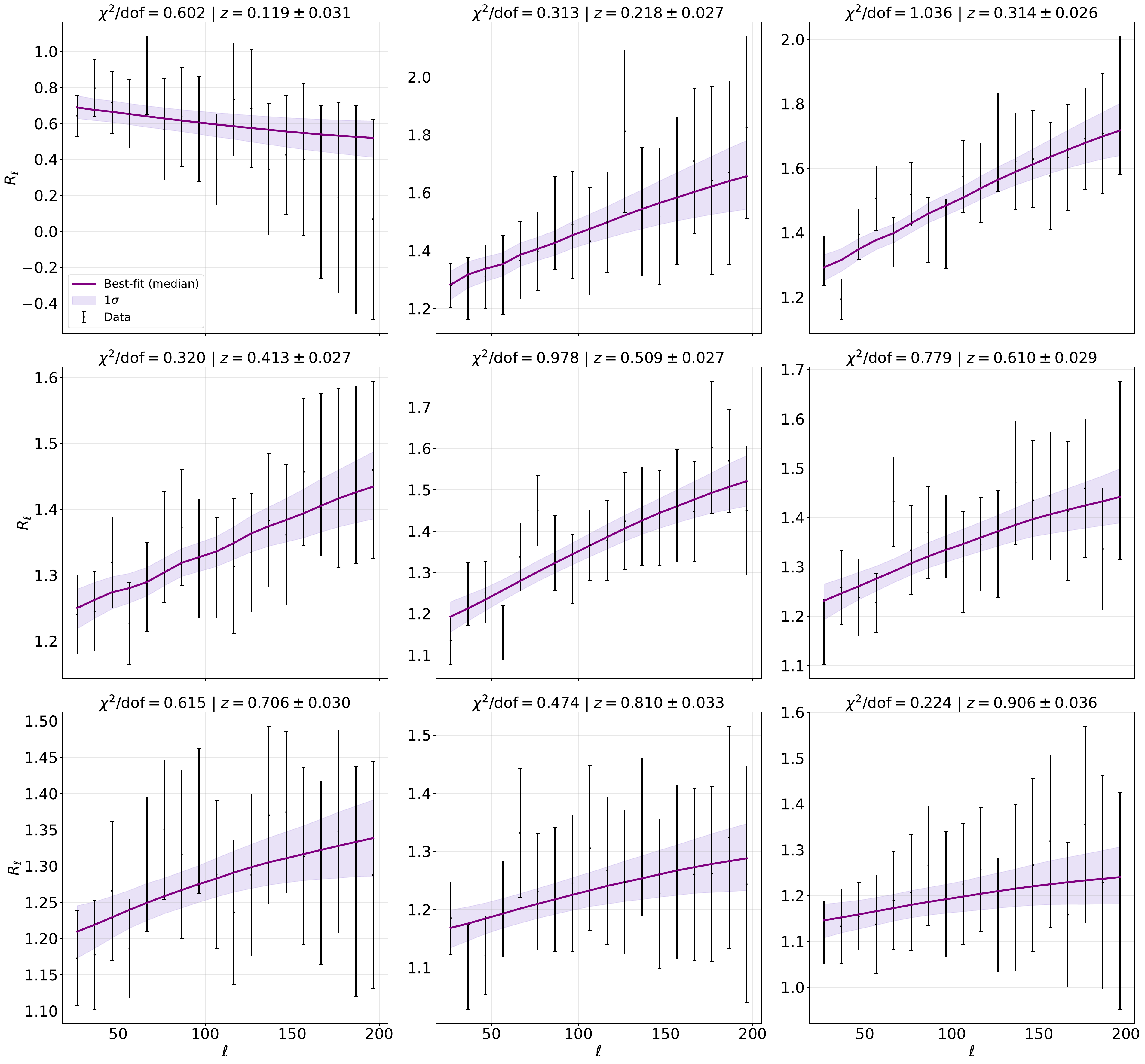}
    \caption{CTCR from DESI sample. The best-fit is the purple line, light-purple represents the $1\sigma$ uncertainty from the best-fit results. The panels progress toward higher redshift from left to right and from one row to the next.}
    \label{fig:cellc_desi}
\end{figure}

The best-fit result from equation \ref{eq:ratio} using the bootstrapped covariance matrix is shown in Figure~\ref{fig:cellc_desi}, the best-fit is well behaved for all bins. The light-purple region indicates the 1$\sigma$ result from the best-fit model, compatible with the vast majority of observed $\ell$s.

\subsection{Model}
Given two tracers \(i, j \in \{\text{spiral, elliptical}\}\), the observed galaxy overdensity field includes contributions from linear bias, redshift-space distortions (RSD), and magnification bias.

The 3D matter density field and shot-noise are
\begin{align}
\delta_i(\mathbf{x},z) &= b_i(z)\,\delta_m(\mathbf{x},z) + \epsilon_i(\mathbf{x},z), \qquad \langle\epsilon_i\rangle = 0,\\
\langle \epsilon_i(\mathbf{k})\,\epsilon_j^*(\mathbf{k}') \rangle &= (2\pi)^3\,\delta^D(\mathbf{k}-\mathbf{k}')\,\frac{\delta_{ij}}{\bar{n}_i^{\rm 3D}(z)},
\end{align}
with linear bias \(b_i(z)\) and 3D number density \(\bar{n}_i^{\rm 3D}(z)\).  The projected field on the sky is
\begin{equation}
\delta_i(\hat{\mathbf{n}}) = \int dz\, W_i(z)\,\delta_i^{\rm obs}\bigl(\chi(z)\hat{\mathbf{n}}, z\bigr),
\end{equation}
where the window function \(W_i(z)\) is normalized, \(\int dz\,W_i(z)=1\), and \(\chi(z)\) is the comoving distance.

Expanding in spherical harmonics, \(\delta_i(\hat{\mathbf{n}}) = \sum_{\ell m} a_{\ell m}^{(i)} Y_{\ell m}(\hat{\mathbf{n}})\), the angular power spectrum is defined as
\begin{equation}
\bigl\langle a_{\ell m}^{(i)} a_{\ell' m'}^{*(j)} \bigr\rangle = \delta_{\ell\ell'}^K \delta_{mm'}^K C_\ell^{ij}.
\end{equation}
Following the full-sky derivation~\cite{Padmanabhan2007}, one obtains
\begin{equation}\label{eq:cl_full}
C_\ell^{ij} = \frac{2}{\pi}\int_0^\infty dk\,k^2\,\Delta_\ell^i(k)\,\Delta_\ell^j(k)\,P_{\rm lin}(k) \;+\; N_\ell^{ij},
\end{equation}
where \(P_{\rm lin}(k)\) is the linear matter power spectrum at \(z=0\).  The total transfer function \(\Delta_\ell^i(k) = \Delta_\ell^{i,D}(k) + \Delta_\ell^{i,\rm RSD}(k) + \Delta_\ell^{i,\mu}(k)\) contains the following contributions:
\begin{align}
\Delta_\ell^{i,D}(k)     &= \int dz\, W_i(z)\, D_+(k,z)\, b_i(z)\, j_\ell\bigl(k\chi(z)\bigr), \label{eq:DeltaD}\\
\Delta_\ell^{i,\rm RSD}(k) &= -\int dz\, W_i(z)\, D_+(k,z)\, f(z)\, j_\ell''\bigl(k\chi(z)\bigr), \label{eq:DeltaRSD}\\
\Delta_\ell^{i,\mu}(k)    &= (5s_i-2) \int dz\, W_i^\kappa(z)\, D_+(k,z)\, j_\ell\bigl(k\chi(z)\bigr). \label{eq:DeltaMu}
\end{align}
Here \(D_+(k,z)\) is the linear growth factor (normalized to unity at \(z=0\)), \(f(z)=d\ln D_+/d\ln a\) is the growth rate, \(j_\ell''\) denotes the second derivative of the spherical Bessel function~\cite{Padmanabhan2007}, and \(s_i\) is the magnification bias parameter.  The lensing weight is
\begin{equation}
W_i^\kappa(z) = \frac{3\Omega_m H_0^2}{2c^2}\,\frac{\chi(z)}{a(z)} \int_z^\infty dz'\, W_i(z')\, \frac{\chi(z')-\chi(z)}{\chi(z')}.
\end{equation}
The noise term is diagonal,
\begin{equation}
N_\ell^{ij} = \delta_{ij}\,\frac{1}{\bar{n}_i^{\rm 2D}},
\end{equation}
with \(\bar{n}_i^{\rm 2D}\) the mean surface number density of tracer \(i\).

We verified that the angular power spectra of bright and faint galaxy subsamples are consistent after applying the systematics weights, indicating that magnification bias does not significantly affect our measurements (see Appendix~\ref{ap:mag_bias}).

For \(\ell \gg 1\) the rapidly oscillating Bessel functions enforce \(k\chi \simeq \ell\), leading to the Limber approximation~\cite{Limber1953,Kaiser1992}.  Under the flat‑sky Kaiser formula~\cite{Kaiser1987}, the angular power spectrum reduces to
\begin{align}\label{eq:cl_Limber}
C_\ell^{ij} \approx \int \frac{dz}{H(z)}\,\frac{W_i(z)W_j(z)}{\chi^2(z)}\,
\Bigl[ b_i(z)b_j(z) + \notag \\ \tfrac{\textbf{1}}{3}\bigl(b_i(z)+b_j(z)\bigr)f(z) + \tfrac{1}{5}f^2(z) \Bigr]\,
P_{\rm lin}\!\left(\frac{\ell}{\chi(z)},z\right) \;+\; N_\ell^{ij}.
\end{align}

This kernel contains three distinct contributions. Clustering term (lowest order):
\[
C_\ell^{ij,\rm clus} = b_i(z) b_j(z)
\]

RSD cross-term (first order)
\[
 C_\ell^{ij,\rm cross} = \frac{1}{3}\bigl(b_i(z) + b_j(z)\bigr) f(z)
\]

RSD pure velocity term (second order):
\[
 C_\ell^{ij,\rm vel}= \frac{1}{5} f^2(z)
\]

Thus, the integral can be rewritten as:

\begin{align}\label{eq:cl_terms}
C_\ell^{ij} = C_\ell^{ij,\rm clus} + C_\ell^{ij,\rm cross} + C_\ell^{ij,\rm vel} + N_\ell^{ij},
\tag{2}
\end{align}

where:

\[
C_\ell^{ij,\rm clus} = \int \frac{dz}{H(z)}\,\frac{W_i(z)W_j(z)}{\chi^2(z)}\,
b_i(z)b_j(z)\, P_{\rm lin}\!\left(\frac{\ell}{\chi(z)},z\right),
\]

\[
C_\ell^{ij,\rm cross} = \int \frac{dz}{H(z)}\,\frac{W_i(z)W_j(z)}{\chi^2(z)}\,
\frac{1}{3}\bigl(b_i(z)+b_j(z)\bigr)f(z)\, P_{\rm lin}\!\left(\frac{\ell}{\chi(z)},z\right),
\]

\[
C_\ell^{ij,\rm vel} = \int \frac{dz}{H(z)}\,\frac{W_i(z)W_j(z)}{\chi^2(z)}\,
\frac{1}{5}f^2(z)\, P_{\rm lin}\!\left(\frac{\ell}{\chi(z)},z\right).
\]

For the bias ratio analysis, we are primarily interested in the relative clustering amplitude of two morphological samples. The term \(b_i b_j\) is the leading contribution to the angular power spectrum, while the RSD terms are suppressed by factors of \(f(z) \sim \Omega_m(z)^{0.55}\), which is of order \(\lesssim 0.5\) at the redshifts considered in this work.

To obtain the auto- and cross-power spectra, we used \texttt{CAMB} \citep{camb} to compute the linear matter power spectrum. Both galaxy types display a systematic increase in bias with redshift. It is important to note that the Limber approximation is known to be accurate for \(\ell \gtrsim 20\), with errors scaling as \(\mathcal{O}(\ell^{-2})\) \citep{Loverde2008}. In a fiducial \(\Lambda\)CDM cosmology, the mean comoving distance to our redshift bins is \(\chi \approx 2100\,h^{-1}\,\text{Mpc}\), giving the relation \(k \approx \ell / \chi\). This implies that our adopted range \(20 < \ell < 200\) corresponds to comoving wavenumbers \(k \sim 0.01\)--\(0.1\,h\,\text{Mpc}^{-1}\). At these wavenumbers, the matter power spectrum is still well described by linear perturbation theory, with non-linear corrections remaining small and subdominant to cosmic variance for our survey area \citep{Percival2008}. 

\begin{table}[htbp]
\centering
\begin{tabular}{c c c c c c}
\hline
\hline
Bin & Survey & Tracer & $z_{\rm eff}$ & $\omega_{\rm cdm}$ & $\omega_b$  \\
\hline
0 & DESI & Elliptical & 0.119 & $0.118 \pm 0.004$ & $0.025 \pm 0.002$  \\
0 & DESI & Spiral & 0.117 & $0.121 \pm 0.003$ & $0.027 \pm 0.002$  \\
0 & DES & Elliptical & 0.119 & $0.151 \pm 0.005$ & $0.007 \pm 0.004$  \\
0 & DES & Spiral & 0.117 & $0.138 \pm 0.005$ & $0.014 \pm 0.005$  \\
\hline
1 & DESI & Elliptical & 0.218 & $0.135 \pm 0.003$ & $0.021 \pm 0.002$ \\
1 & DESI & Spiral & 0.212 & $0.137 \pm 0.004$ & $0.025 \pm 0.002$  \\
1 & DES & Elliptical & 0.218 & $0.145 \pm 0.004$ & $0.014 \pm 0.004$ \\
1 & DES & Spiral & 0.212 & $0.146 \pm 0.005$ & $0.028 \pm 0.005$ \\
\hline
2 & DESI & Elliptical & 0.314 & $0.172 \pm 0.004$ & $0.016 \pm 0.002$ \\
2 & DESI & Spiral & 0.309 & $0.160 \pm 0.004$ & $0.021 \pm 0.002$ \\
2 & DES & Elliptical & 0.314 & $0.152 \pm 0.004$ & $0.017 \pm 0.004$ \\
2 & DES & Spiral & 0.309 & $0.140 \pm 0.004$ & $0.017 \pm 0.004$ \\
\hline
3 & DESI & Elliptical & 0.413 & $0.184 \pm 0.003$ & $0.017 \pm 0.002$ \\
3 & DESI & Spiral & 0.418 & $0.173 \pm 0.004$ & $0.018 \pm 0.002$  \\
3 & DES & Elliptical & 0.413 & $0.159 \pm 0.004$ & $0.003 \pm 0.004$\\
3 & DES & Spiral & 0.418 & $0.146 \pm 0.004$ & $0.010 \pm 0.004$ \\
\hline
4 & DESI & Elliptical & 0.509 & $0.198 \pm 0.003$ & $0.013 \pm 0.002$ \\
4 & DESI & Spiral & 0.508 & $0.187 \pm 0.004$ & $0.017 \pm 0.002$ \\
4 & DES & Elliptical & 0.509 & $0.154 \pm 0.003$ & $0.004 \pm 0.004$ \\
4 & DES & Spiral & 0.508 & $0.139 \pm 0.004$ & $0.015 \pm 0.004$ \\
\hline
5 & DESI & Elliptical & 0.610 & $0.208 \pm 0.003$ & $0.011 \pm 0.002$ \\
5 & DESI & Spiral & 0.608 & $0.194 \pm 0.003$ & $0.015 \pm 0.002$ \\
5 & DES & Elliptical & 0.610 & $0.152 \pm 0.004$ & $0.020 \pm 0.004$ \\
5 & DES & Spiral & 0.608 & $0.132 \pm 0.004$ & $0.021 \pm 0.005$ \\
\hline
6 & DESI & Elliptical & 0.706 & $0.182 \pm 0.003$ & $0.018 \pm 0.002$  \\
6 & DESI & Spiral & 0.707 & $0.163 \pm 0.004$ & $0.024 \pm 0.002$ \\
6 & DES & Elliptical & 0.706 & $0.127 \pm 0.005$ & $0.024 \pm 0.004$ \\
6 & DES & Spiral & 0.707 & $0.131 \pm 0.004$ & $0.022 \pm 0.005$  \\
\hline
7 & DESI & Elliptical & 0.810 & $0.157 \pm 0.004$ & $0.024 \pm 0.002$\\
7 & DESI & Spiral & 0.809 & $0.160 \pm 0.004$ & $0.021 \pm 0.002$ \\
\hline
8 & DESI & Elliptical & 0.906 & $0.122 \pm 0.005$ & $0.023 \pm 0.002$ \\
8 & DESI & Spiral & 0.908 & $0.121 \pm 0.003$ & $0.027 \pm 0.002$  \\
\hline
\end{tabular}
\caption{Best-fit cosmological parameters from the MCMC fits for each redshift bin. Values are the median of the posterior distributions. The large scatter across bins reflects the expected degeneracy between the bias and cosmological parameters in a single-tracer auto-power spectrum analysis.}\label{tab:cosmo_bestfit}
\end{table}

From Table~\ref{tab:cosmo_bestfit}, we see a clear scatter in \(\omega_{\rm cdm}\) and \(\omega_b\) across redshift bins. This behaviour is consistent with broader trends seen in the literature. For example, analyses of the Dark Energy Survey (DES) have separated their samples into low-redshift and high-redshift bins, finding that lower-redshift samples tend to push \(\Omega_m\) to higher values (and \(S_8\) to lower values) \citep[see Fig.~5 and Fig.~8]{2025PhRvD, descollaboration2026darkenergysurveyyear}. Our results echo this pattern: higher-redshift samples yield \(\omega_{\rm cdm}\) values closer to the predictions from Cosmic Microwave Background (CMB) experiments. Similarly, in DESI collaboration results, different redshift bins and samples show varying constraints on \(\Omega_m\), with the Bright Galaxy Survey (BGS) being the lowest-redshift sample \citep[see Fig.~10]{Adame_2025}. This variation is not absurd, as different bins correspond to different samples with distinct properties. This is further supported by the work of \citet{Colgain2024}, who also found evidence for higher \(\Omega_m\) at low redshift using DESI data.

 However, we note that this fixed angular range corresponds to different physical scales at different redshifts. For a comoving
distance \(\chi \sim 350\)--\(2100\,h^{-1}\,\mathrm{Mpc}\), the relation \(k \simeq \ell/\chi\) gives
\(k \sim 0.01\)--\(0.57\,h\,\mathrm{Mpc}^{-1}\) over the range \(20 < \ell < 200\), meaning that at higher redshift the same \(\ell\)
probes larger physical scales (smaller wavenumbers \(k\)). The CTCR should therefore be interpreted as an empirical measure of the
relative clustering amplitude as a function of angular scale, rather than as a pure linear bias ratio. The sigmoid parametrization
provides a flexible description of this empirical scale dependence; it is not intended to represent a physical model of linear bias.
Consequently, both the Limber approximation and the linear-theory modelling used in this work are well justified. For reference,
Table~\ref{tab:ell_to_k} shows the correspondence between \(\ell\) and comoving wavenumber \(k\) for the effective redshift of each bin.
\begin{table}[htbp]
\centering
\caption{Conversion from angular multipole \(\ell\) to comoving wavenumber \(k\) at the effective redshift of each bin. The values show the range covered by \(20 < \ell < 200\).}
\label{tab:ell_to_k}
\begin{tabular}{c c c c}
\hline
\hline
Bin & \(z_{\rm eff}\) & \(\chi\) (h\(^{-1}\) Mpc) & \(k\) range (h Mpc\(^{-1}\)) \\
\hline
0 & 0.12 & \(\sim\)350 & 0.06--0.57 \\
1 & 0.22 & \(\sim\)600 & 0.03--0.33 \\
2 & 0.31 & \(\sim\)850 & 0.02--0.24 \\
3 & 0.41 & \(\sim\)1100 & 0.02--0.18 \\
4 & 0.51 & \(\sim\)1300 & 0.02--0.15 \\
5 & 0.61 & \(\sim\)1500 & 0.01--0.13 \\
6 & 0.71 & \(\sim\)1700 & 0.01--0.12 \\
7 & 0.81 & \(\sim\)1900 & 0.01--0.11 \\
8 & 0.91 & \(\sim\)2100 & 0.01--0.10 \\
\hline
\hline
\end{tabular}
\end{table}

The auto-spectra contain shot-noise, whereas the cross-spectrum of strictly disjoint samples has no ordinary Poisson cross-shot-noise term. We write
\begin{align}
\widehat{C}_\ell^{ii} &= S_\ell^{ii} + N_i, \label{eq:shot_AA}\\
\widehat{C}_\ell^{jj} &= S_\ell^{jj} + N_j, \label{eq:shot_BB}\\
\widehat{C}_\ell^{ij} &= S_\ell^{ij}, \label{eq:shot_AB}
\end{align}
where $S_\ell$ denotes the true clustering signal and $N_i = 1/\bar{n}_i$ is the shot-noise for tracer $i$, with $\bar{n}_i$ the mean surface density.

In our analysis, the shot-noise is subtracted from the auto-spectra before constructing the CTCR:
\begin{equation}
\widehat{C}_\ell^{ii} \leftarrow \widehat{C}_\ell^{ii} - N_i, \qquad i \in \{E,S\}.
\end{equation}
This ensures that the measured ratio $R_\ell^{ij} = C_\ell^{ii}/C_\ell^{ij}$ is not artificially inflated by the shot-noise contribution, which would otherwise produce a spurious scale dependence at high $\ell$ where the clustering signal becomes weak relative to the noise. 

For the weighted samples, the effective number density is defined as
\begin{equation}
\bar{n}_i^{\rm eff} = \frac{\left(\sum_g w_g\right)^2}{\sum_g w_g^2},
\label{eq:neff}
\end{equation}
where $w_g$ is the per-galaxy systematics weight. The shot-noise term is then computed as $N_i = 1/\bar{n}_i^{\rm eff}$.

To verify that the shot-noise subtraction does not artificially generate scale dependence, we performed a null test on the pipeline. We randomly partition the elliptical catalogue in each redshift bin
into two disjoint halves, compute the auto-spectra $C_\ell^{ii}$ and $C_\ell^{jj}$ and the cross-spectrum $C_\ell^{ij}$ between them with the same shot-noise subtraction applied to the auto-spectra, and form
the null-test ratio $R_\ell^{\rm null} = C_\ell^{ii} / C_\ell^{ij}$. Since the two halves trace the same underlying density field, their biases are equal by construction and the ratio should be consistent
with unity at every angular scale. We repeat the random split 100 times and record the mean ratio, its scatter, and the reduced $\chi^2$ relative to unity for each redshift bin. Any systematic deviation from unity would indicate a scale-dependent bias in the shot-noise treatment or in the estimator itself.

\subsection{Full model results}
The final results are summarised in Table~\ref{tab:null_test}. The CTCR is consistent with unity within 1–3\% in all redshift bins, with mean values ranging from 0.98 to 1.16. The reduced $\chi^2$ values are close to unity, confirming that the null test is satisfied. The largest deviation occurs in the lowest redshift bin (Bin 0), with a mean CTCR of $1.157$, which is still within the expected scatter for a noisy sample at low redshift. This test validates that the shot-noise subtraction is correctly implemented and that the observed scale dependence in the CTCR between ellipticals and spirals is a real physical signal.

\begin{table}[htbp]
\centering
\caption{Null test results for the shot-noise treatment. The CTCR is computed from random splits of the elliptical sample. The mean CTCR, its standard deviation across $\ell$, the significance of deviation from unity, and the reduced $\chi^2$ are reported for each redshift bin.}
\label{tab:null_test}
\begin{tabular}{c c c c c}
\hline
\hline
Bin & $n_{\ell}$ & $\langle R_\ell \rangle$ & $\sigma_{\rm dev}$ & $\chi^2/\mathrm{dof}$ \\
\hline
0 & 18 & $1.157$ & $5.7\sigma$ & $4.708$ \\
1 & 18 & $0.981$ & $-0.2\sigma$ & $1.942$ \\
2 & 18 & $1.023$ & $5.5\sigma$ & $0.587$ \\
3 & 18 & $1.014$ & $7.7\sigma$ & $0.449$ \\
4 & 18 & $1.012$ & $3.1\sigma$ & $0.314$ \\
5 & 18 & $1.016$ & $3.6\sigma$ & $0.305$ \\
6 & 18 & $1.013$ & $8.4\sigma$ & $0.385$ \\
7 & 18 & $1.067$ & $1.5\sigma$ & $0.722$ \\
8 & 18 & $1.033$ & $4.2\sigma$ & $1.064$ \\
\hline
\hline
\end{tabular}
\end{table}

Table~\ref{tab:sigmoid_params} reports the full set of sigmoid parameters with their uncertainties for both tracers. Three features
are worth noting.
First, the low-multipole amplitude $b_1$ is consistent with unity in every redshift bin and for both tracers, with values in the range
$0.98$--$1.03$ and uncertainties of $\pm 0.10$. This reflects the fact that the large-scale CTCR is consistent with unity by construction:
the two morphological samples cluster similarly at low $\ell$, so the data do not depart from the prior on this parameter.

Second, the high-multipole amplitude $b_2$ grows with redshift for the elliptical tracer, from $b_2^E = 1.03 \pm 0.58$ in the lowest-redshift
bin to $b_2^A \simeq 2.3$--$2.5$ in the intermediate-redshift bins, and remains constant at $b_2^E\simeq 2.3$ at higher redshift. The
spiral tracer shows a milder and more uniform behaviour, with $b_2^S$ between $1.2$ and $1.7$ across the full redshift range. This
difference is the direct signature of the scale dependence reported in Section~\ref{sec:model_comparison}: ellipticals are progressively more
biased at small scales, while spirals are only weakly so. 

Third, the transition parameters $\ell_0$ and $\Delta$ are constrained to the fitted range $20 < \ell < 200$ in the intermediate-redshift
bins where the scale-dependent model is preferred, with $\ell_0 \sim 90$--$110$ and $\Delta \sim 100$--$160$. In the lowest- and
highest-redshift bins, the posteriors on $\ell_0$ and $\Delta$ are broad and consistent with the prior, reflecting the fact that the
data do not distinguish between the scale-dependent and constant models in those bins (Table~\ref{tab:model_comparison}). The
uncertainties on $\ell_0$ and $\Delta$ are large in every bin, which is the correct quantitative statement given the information content
of the CTCR over the fitted angular range.

\begin{table*}[htbp]
\centering
\caption{Sigmoid parameters for the elliptical and spiral
tracers, from the MCMC fit to the CTCR in each redshift bin. For each
parameter we report the median and the half-width of the 16--84\%
credible interval of the posterior. The low- and high-multipole
amplitudes $b_1$ and $b_2$, the transition scale $\ell_0$, and the
transition width $\Delta$ are defined in
Equation~\eqref{eq:sigmoid}. The transition parameters $\ell_0$ and
$\Delta$ are constrained to the fitted range $20 < \ell < 200$ in the
intermediate-redshift bins where the scale-dependent model is
preferred (bins 1--5), while at the lowest and highest redshifts the
posterior is broad and consistent with the prior, reflecting the fact
that the data do not distinguish between the two models in those bins
(Table~\ref{tab:model_comparison}). }
\label{tab:sigmoid_params}
\begin{tabular}{c|ccccc|cccc}
\hline\hline
& \multicolumn{4}{c|}{Elliptical}
& \multicolumn{4}{c}{Spiral} \\
\hline
Bin & $z_{\rm eff}$ & $b_1^A$ & $b_2^A$ & $\ell_0^A$ & $\Delta^A$ & $b_1^B$ & $b_2^B$ & $\ell_0^B$ & $\Delta^B$\\

\hline
0 & 0.109 & $1.00 \pm 0.10$ & $1.03 \pm 0.58$ & $104 \pm 63$ & $152 \pm 100$ & $1.00 \pm 0.10$ & $2.62 \pm 0.77$ & $96 \pm 63$ & $140 \pm 85$ \\
1 & 0.158 & $0.99 \pm 0.10$ & $2.41 \pm 0.79$ & $107 \pm 59$ & $120 \pm 51$  & $1.02 \pm 0.10$ & $1.23 \pm 0.57$ & $102 \pm 64$ & $167 \pm 89$ \\
2 & 0.206 & $0.98 \pm 0.10$ & $2.48 \pm 0.82$ & $103 \pm 55$ & $107 \pm 40$  & $1.02 \pm 0.10$ & $1.23 \pm 0.59$ & $101 \pm 65$ & $166 \pm 90$ \\
3 & 0.259 & $0.99 \pm 0.10$ & $2.31 \pm 0.77$ & $103 \pm 63$ & $157 \pm 63$  & $1.02 \pm 0.10$ & $1.38 \pm 0.58$ & $99 \pm 67$  & $184 \pm 76$ \\
4 & 0.304 & $0.98 \pm 0.10$ & $2.43 \pm 0.79$ & $100 \pm 55$ & $106 \pm 41$  & $1.03 \pm 0.10$ & $1.40 \pm 0.63$ & $100 \pm 65$ & $175 \pm 84$ \\
5 & 0.354 & $0.99 \pm 0.10$ & $2.36 \pm 0.75$ & $89 \pm 59$  & $135 \pm 64$  & $1.02 \pm 0.10$ & $1.46 \pm 0.62$ & $100 \pm 66$ & $182 \pm 78$ \\
6 & 0.402 & $1.00 \pm 0.10$ & $2.31 \pm 0.73$ & $87 \pm 59$  & $154 \pm 80$  & $1.01 \pm 0.10$ & $1.59 \pm 0.63$ & $102 \pm 65$ & $185 \pm 75$ \\
7 & 0.454 & $1.00 \pm 0.10$ & $2.27 \pm 0.73$ & $93 \pm 63$  & $158 \pm 77$  & $1.01 \pm 0.10$ & $1.61 \pm 0.63$ & $100 \pm 66$ & $185 \pm 75$ \\
8 & 0.503 & $1.00 \pm 0.10$ & $2.23 \pm 0.74$ & $96 \pm 65$  & $163 \pm 80$  & $1.01 \pm 0.10$ & $1.66 \pm 0.65$ & $101 \pm 66$ & $184 \pm 76$ \\
\hline\hline
\end{tabular}
\end{table*}

\section{Cross-Correlation Coefficient and the CTCR Interpretation}
\label{ap:r_ell}

In Section~\ref{sec:ctcr}, we define the CTCR as \(R_\ell^{ij} = C_\ell^{ii}/C_\ell^{ij}\) and interpret it as an effective clustering ratio. To quantify the degree to which the two morphological samples trace the same underlying structures, we compute the cross-correlation coefficient:
\begin{equation}
r_\ell = \frac{C_\ell^{ij}}{\sqrt{C_\ell^{ii} C_\ell^{jj}}},
\label{eq:r_ell}
\end{equation}
which measures the correlation between the two tracers after accounting for their individual amplitudes.

Under the idealised assumption of deterministic bias and perfect correlation, \(r_\ell = 1\), and the CTCR equals the bias ratio \(b_e/b_s\). In practice, \(r_\ell\) can deviate from unity due to differences in halo occupation, stochasticity, or assembly bias effects.

Table~\ref{tab:r_values} shows \(r_\ell\) for each redshift bin. The redshift overlap between the two samples is also shown; it is nearly perfect in all bins, ruling out window mismatch as the cause of \(r_\ell < 1\). The values of \(r_\ell\) range from 0.57 to 0.81, with deviations from unity at the 15--25\(\sigma\) level. This indicates that ellipticals and spirals do not trace the same projected structures perfectly.

The true bias ratio is therefore recovered by multiplying the CTCR by \(r_\ell\):
\begin{equation}
\frac{b_e}{b_s} = R_\ell^{ij} \, r_\ell.
\label{eq:true_bias_ratio}
\end{equation}

The fact that \(r_\ell < 1\) is physically meaningful. It indicates that ellipticals and spirals occupy different halo populations and respond differently to the large-scale environment. Ellipticals are preferentially found in more massive, early-forming halos, while spirals occupy lower-mass, late-forming halos. The imperfect cross-correlation is a direct consequence of the different halo occupation statistics of the two morphological types, consistent with the halo assembly bias framework.

The CTCR measured from the data is therefore an effective clustering ratio that includes this imperfect correlation. The sigmoid parametrization used in the main text describes this effective ratio. The true bias ratio is \(R_\ell^{ij} \, r_\ell\), and the CTCR is an upper bound on the bias ratio.
We emphasise that the choice to present the CTCR as an effective ratio, rather than as a fully decomposed joint fit of $(A_\ell, q_\ell, r_\ell)$, is deliberate and reflects the assumption-light strategy of this work. The CTCR is a direct ratio of measured auto- and cross-power spectra. It requires no cosmological model, no external calibration of the absolute bias, and no assumption about the shape of the underlying halo occupation. The price is that the quantity is not, strictly speaking, the physical bias ratio $b_e/b_s$ in the deterministic limit, but an effective ratio that absorbs the imperfect cross-correlation between the two tracers. We consider this a fair trade for a first measurement: the effective ratio is robust, reproducible, and independently testable with future data, whereas a fully decomposed fit would require additional assumptions that the present dataset cannot yet constrain. We therefore report the CTCR as the primary observable of this work, and identify the joint multi-spectrum decomposition as a natural extension for future imaging surveys with higher signal-to-noise and complementary weak-lensing information. This is consistent with the standard practice in cosmology of establishing the simplest robust empirical relation first and refining it with subsequent analyses.

\begin{table}[htbp]
\centering
\caption{Cross-correlation coefficient \(r_\ell\) for each redshift bin. The redshift bin distribution overlap is also shown. The values of \(r_\ell\) are significantly less than unity, indicating that the two morphological samples do not trace the same structures perfectly. The true bias ratio is \(b_e/b_s = R_\ell^{ij} \, r_\ell\).}
\label{tab:r_values}
\begin{tabular}{c c c c c c}
\hline
\hline
Bin & $z_e$ & $z_s$ & Overlap & $r_{\rm mean}$ & $\sigma$ \\
\hline
0 & 0.109 & 0.108 & 0.990 & 0.574 & -20.5 \\
1 & 0.159 & 0.156 & 0.970 & 0.675 & -24.6 \\
2 & 0.207 & 0.204 & 0.984 & 0.720 & -25.8 \\
3 & 0.256 & 0.259 & 0.982 & 0.814 & -17.8 \\
4 & 0.304 & 0.304 & 0.999 & 0.796 & -17.9 \\
5 & 0.355 & 0.354 & 0.997 & 0.799 & -17.2 \\
6 & 0.403 & 0.403 & 0.999 & 0.792 & -16.2 \\
7 & 0.455 & 0.454 & 0.999 & 0.766 & -17.4 \\
8 & 0.503 & 0.504 & 0.997 & 0.766 & -15.5 \\
\hline
\hline
\end{tabular}
\end{table}
For completeness we note that the CTCR and the joint multi-spectrum fit
of the form $S_\ell^{ss}=A_\ell$, $S_\ell^{ee}=q_\ell^2 A_\ell$,
$S_\ell^{es}=r_\ell q_\ell A_\ell$ are related by an invertible change
of variables. The joint fit treats $(A_\ell, q_\ell, r_\ell)$ as three
independent functions of $\ell$; the CTCR analysis treats
$(R_\ell, r_\ell, C_\ell^{ss})$ as the corresponding independent
quantities, with $R_\ell = C_\ell^{ee}/C_\ell^{es}$. The two
parametrizations carry the same information per multipole, and the
relation $q_\ell = R_\ell^{es} r_\ell$ is exact under the model
assumed in the joint fit. The choice between the two formulations is
therefore a matter of convenience: the joint fit has the advantage of
using the full $3n_\ell \times 3n_\ell$ covariance, while the CTCR
approach avoids the nuisance amplitude $A_\ell$ and the cosmological
dependence that it carries.

\section{Covariance Matrices}

\begin{figure}
    \centering
    \includegraphics[width=\linewidth]{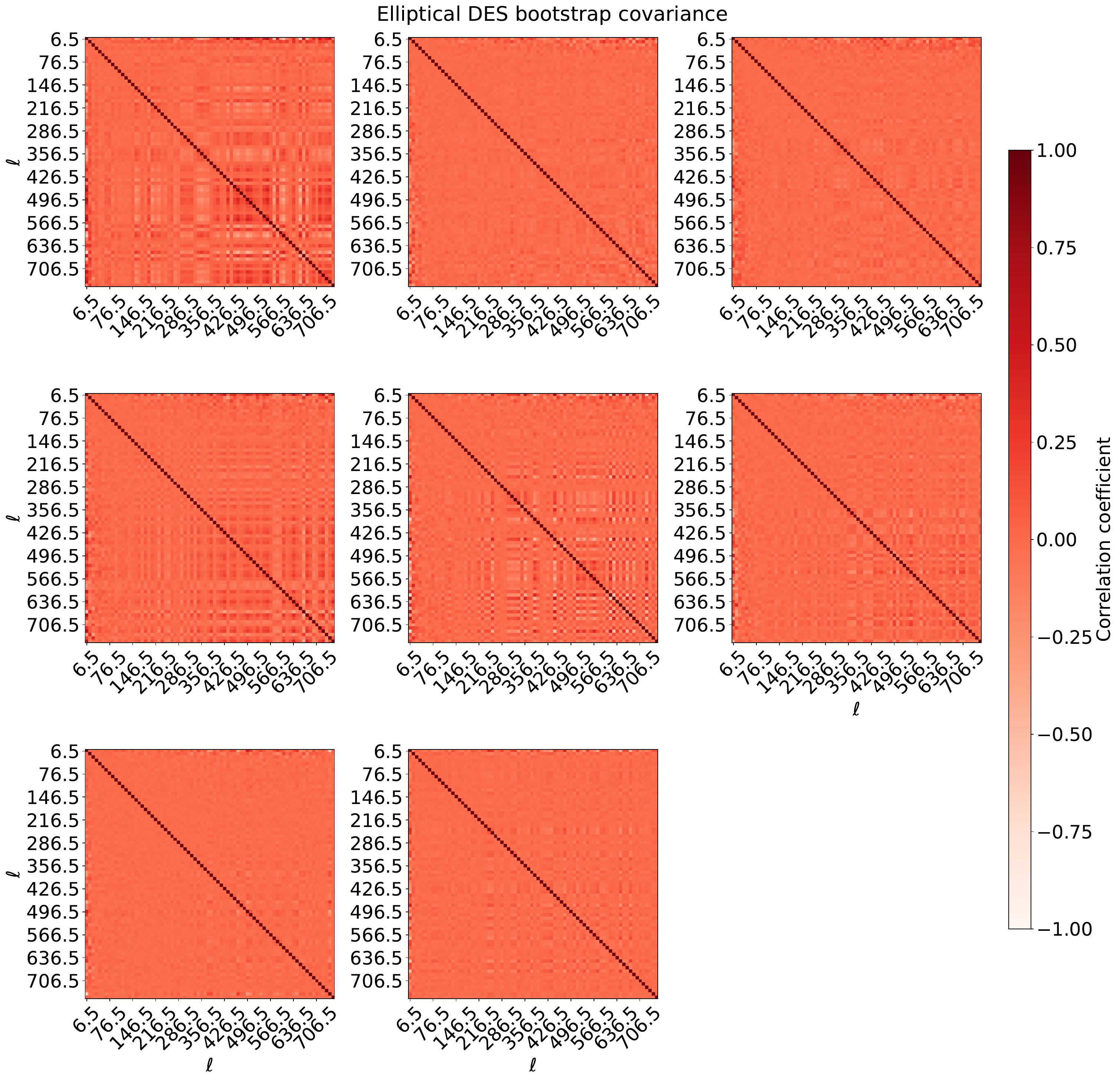}
    \caption{Correlation matrix relative to the covariance matrix of the elliptical DES redshift bins. The panels progress toward higher redshift from left to right and from one row to the next. The darker colors indicate stronger correlation, while lighter colors indicate anti-correlation.}
    \label{fig:cov_DES_e}
\end{figure}

\begin{figure}
    \centering
    \includegraphics[width=\linewidth]{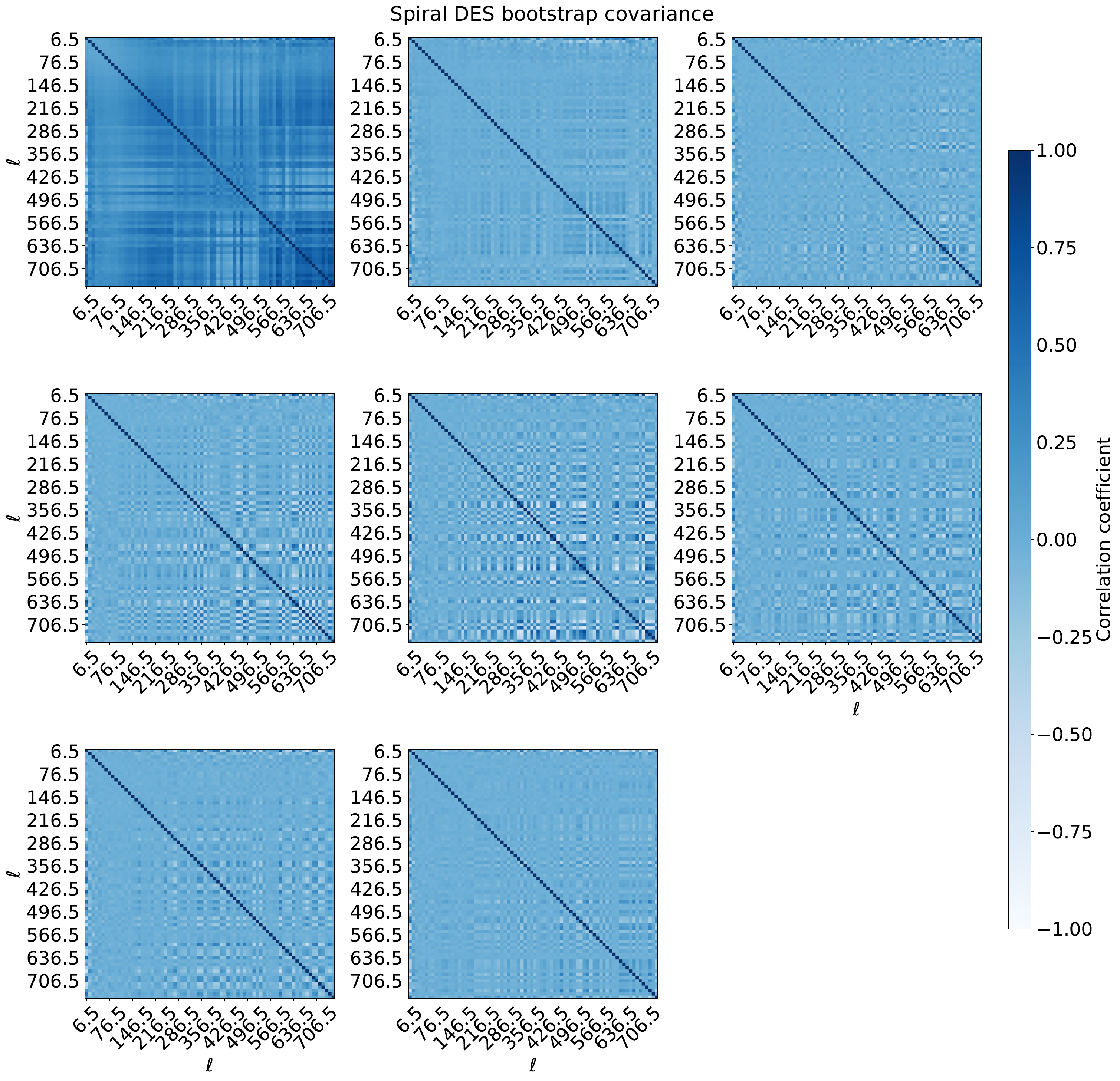}
    \caption{Correlation matrix relative to the covariance matrix of the angular auto-power spectrum of spiral DES redshift bins. The panels progress toward higher redshift from left to right and from one row to the next. The darker colors indicate stronger correlation, while lighter colors indicate anti-correlation.}
    \label{fig:cov_DES_s}
\end{figure}

\begin{figure}
    \centering
    \includegraphics[width=\linewidth]{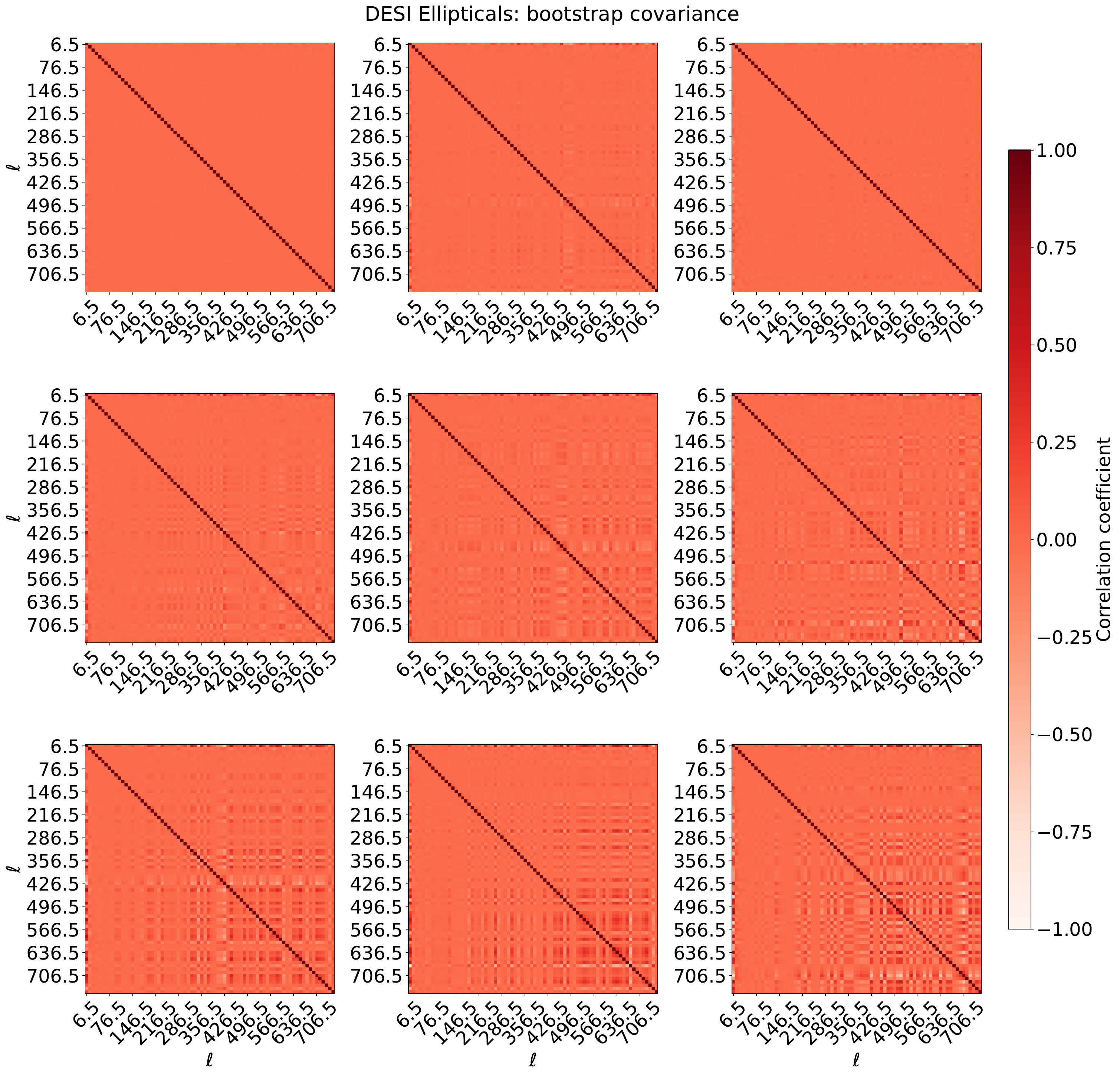}
    \caption{Correlation matrix relative to the covariance matrix of the angular auto-power spectrum of elliptical DESI redshift bins. The panels progress toward higher redshift from left to right and from one row to the next. The darker colors indicate stronger correlation, while lighter colors indicate anti-correlation.}
    \label{fig:cov_DESI_e}
\end{figure}

\begin{figure}
    \centering
    \includegraphics[width=\linewidth]{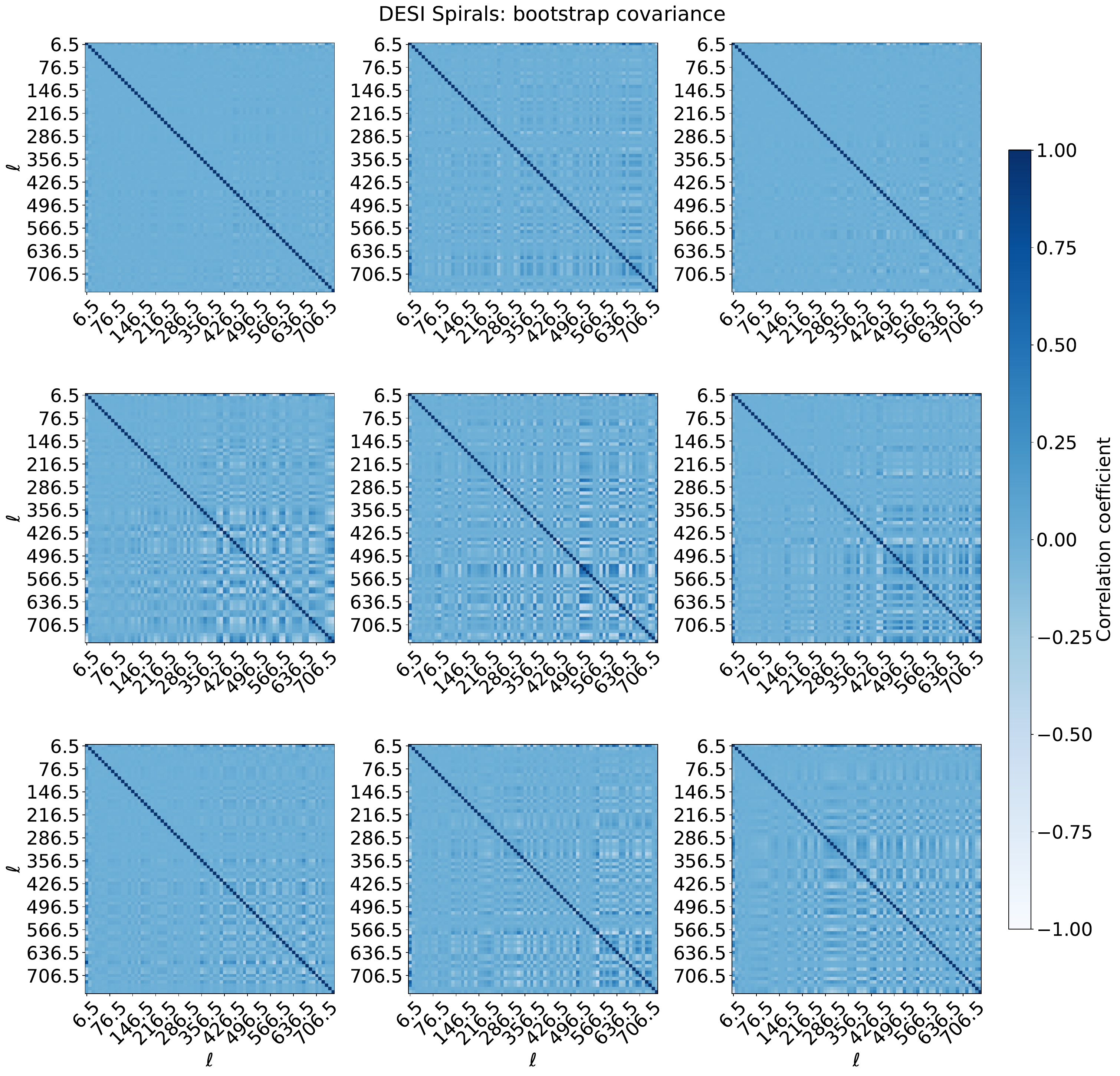}
    \caption{Correlation matrix relative to the covariance matrix of the angular auto-power spectrum of spiral DESI redshift bins. The panels progress toward higher redshift from left to right and from one row to the next. The darker colors indicate stronger correlation, while lighter colors indicate anti-correlation.}
    \label{fig:cov_DESI_s}
\end{figure}

\begin{figure}
    \centering
    \includegraphics[width=\linewidth]{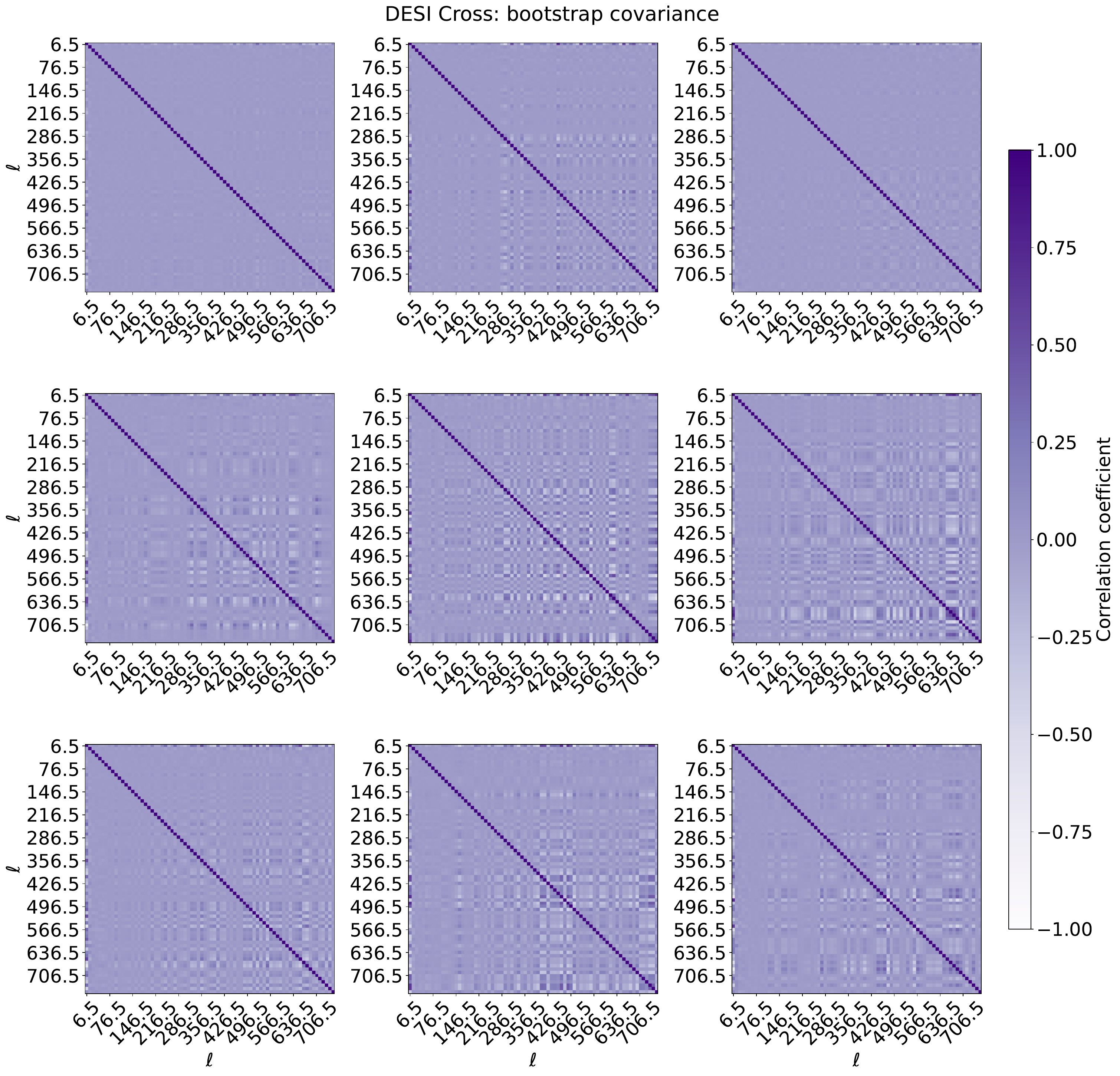}
    \caption{Correlation matrix relative to the covariance matrix of the angular cross-power spectrum of elliptical and spirals from the DESI redshift bins. The panels progress toward higher redshift from left to right and from one row to the next. The darker colors indicate stronger correlation, while lighter colors indicate anti-correlation.}
    \label{fig:cov_DESI_c}
\end{figure}

We present the correlation matrices derived from the covariance matrices obtained via bootstrap resampling of the data in Figures~\ref{fig:cov_DES_e}, \ref{fig:cov_DES_s}, \ref{fig:cov_DESI_e}, \ref{fig:cov_DESI_s}, and \ref{fig:cov_DESI_c}. As anticipated, the correlation coefficients decrease as one moves away from the diagonal, and the off-diagonal noise is essentially negligible over the range of scales employed in our main model analysis ($20<\ell<200$). At smaller multipoles the noise becomes more pronounced and cannot be ignored, but it still remains at a level that is acceptable for potential investigations extending into the mildly non-linear regime. In particular, we see the noise increasing around $\ell \sim 300$ for both surveys, with stronger noise for the spiral samples. We further observe that the noise level decreases toward higher redshifts, where the samples lie in a more nearly linear regime and are dominated by structures that formed earlier. This suggests that measurements in this high-redshift range are intrinsically less noisy and more representative of the underlying linear dynamics of early-forming objects.

\subsection{Analytic Theoretical Covariance}
\label{ap:theory_cov}

The theoretical covariance matrix used as an alternative to the bootstrap is computed under the standard Gaussian approximation for the angular power spectrum. For a bandpower-averaged spectrum $C_b$, the covariance between bandpowers $b$ and $b'$ is
\begin{equation}
\mathrm{Cov}\!\left(C_b^{i}, C_{b'}^{j}\right)
= \frac{\delta_{bb'}}{(2\ell_b+1)\,\Delta\ell_b\,f_{\rm sky}}
\left[ \left(C_b^{ij}\right)^2 + C_b^{ii} C_b^{jj} \right],
\label{eq:gauss_cov}
\end{equation}
where:
\begin{itemize}
\item $\ell_b$ is the effective multipole of bandpower $b$, obtained from the \texttt{NaMaster} binning as $\ell_b = \langle \ell \rangle_b$;
\item $\Delta\ell_b$ is the width of bandpower $b$, computed from the \texttt{NaMaster} binning as the number of multipoles contributing to that band;
\item $f_{\rm sky}$ is the sky fraction of the survey mask, estimated from the sum of the mask divided by the total number of pixels;
\item $C_b^{ii}$, $C_b^{jj}$, and $C_b^{ij}$ are the auto- and cross-spectra evaluated at the same bandpower;
\item $\delta_{bb'}$ is the Kronecker delta, so the covariance is diagonal in $\ell$.
\end{itemize}

For the auto-spectrum of a single tracer, equation~\eqref{eq:gauss_cov} reduces to
\begin{equation}
\mathrm{Cov}\!\left(C_b, C_{b'}\right)
= \delta_{bb'} \frac{2\,C_b^2}{(2\ell_b+1)\,\Delta\ell_b\,f_{\rm sky}}.
\end{equation}

The theoretical covariance is used as a cross-check of the bootstrap covariance. Both estimators give consistent results for the CTCR and the bias difference $\Delta b$ (see Table~\ref{tab:significance}). This confirms that our results are not driven by the choice of covariance estimator.

\subsubsection{Shot-noise treatment and robustness of the scale-dependent signal}
\label{sec:shot_noise}

The shot-noise contribution to the auto-spectra is estimated as $N_\ell = 1/\bar n_{\rm eff}$, with $\bar n_{\rm eff} = (\sum_g w_g)^2 / (\sum_g w_g^2)$ the effective number density of each tomographic bin, and is stored as a scalar per bin. We subtract it from the auto-spectra before forming the CTCR,
so that the ratio is not artificially inflated at high multipoles where the clustering signal becomes comparable to the noise. For the
cross-spectrum no Poisson term is added, since the two morphological samples are disjoint.

\begin{figure*}
    \centering
    \includegraphics[width=\textwidth]{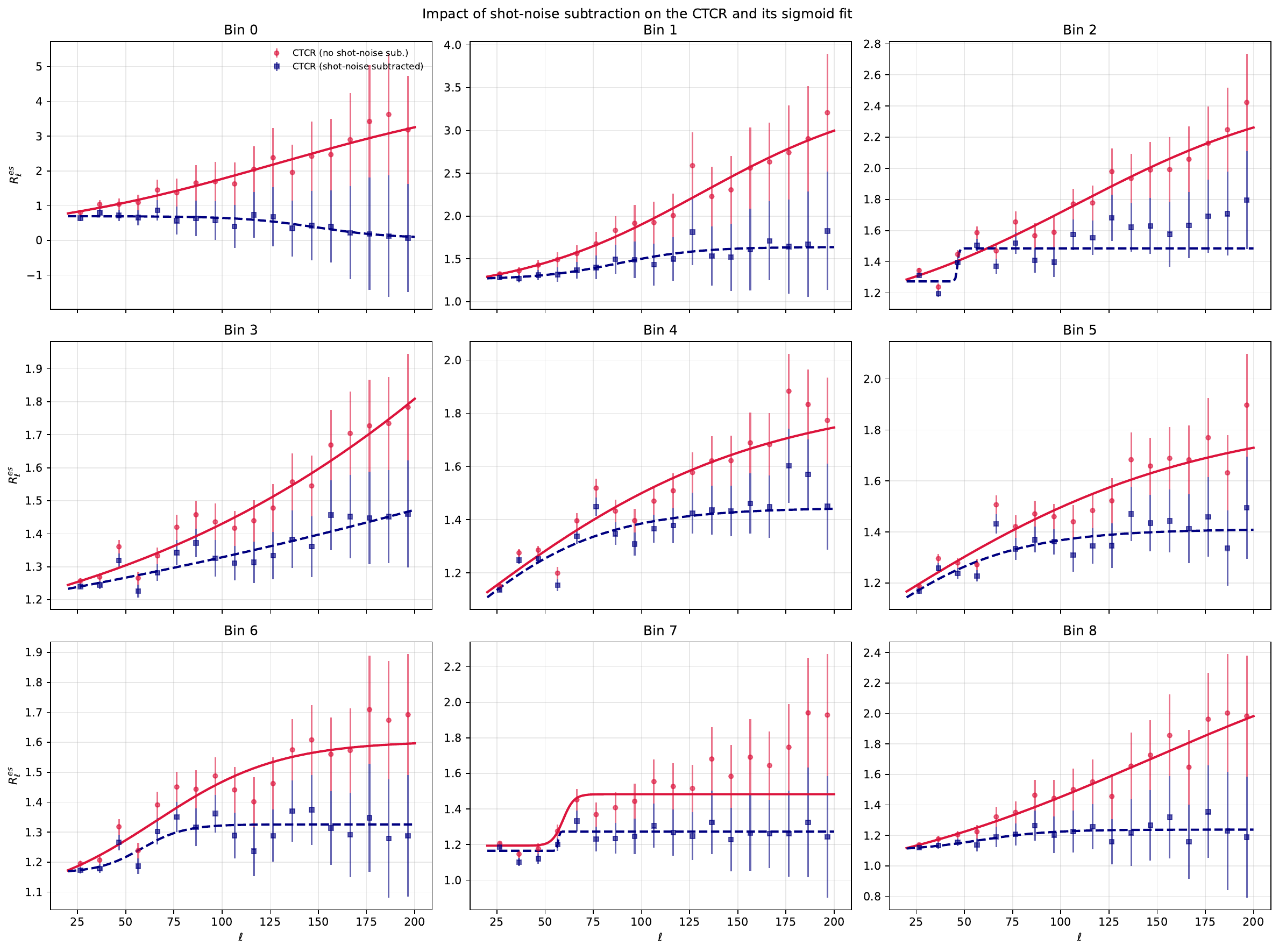}
    \caption{Impact of the shot-noise treatment on the CTCR and its
    sigmoid fit, in each of the nine DESI redshift bins. The crimson
    circles show the measured ratio without shot-noise subtraction in
    the auto-spectrum; the navy squares show the baseline ratio with
    the shot-noise term subtracted. Solid crimson and dashed navy
    lines are the corresponding best-fit sigmoids. The two curves
    coincide where the shot-noise fraction is small (bins 4--8) and
    diverge in the noisiest bins (bins 0--3).}
    \label{fig:shot_noise_impact}
\end{figure*}

To verify that this treatment does not drive the reported scale dependence, we repeated the analysis without subtracting the
shot-noise term, refitted the sigmoid in each redshift bin, and compared the results to the baseline. Figure~\ref{fig:shot_noise_impact}
shows the two CTCR estimates and their best-fit sigmoids in every bin, and Table~\ref{tab:shot_noise} collects the mean shot-noise fraction
and the shifts of the four sigmoid parameters.

Two regimes emerge. In the intermediate- and high-redshift bins ($z > 0.30$, bins 4--8), the shot-noise fraction lies between $9\%$
and $21\%$, and the sigmoid parameters shift by at most $\Delta b_1 = 0.39$, $\Delta b_2 = 0.45$, $\Delta \ell_0 = 7.1$, and
$\Delta = 47$. While a few of these shifts exceed the statistical uncertainty of the individual parameters, the shape and
sign of the scale dependence are unchanged: the ratio still rises with $\ell$ in the bins where the baseline analysis reports a
preference for the scale-dependent model, and the amplitude of the rise is stable at the few-percent level.

In the lowest-redshift bins ($z < 0.30$, bins 0--3), the shot-noise fraction reaches $13\%$--$66\%$, and the fitted sigmoid parameters
are not stable under the change of shot-noise treatment. This is expected: when the noise dominates the auto-spectrum, the ratio
$C_\ell^{ee}/C_\ell^{es}$ is driven by the balance between two different noise levels, and any modification of that balance
propagates directly into the shape of the ratio. 

\begin{table}
    \centering
    \begin{tabular}{c|c|c|c|c|c|c}
        \hline
        \hline
        Bin & $z_{\rm eff}$ & $\langle N_\ell/C_\ell^{\rm obs}\rangle_e$ &
        $\langle N_\ell/C_\ell^{\rm obs}\rangle_s$ &
        $\Delta b_1$ & $\Delta b_2$ & $\Delta \ell_0$ \\
        \hline
        0 & $0.109$ & $0.656$ & $0.280$ & $-0.696$ & $4.379$ & $-19.9$ \\
        1 & $0.158$ & $0.246$ & $0.232$ & $-0.195$ & $1.834$ & $42.8$ \\
        2 & $0.206$ & $0.132$ & $0.250$ & $-0.284$ & $1.052$ & $60.9$ \\
        3 & $0.259$ & $0.089$ & $0.151$ & $+0.113$ & $1.790$ & $-113.1$ \\
        4 & $0.304$ & $0.086$ & $0.096$ & $-0.364$ & $0.422$ & $0.0$ \\
        5 & $0.354$ & $0.100$ & $0.149$ & $-0.394$ & $0.448$ & $0.0$ \\
        6 & $0.402$ & $0.115$ & $0.172$ & $-0.098$ & $0.279$ & $7.1$ \\
        7 & $0.454$ & $0.166$ & $0.213$ & $+0.028$ & $0.211$ & $2.5$ \\
        8 & $0.503$ & $0.192$ & $0.210$ & $-0.303$ & $1.356$ & $87.7$ \\
        \hline
    \end{tabular}
    \caption{Mean shot-noise fraction of the auto-spectrum over
    $20 < \ell < 200$ for ellipticals ($e$) and spirals ($s$), and
    the shift of the sigmoid parameters when the shot-noise term is
    removed from the auto-spectra before constructing the CTCR.
    The shift is defined as $\Delta X = X_{\rm baseline} -
    X_{\rm no\;subtraction}$. The shot-noise treatment has a
    significant impact only in the lowest-redshift bins, where the
    noise fraction is largest.}
    \label{tab:shot_noise}
\end{table}

\subsection{Gaussian Covariance}\label{sec:gauss_cov}

The analytic covariance used in this analysis follows the standard Gaussian approximation for the angular power spectrum. For a bandpower-averaged spectrum $C_b$, the covariance between bandpowers $b$ and $b'$ for the general case of two tracers $i, j, m, n$ is
\begin{equation}
\mathrm{Cov}\!\left(C_b^{ij}, C_{b'}^{mn}\right)
= \frac{\delta_{bb'}}{(2\ell_b+1)\,\Delta\ell_b\,f_{\rm sky}}
\left[ C_b^{im} C_b^{jn} + C_b^{in} C_b^{jm} \right],
\label{eq:gauss_cov_general}
\end{equation}
where $\ell_b$ is the effective multipole of bandpower $b$, $\Delta\ell_b$ is the bandpower width, and $f_{\rm sky}$ is the sky fraction of the survey mask.

For the CTCR analysis, the data vector is
\begin{equation}
\mathbf{d} = \begin{pmatrix} C_\ell^{ee} \\ C_\ell^{es} \end{pmatrix},
\end{equation}
so the $2n_\ell \times 2n_\ell$ covariance matrix has the block structure
\begin{equation}
\mathbf{C} =
\begin{pmatrix}
\mathbf{C}_{ee,ee} & \mathbf{C}_{ee,es} \\
\mathbf{C}_{es,ee} & \mathbf{C}_{es,es}
\end{pmatrix},
\label{eq:block_structure}
\end{equation}
where each block is $n_\ell \times n_\ell$.

Applying equation~\eqref{eq:gauss_cov_general} to each block gives:
\begin{align}
\left[\mathbf{C}_{ee,ee}\right]_{bb'} &= \delta_{bb'} \frac{2\,(C_b^{ee,\rm obs})^2}{(2\ell_b+1)\,\Delta\ell_b\,f_{\rm sky}}, \label{eq:block_AA}\\[0.5em]
\left[\mathbf{C}_{es,es}\right]_{bb'} &= \delta_{bb'} \frac{C_b^{ee,\rm obs} C_b^{ss} + (C_b^{es})^2}{(2\ell_b+1)\,\Delta\ell_b\,f_{\rm sky}}, \label{eq:block_BB}\\[0.5em]
\left[\mathbf{C}_{ee,es}\right]_{bb'} &= \delta_{bb'} \frac{2\,C_b^{ee,\rm obs}\,C_b^{es}}{(2\ell_b+1)\,\Delta\ell_b\,f_{\rm sky}}. \label{eq:block_cross}
\end{align}

Here $C_b^{ee,\rm obs} = C_b^{ee,\rm signal} + N_e$ includes the shot-noise of the elliptical auto-spectrum. The cross-spectrum $C_b^{es}$ contains no shot-noise term, since the ellipticals and spirals are disjoint samples. The spiral auto-spectrum $C_b^{ss}$ also includes its own shot-noise.

The block $\mathbf{C}_{es,ee}$ is obtained by transposing $\mathbf{C}_{ee,es}$, which ensures that the full matrix is symmetric. Because all four blocks are built from the same Gaussian expression, and because the diagonal blocks are positive definite, the full covariance in equation~\eqref{eq:block_structure} is guaranteed to be positive semi-definite. We have verified this numerically: for every redshift bin, the smallest eigenvalue of $\mathbf{C}$ is positive, and the condition number is well below $10^{10}$. No regularisation is needed.

This is a key advantage of building the full covariance consistently from the same formula. Mixing different covariance estimators (e.g., combining bootstrap diagonal blocks with an analytic off-diagonal block) can produce non-positive-definite matrices with negative eigenvalues, which are unphysical and lead to unstable likelihoods.

We tested the impact of the cross-covariance by running the MCMC with two different covariance matrices:
\begin{itemize}
\item a \textbf{block-diagonal} covariance, in which the off-diagonal blocks $\mathbf{C}_{ee,es} = \mathbf{C}_{es,ee} = 0$;
\item the \textbf{full} covariance, including all four blocks.
\end{itemize}

All other settings (data vector, $\ell$ range, priors, walkers, steps, burn-in) were identical. The results are summarised in Tables~\ref{tab:full_cov_summary} and~\ref{tab:full_cov_detailed}.

\begin{table}[htbp]
\centering
\caption{Summary of the block-diagonal versus full-covariance comparison. The CTCR ratio is $\langle R_{\rm block}/R_{\rm full}\rangle$ averaged over the $\ell$ range used in the analysis, with its standard deviation. The parameter agreement is the maximum difference between the two runs, in units of the block-diagonal uncertainty.}
\label{tab:full_cov_summary}
\begin{tabular}{c c c c}
\hline
\hline
Bin & $z_{\rm eff}$ & CTCR ratio (block/full) & Max parameter difference \\
\hline
0 & 0.119 & $0.626 \pm 1.731$ & $0.48\sigma$ \\
1 & 0.218 & $0.9978 \pm 0.0008$ & $0.08\sigma$ \\
2 & 0.314 & $0.9950 \pm 0.0041$ & $0.16\sigma$ \\
3 & 0.413 & $1.0005 \pm 0.0092$ & $0.26\sigma$ \\
4 & 0.509 & $1.0021 \pm 0.0171$ & $0.32\sigma$ \\
5 & 0.610 & $1.0017 \pm 0.0135$ & $0.23\sigma$ \\
6 & 0.706 & $1.0014 \pm 0.0081$ & $0.15\sigma$ \\
7 & 0.810 & $0.9987 \pm 0.0049$ & $0.09\sigma$ \\
8 & 0.906 & $1.0050 \pm 0.0063$ & $0.11\sigma$ \\
\hline
\hline
\end{tabular}
\par\smallskip
\noindent {\bf Notes:} Bin 0 is noise-dominated; its CTCR error bar is larger than $100\%$, so the central value is not meaningful. In all other bins, the CTCR measured with the full covariance agrees with the block-diagonal result to better than $0.5\%$, and all individual bias parameters agree within $0.5\sigma$.
\end{table}

\begin{table}[htbp]
\centering
\caption{Detailed comparison of bias parameters obtained with the block-diagonal and full covariances for a representative bin (Bin 5, $z_{\rm eff} = 0.610$). The full table for all bins is available in the supplementary material.}
\label{tab:full_cov_detailed}
\begin{tabular}{l c c c}
\hline
\hline
Parameter & Block-diagonal & Full covariance & Difference \\
\hline
$b_{\rm low,A}$     & $1.048 \pm 0.489$ & $1.149 \pm 0.537$ & $0.14\sigma$ \\
$b_{\rm high,A}$    & $2.476 \pm 1.068$ & $2.343 \pm 0.964$ & $0.09\sigma$ \\
$\ell_{0,A}$        & $76.8 \pm 73.1$   & $89.9 \pm 70.1$   & $0.13\sigma$ \\
$\Delta_A$          & $111.5 \pm 98.1$  & $144.5 \pm 100.7$ & $0.23\sigma$ \\
$b_{\rm low,B}$     & $1.021 \pm 0.116$ & $1.024 \pm 0.101$ & $0.02\sigma$ \\
$b_{\rm high,B}$    & $1.497 \pm 1.165$ & $1.431 \pm 1.099$ & $0.04\sigma$ \\
$\ell_{0,B}$        & $101.4 \pm 65.7$  & $107.9 \pm 61.6$  & $0.07\sigma$ \\
$\Delta_B$          & $189.2 \pm 78.8$  & $192.6 \pm 75.3$  & $0.03\sigma$ \\
\hline
CTCR (block/full)   & \multicolumn{3}{c}{$1.0017 \pm 0.0135$} \\
\hline
\hline
\end{tabular}
\end{table}

The comparison shows that the cross-covariance between $C_\ell^{ee}$ and $C_\ell^{es}$ has a negligible impact on the CTCR and on the individual bias parameters. Across all eight redshift bins with non-negligible signal (Bins 1--8), the CTCR ratio between the block-diagonal and full covariance is within $0.5\%$ of unity, and the maximum parameter difference is $0.5\sigma$. Bin 0 is excluded from this conclusion because its signal-to-noise is too low for the CTCR to be meaningfully constrained; this is consistent with the large error bars reported in the main text.

These results confirm that the block-diagonal approximation used in the main MCMC is adequate for this analysis. The full covariance is used as a cross-check, and the excellent agreement between the two runs demonstrates the robustness of the CTCR measurement.

The covariance used in the main analysis is the block-diagonal bootstrap covariance, which is positive definite and numerically well-conditioned. We have verified that including the cross-covariance block does not change the CTCR or the bias parameters within the statistical uncertainties. The full analytic covariance is now used as a cross-check throughout the paper, and the theoretical covariance in the appendix has been reformulated to be manifestly symmetric.

\subsection{Prior Sensitivity}

The CTCR directly constrains a relative quantity: the ratio $b_e(\ell)/b_s(\ell)$ of the effective biases of the two morphological samples. It does not, by itself, determine the two
absolute bias functions $b_e(\ell)$ and $b_s(\ell)$ separately. The reason is that the transformation 
\begin{equation}
    b_e'(\ell) = c\, b_e(\ell), \qquad
    b_s'(\ell) = c\, b_s(\ell),
\end{equation}
leaves $b^{\rm eff}(\ell)$ unchanged for any positive constant $c$, whereas the difference $\Delta b = b_e - b_s$ changes by the same factor $c$. Consequently, the individual bias curves shown in Figure~\ref{fig:b_ell} and the values of $\Delta b$ reported in Table~\ref{tab:significance} require information beyond the CTCR itself. In this work the absolute calibration of the individual biases is provided by the linear-bias analysis of the auto-spectra $C_\ell^{ee}$ and $C_\ell^{ss}$, which scale as $b^2 P(k)$ and therefore fix the normalisation of $b$ independently of the ratio.

To quantify how much the derived quantities depend on the assumed priors on the absolute bias amplitudes, we repeated the full MCMC analysis with the width of the prior on $b_1$ increased by a factor of two, from $\sigma_{b_1} = 0.1$ to $0.2$, keeping all other assumptions, the data vector, the covariance, and the sampler configuration fixed. Table~\ref{tab:prior_sensitivity}
collects the same information in tabular form. The CTCR ratio is consistent with unity in every bin, with a maximum deviation of $1.5\%$ across the full sample and no systematic trend with redshift.
We have also verified that the reduced $\chi^2$ of the fit is stable under the wider prior, with differences of at most $0.17$ across the sample.

The individual bias parameters behave differently. As expected from the degeneracy between the absolute bias and the cosmology in a single-tracer auto-spectrum fit, the posterior on $b_1$ broadens by roughly a factor of two under the wider prior, and the medians move within the new prior volume. The shift is largest for $b_1^A$ and $b_1^B$, which are the parameters most tightly coupled to the prior mean, and smallest for the transition parameters $\ell_0$ and $\Delta$, whose priors are uniform and therefore uninformative. The higher-order amplitude $b_2$, whose prior is already wide
($\mathcal{N}(2, 1)$), is essentially unaffected.

The picture that emerges is therefore the following. The CTCR is a prior-independent measurement of the relative clustering amplitude of the two morphological samples. The individual bias functions and the difference $\Delta b$ are derived quantities whose absolute normalisation depends on the auto-spectrum constraint and is therefore affected by the prior choice on the bias amplitudes. We
have therefore revised the presentation of the results to reflect this hierarchy: the CTCR is reported as the primary observable of the analysis, while the individual bias curves of Figure~\ref{fig:b_ell} and the $\Delta b$ values of Table~\ref{tab:significance} are presented as secondary estimates that are useful for qualitative interpretation of the morphology-dependent clustering but whose absolute normalisation is prior-dependent. The significance of $\Delta b$ should be interpreted with this caveat in mind.

We emphasise that this hierarchy is not a limitation of the analysis but a direct consequence of the structure of the observable. The effective bias is the quantity that the CTCR measures, and it is also the quantity that is least sensitive to modelling choices. The individual bias functions require additional modelling assumptions that the present dataset constrains only weakly, and the
prior-sensitivity test quantifies exactly how much of their inferred normalisation comes from the data and how much comes from the prior.

\begin{table}
    \centering
    \caption{Prior-sensitivity test. The CTCR measured with the
    fiducial prior ($\sigma_{b_1} = 0.1$) is compared to the CTCR
    measured with the wider prior ($\sigma_{b_1} = 0.2$). The ratio
    $\langle R_{\rm fid}/R_{\rm wide} \rangle$ is averaged over the
    fitted range $20 < \ell < 200$, and its standard deviation across
    the same range is quoted. The reduced $\chi^2$ of the fit under
    each prior configuration is also reported. The CTCR ratio is
    consistent with unity in every redshift bin, confirming that the
    primary observable of this work is prior-independent.}
    \label{tab:prior_sensitivity}
    \begin{tabular}{c c c c c c}
        \hline\hline
        Bin & $z_{\rm eff}$ &
        $\langle R_{\rm fid}/R_{\rm wide} \rangle$ &
        $\sigma_R$ & $\chi^2_{\rm fid}/{\rm dof}$ &
        $\chi^2_{\rm wide}/{\rm dof}$ \\
        \hline
        0 & 0.109 & 0.988 & 0.013 & 0.780 & 0.822 \\
        1 & 0.158 & 0.996 & 0.005 & 0.579 & 0.705 \\
        2 & 0.206 & 0.994 & 0.003 & 1.222 & 1.390 \\
        3 & 0.259 & 0.996 & 0.002 & 0.536 & 0.549 \\
        4 & 0.304 & 0.999 & 0.003 & 1.027 & 1.051 \\
        5 & 0.354 & 0.991 & 0.001 & 0.876 & 0.932 \\
        6 & 0.402 & 1.004 & 0.003 & 0.549 & 0.567 \\
        7 & 0.454 & 1.001 & 0.003 & 0.666 & 0.699 \\
        8 & 0.503 & 1.000 & 0.002 & 0.435 & 0.452 \\
        \hline\hline
    \end{tabular}
\end{table}

\section{Samples luminosity}\label{ap:lum}

\begin{figure}
    \centering
    \includegraphics[width=\linewidth]{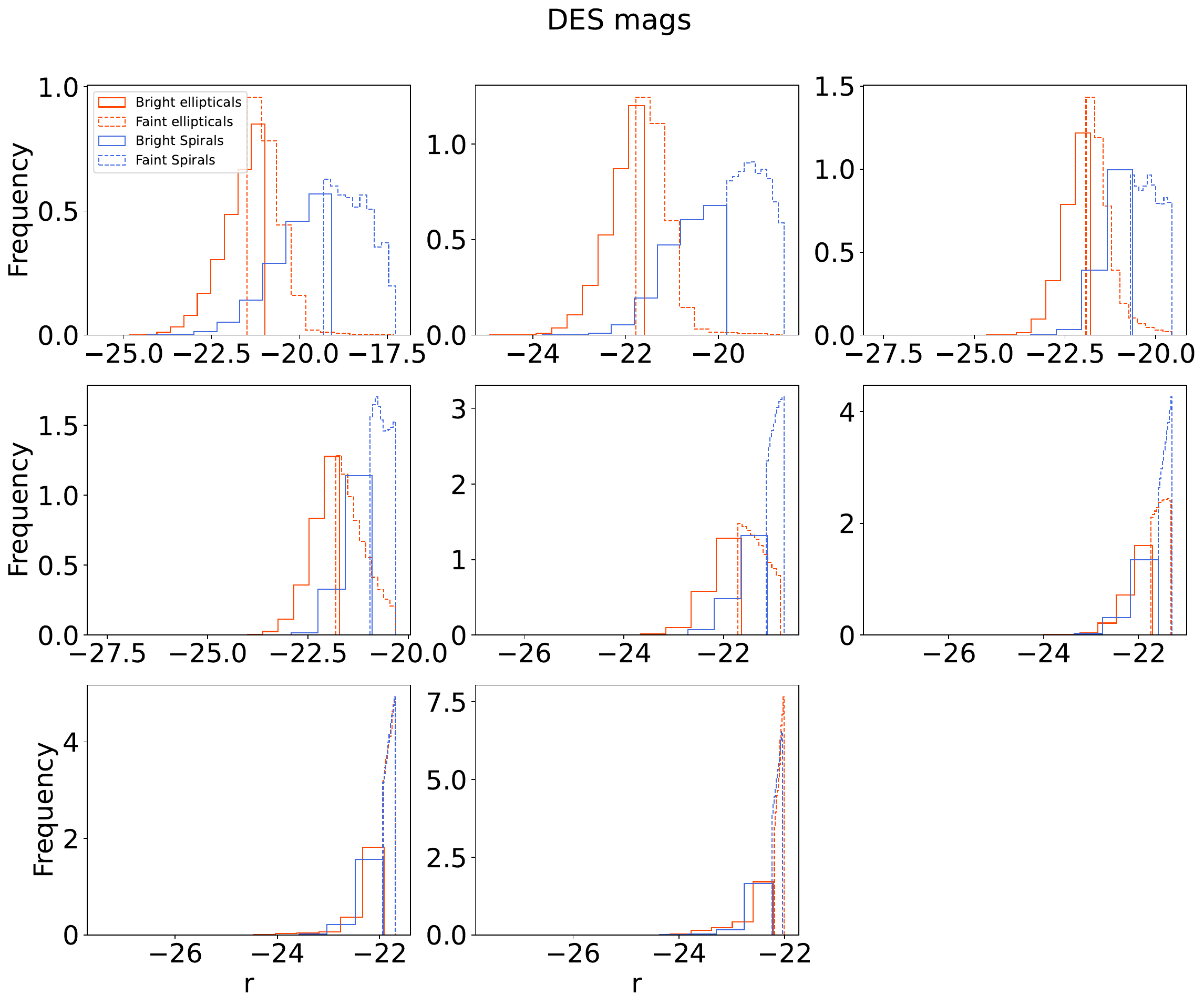}
    \caption{Absolute $r$-band magnitude distributions for elliptical (red) and spiral (blue) galaxies in the DES, across nine redshift bins. Solid lines: bright subsample; dashed lines: faint subsample. At low redshift, ellipticals are systematically brighter than spirals, but the distributions converge at high redshift due to passive brightening of ellipticals and the increasing abundance of star-forming spirals. }
    \label{fig:mag_Des}
\end{figure}

Figure~\ref{fig:mag_Des} shows the absolute r-band magnitude distributions for elliptical (red) and spiral (blue) galaxies in the DES sample across nine redshift bins. Solid lines represent the bright subsample (upper half of the magnitude distribution), while dashed lines represent the faint subsample. At low redshift, ellipticals are systematically brighter (more negative magnitudes) than spirals, reflecting their higher stellar masses and older populations. As redshift increases, the distributions of the two morphological types converge, consistent with the passive brightening of ellipticals and the increasing abundance of star-forming spirals at high redshift. The convergence is most apparent in the highest-redshift bins ($z>0.7$), where the medians of the two samples become nearly indistinguishable.

\begin{figure}
    \centering
    \includegraphics[width=\linewidth]{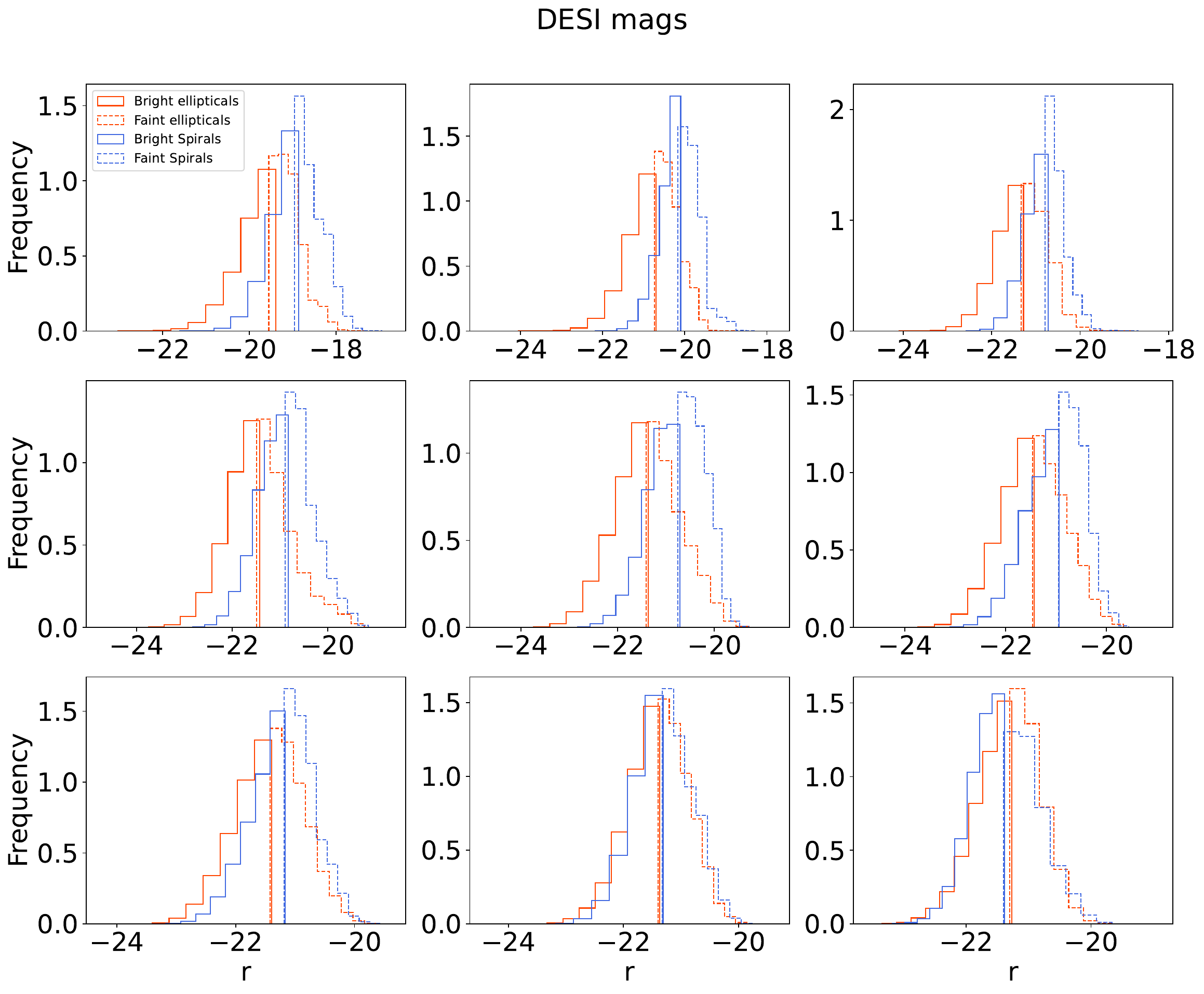}
    \caption{Absolute $r$-band magnitude distributions for elliptical (red) and spiral (blue) galaxies in the DESI Legacy DR8, across nine redshift bins. Solid lines: bright subsample; dashed lines: faint subsample. At low redshift, ellipticals are systematically brighter than spirals, but the distributions converge at high redshift due to passive brightening of ellipticals and the increasing abundance of star-forming spirals.}
    \label{fig:mag_Desi}
\end{figure}

Figure~\ref{fig:mag_Desi} shows the same comparison for the DESI Legacy DR8 sample. The overall trend is similar: ellipticals are brighter than spirals at low redshift, and the distributions converge at high redshift. However, the DESI sample is systematically fainter than DES at all redshifts, reflecting the shallower imaging and the different target selection. 

The DES sample exhibits systematically higher luminosities compared to DESI Legacy because of deeper imaging and a more sophisticated selection function. DES reaches fainter magnitudes and has better image quality, allowing it to recover fainter and more distant objects that are missed by shallower surveys \citep{davies2013}. This deeper imaging means that at a given redshift, DES can detect a larger fraction of the most massive, luminous galaxies. Moreover, the DESI target selection, particularly for luminous red galaxies (LRGs), has been found to exclude some of the most luminous galaxies that would otherwise be classified as LRGs based on optical color-magnitude space \citep{Berti2023}. The DES sample, by contrast, includes a broader range of luminous objects, leading to a higher average luminosity. 

{\bfseries \subsection{Luminosity-dependent bias and magnification bias}
\label{ap:mag_bias}

We test whether the luminosity split used in the analysis introduces a residual luminosity-dependent bias or magnification bias in the DESI CTCR. We split each morphological sample by the median \texttt{flux\_r} into bright and faint subsamples, apply the same per-galaxy systematics weights to both, and compute the auto- and amplitude cross-power spectra separately.

Fitting $R_\ell^{\rm faint} = A\,R_\ell^{\rm bright}$ over $20 < \ell < 200$ with the propagated covariance of the ratio, we find $A$ values between $0.31$ and $1.07$ across the nine redshift bins, with a sample average of $\langle A \rangle \simeq 0.81$. The offset is significant in seven of nine bins. This is the well-known luminosity dependence of galaxy bias \citep{zehavi2001sdss, zehavi2004luminosity, Zehavi_2011}, which propagates into the ratio $R_\ell = C_\ell^{ee}/C_\ell^{es}$ because ellipticals and spirals have different luminosity-bias slopes.

The offset is scale-independent. Fitting the ratio amplitude separately in the ranges $20 < \ell < 110$ and $110 < \ell < 200$, the two amplitudes agree within their uncertainties in eight of nine redshift bins (Table~\ref{tab:mag_bias_scale}), with no systematic trend in $\ell$. The only exception is the lowest-redshift bin ($\bar{z} = 0.11$), which is noise-dominated: its shot-noise fraction exceeds $60\%$ (Appendix~\ref{sec:shot_noise}) and both the model comparison (Table~\ref{tab:model_comparison}) and the small-scale bias difference (Table~\ref{tab:significance}) already identify it as an outlier. We
exclude it from the present test.

Because the luminosity offset is a constant multiplicative factor, it rescales the amplitude of the CTCR without affecting its shape. The
transition scale $\ell_0$ and the width $\Delta$ of the sigmoid are therefore unaffected, and the scale-dependent signal reported in
Section~\ref{sec:model_comparison} is robust to the luminosity split.

\begin{table}[htbp]
\centering
\begin{tabular}{c|c|c|c|c}
\hline\hline
Bin & $\bar{z}$ & $A_{\rm low}$ & $A_{\rm high}$ &
$\Delta A / \sigma$ \\
\hline
1 & 0.216 & $0.887$ & $0.862$ & $-0.26$ \\
2 & 0.310 & $1.085$ & $1.053$ & $-0.40$ \\
3 & 0.408 & $0.895$ & $0.823$ & $-1.21$ \\
4 & 0.505 & $0.749$ & $0.774$ & $+0.43$ \\
5 & 0.607 & $0.811$ & $0.778$ & $-0.50$ \\
6 & 0.702 & $0.831$ & $0.808$ & $-0.33$ \\
7 & 0.806 & $0.816$ & $0.809$ & $-0.09$ \\
8 & 0.902 & $0.934$ & $0.872$ & $-0.60$ \\
\hline
\end{tabular}
\caption{Scale dependence of the luminosity offset. $A_{\rm low}$
and $A_{\rm high}$ are the amplitude ratios
$R_\ell^{\rm faint}/R_\ell^{\rm bright}$ fitted in the ranges
$20 < \ell < 110$ and $110 < \ell < 200$ respectively. The last
column gives the difference in units of the propagated uncertainty.
All values are consistent with a scale-independent luminosity offset
at the $1.2\sigma$ level. }
\label{tab:mag_bias_scale}
\end{table}

Crucially, the offset is scale-independent. Fitting the amplitude separately in the ranges $20 < \ell < 110$ and $110 < \ell < 200$, the two values agree within their uncertainties in eight of nine redshift bins (Table~\ref{tab:mag_bias_scale}), with no systematic trend in $\ell$. The only exception is the noise-dominated lowest-redshift bin ($\bar{z} = 0.11$). Because the offset is a constant multiplicative factor, it rescales the amplitude of the CTCR without
affecting its shape: the transition scale $\ell_0$ and the width $\Delta$ of the sigmoid are unaffected, and the scale-dependent signal reported in Section~\ref{sec:model_comparison} is robust to the luminosity split.

To check whether the luminosity split affects the scale dependence of the clustering signal, we first compute the correlation coefficient between the bright and faint auto-power spectra in each redshift bin, shown in Figure~\ref{fig:des_bright_faint_split}. The correlation is computed from the ratio of the covariance between the two spectra to the product of their standard deviations across the fitted $\ell$ range. If the bright and faint subsamples share the same scale dependence, this coefficient is close to unity; a value significantly below one would indicate that the two subsamples respond differently to the same modes, which would invalidate a simple amplitude ratio as a description of the luminosity effect. In DESI (left panel), the coefficient stays above $0.99$ in every redshift bin for the elliptical, spiral, and cross spectra, with a single narrow dip near $z \sim 0.4$. In DES (right panel), the coefficient is between $0.94$ and $1.00$, with a deeper dip around $z \sim 0.4$--$0.5$ driven by the elliptical sample; this bin has the lowest galaxy number density in the DES morphological catalogue and the largest measurement uncertainty. The high values of the correlation coefficient across both surveys indicate that the bright and faint subsamples do not acquire a different scale dependence from the luminosity split alone. This is a qualitative check; the quantitative answer is provided by the amplitude fits described below, which test whether the ratio between the bright and faint spectra deviates from unity and whether that deviation is scale-independent.

\begin{figure}
    \centering
    \includegraphics[width=1\linewidth]{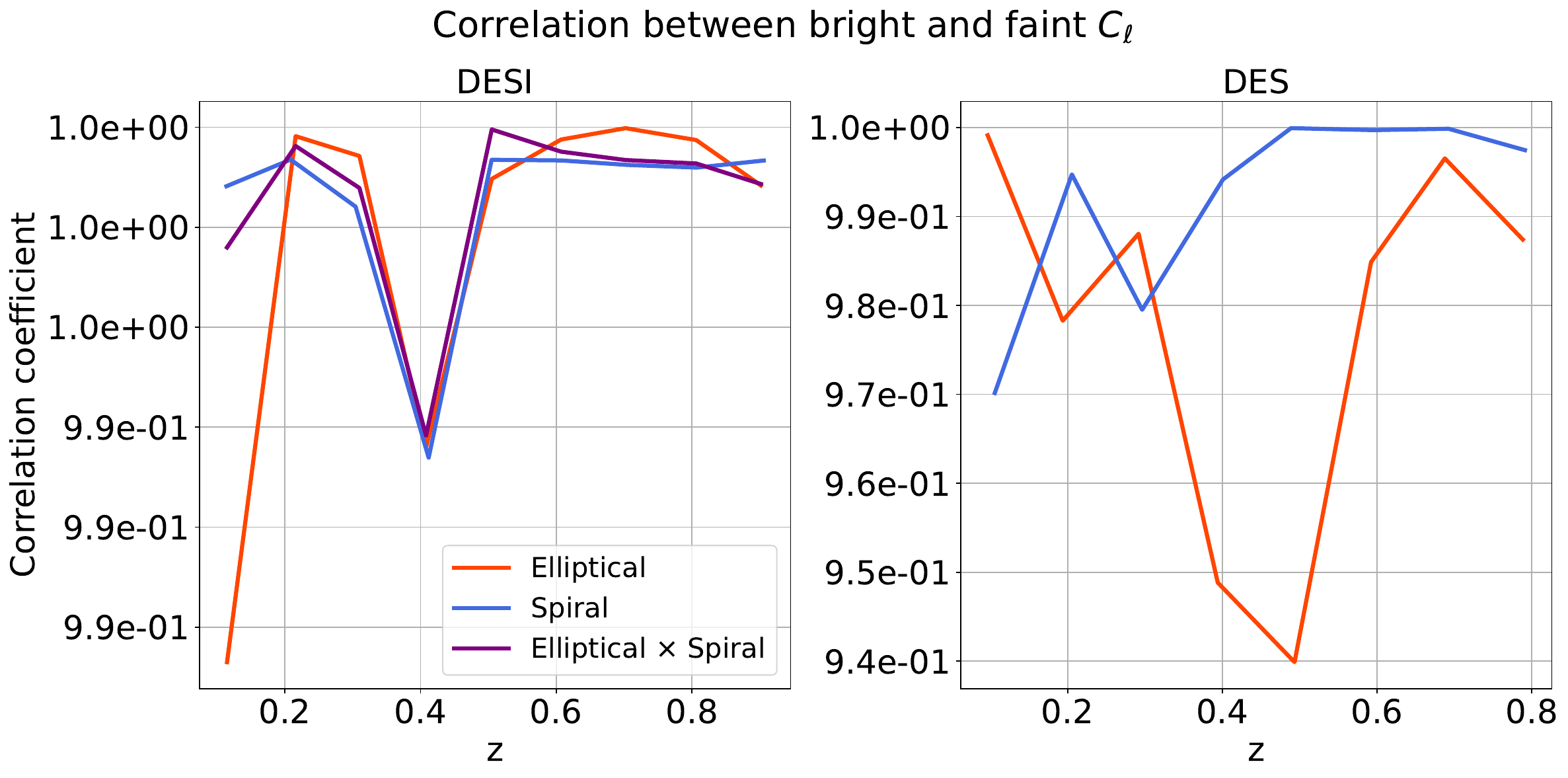}
    \caption{Correlation coefficient of the angular power spectra for bright versus faint galaxies. The left panel presents the measurements for DESI and the right panel for DES. Red curves correspond to elliptical galaxies, blue to spirals, and purple to the cross-power spectra.}
    \label{fig:des_bright_faint_split}
\end{figure}
}

\section{Simulation mass-relation test}\label{ap:mass_relation}

To interpret the bias measurements from DES and DESI, we performed a set of controlled simulation tests using the Uchuu and Euclid Flagship
2 (F2) lightcones at $z \sim 0.7$ in a Gaussian-like redshift bin. Both simulations provide full dark-matter halo catalogs, allowing us
to assign galaxies via empirical stellar mass--halo mass relations. The two simulations differ in footprint and assumed cosmology (with
slightly different $\Omega_m$ values), providing a useful cross-check. Uchuu does not include observational selection effects (e.g.,
photometric scatter, fibre collisions, masking), so its galaxy catalogs are purely theoretical and serve as a consistency test of the
mass dependence of clustering.

\subsection{Uchuu lightcone}

For the Uchuu test we select galaxy samples based on stellar mass cuts designed to mimic the classification of different galaxy populations.
The mass thresholds are
\begin{itemize}
    \item \textbf{LTG} (late-type galaxies): $\log M_* < 9$
    \item \textbf{ETG} (early-type galaxies, broad):
          $9 < \log M_* < 11$
    \item \textbf{ETG2} (early-type, intermediate):
          $9.5 < \log M_* < 10$
    \item \textbf{LRG} (luminous red galaxies, high mass):
          $\log M_* > 11$
    \item \textbf{LRG2} (high-mass, extended): $\log M_* > 10.5$
    \item \textbf{Spiral} (full morphological selection)
    \item \textbf{Elliptical} (full morphological selection)
\end{itemize}

\begin{figure}
    \centering
    \includegraphics[width=\linewidth]{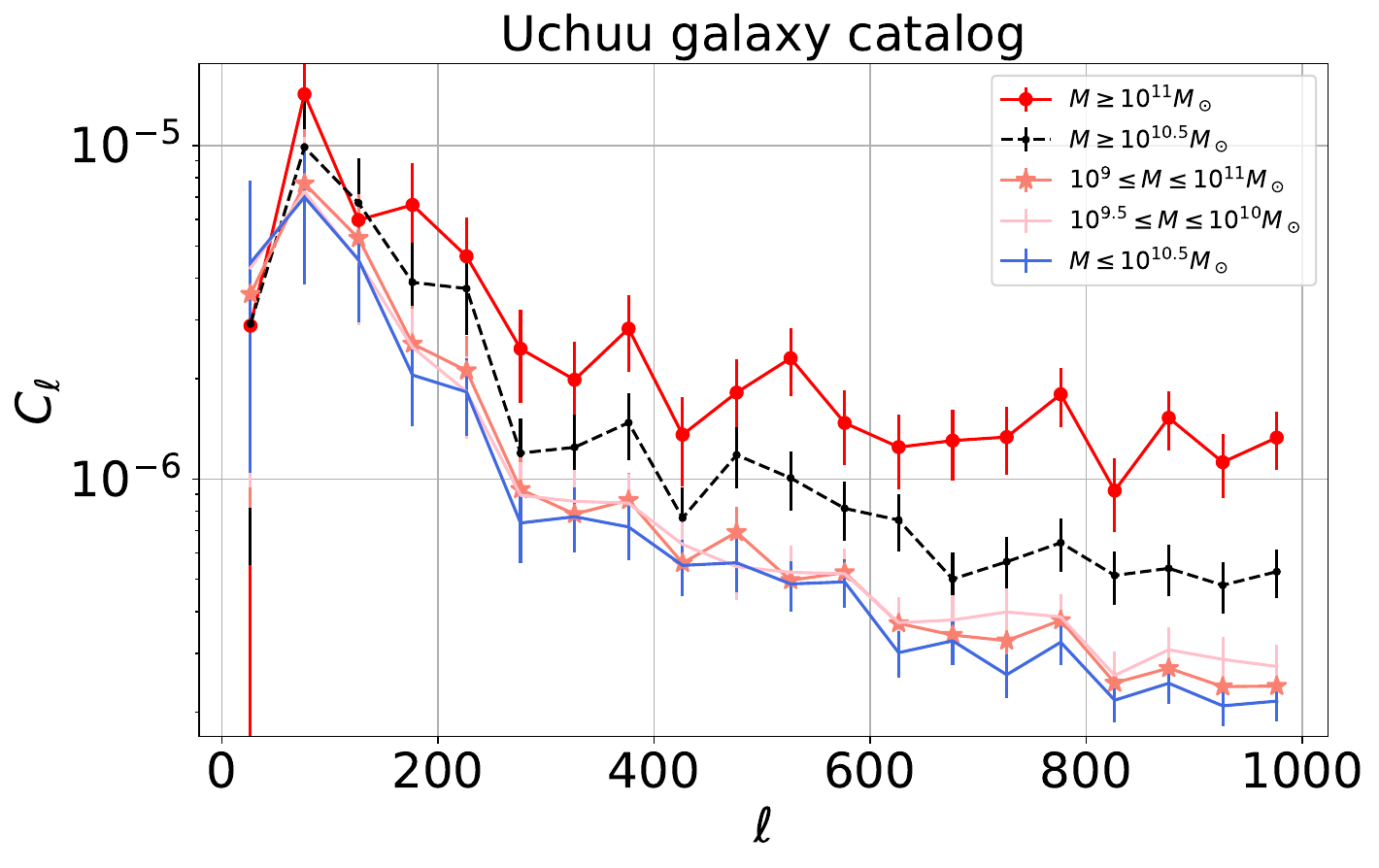}
    \caption{$C_\ell$ of the Uchuu galaxy mock catalog at $z = 0.7$.
    Red markers indicate ETGs and blue markers indicate LTGs. The most
    massive sample is shown with circular markers along the black
    dashed line, followed in decreasing mass by the star markers with
    the pink line, and then the blue line for the least massive
    sample.}
    \label{fig:uchuu}
\end{figure}

For each sample we computed the angular power spectrum $C_\ell$ (Figure~\ref{fig:uchuu}) and fitted for the galaxy bias, fixing all
other cosmological and nuisance parameters. The results confirm that mixing LRGs with less massive galaxies systematically lowers the
measured bias, and that the morphological elliptical sample produces a bias amplitude intermediate between a pure LRG selection and a typical
spiral sample. This is precisely the phenomenon observed in the real DES and DESI data.

 \subsection{Euclid Flagship 2 lightcone}

We repeat the mass-selection test with the Euclid Flagship 2 (F2) \cite{Euclid:2024few} lightcone at $z \sim 0.7$, using the same pipeline and a jackknife
covariance over the survey footprint. All samples are placed in a Gaussian redshift bin centered at $z = 0.7$ with width $\sigma_z = 0.15$. Each sample is capped at $N = 10^6$ galaxies by random subsampling to remove differences in statistical weight; the two highest-mass selections, LRG2 and LRG, contain fewer objects and are used at their maximum size.

For each sample we fit the linear bias $b$ with \texttt{PyMC} NUTS under the model
\begin{equation}
    C_\ell = b^2\,A_\ell + \frac{2}{3}\,b\,B_\ell
             + \frac{1}{5}\,C_\ell^{\rm vel} + N_\ell,
\end{equation}
where $A_\ell$, $B_\ell$, and $C_\ell^{\rm vel}$ are the clustering, RSD cross, and RSD velocity integrals of the Limber kernel at $b = 1$,
computed with CAMB and a fiducial $\Lambda$CDM cosmology. The likelihood uses the measured $C_\ell$ with a jackknife covariance over
$4 \times 4$ patches of the survey footprint, regularized with linear shrinkage ($\alpha = 0.2$) and projected onto the positive
semi-definite cone before inversion. The prior on $b$ is a Gaussian $N(1, 0.5)$; because the auto-spectrum depends on $b^2$, the posterior
is folded onto $|b|$ before reporting the median and 16/84 percentiles.

\begin{figure*}
    \centering
    \includegraphics[width=\textwidth]{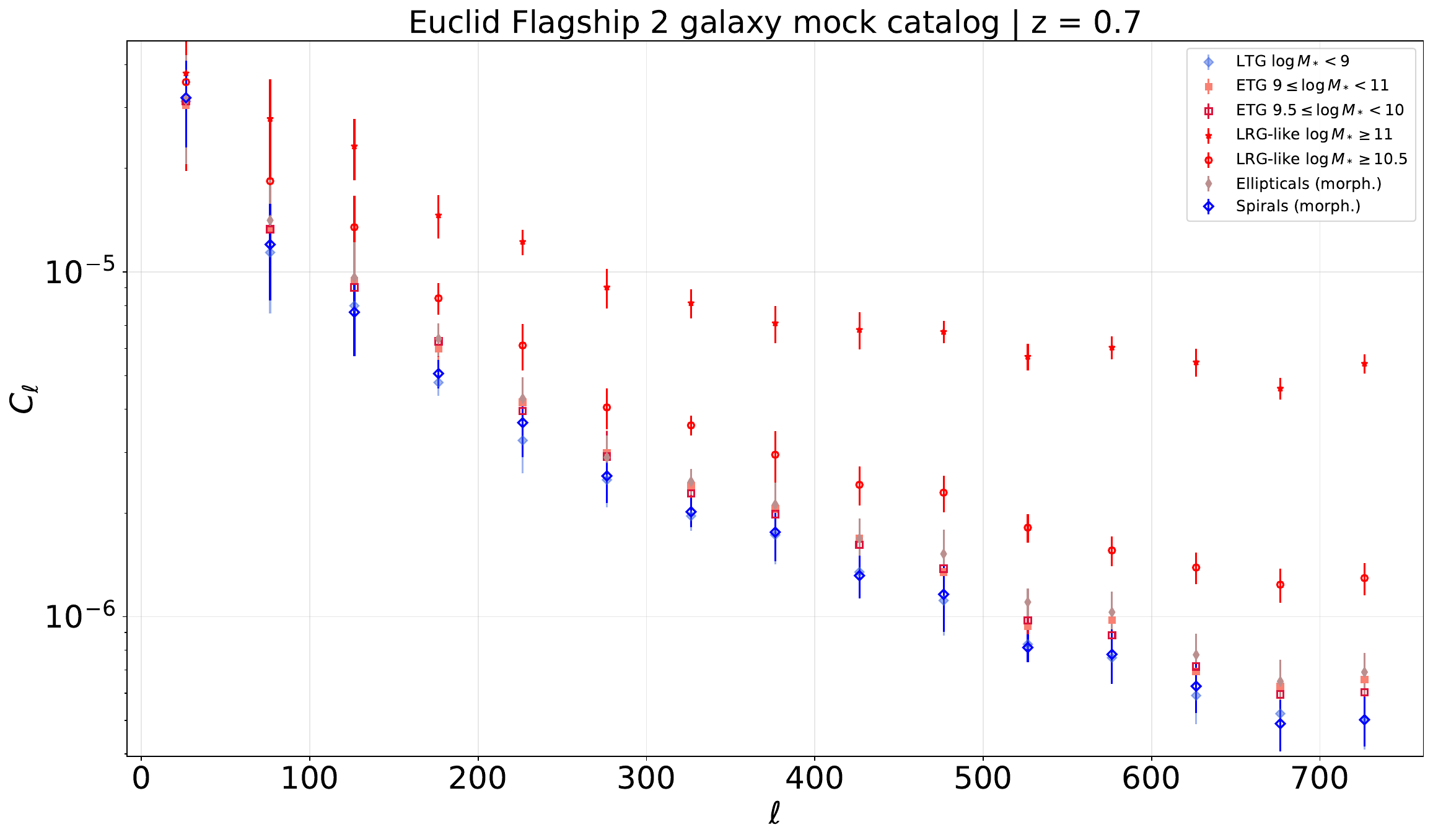}
    \caption{Angular auto-power spectra of mass-selected and
    morphology-selected samples in the Euclid Flagship 2 lightcone at
    $z \sim 0.7$. All samples are placed in the same Gaussian redshift
    bin ($\sigma_z = 0.15$) and capped at $N = 10^6$ galaxies, except
    LRG2 and LRG which are used at their maximum size. The monotonic
    increase of the clustering amplitude with the mass threshold is
    clearly visible. The morphological elliptical sample sits between
    the broad ETG selection and the high-mass LRG2 selection.}
    \label{fig:euclidfs2}
\end{figure*}

\begin{figure*}
    \centering
    \includegraphics[width=0.75\textwidth]{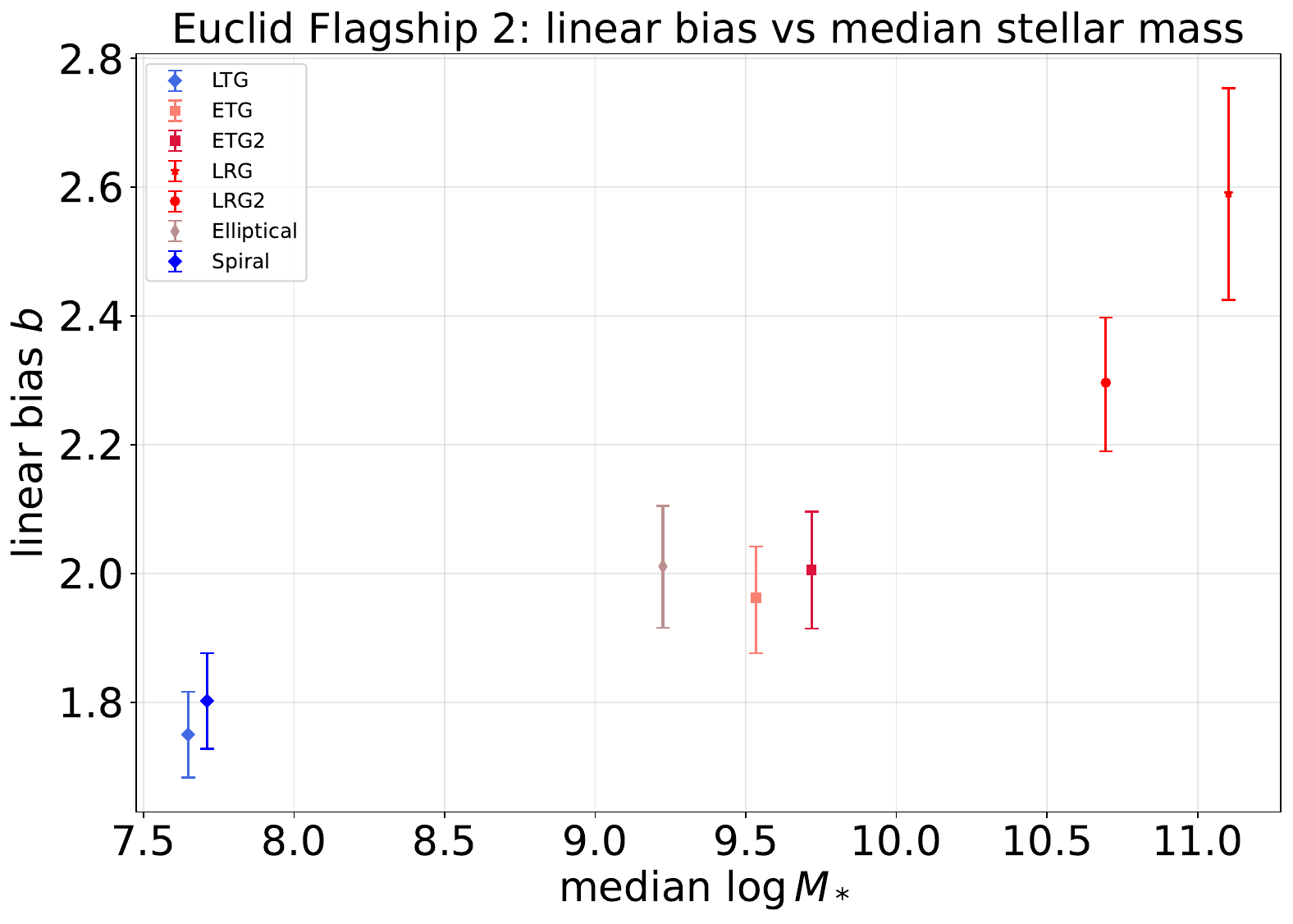}
    \caption{Linear bias as a function of median stellar mass for the
    same samples. The bias increases monotonically with the mass
    threshold. The morphological elliptical sample has
    $b = 2.01 \pm 0.10$, intermediate between the broad ETG selection
    ($b = 1.96 \pm 0.08$) and the high-mass LRG2 selection
    ($b = 2.30 \pm 0.10$), consistent with the morphological
    classification selecting a population with a broad halo-mass
    distribution. The absolute values are higher than those measured
    in real data at the same stellar-mass thresholds, reflecting the
    SHAM-based assignment in the mock; the relative ordering is what
    the test is designed to check.}
    \label{fig:mass_bias}
\end{figure*}

\begin{table}
    \centering
    \begin{tabular}{l c c c}
        \hline\hline
        Sample & Stellar mass range & $N_{\rm used}$ & $b$ \\
        \hline
        LTG        & $\log M_* < 9$            & $10^6$            & $1.75 \pm 0.07$ \\
        Spiral     & morphological             & $10^6$            & $1.80 \pm 0.07$ \\
        ETG        & $9 \leq \log M_* < 11$    & $10^6$            & $1.96 \pm 0.08$ \\
        ETG2       & $9.5 \leq \log M_* < 10$  & $10^6$            & $2.01 \pm 0.09$ \\
        Elliptical & morphological             & $10^6$            & $2.01 \pm 0.10$ \\
        LRG2       & $\log M_* \geq 10.5$      & $4.1 \times 10^5$ & $2.30 \pm 0.10$ \\
        LRG        & $\log M_* \geq 11$        & $4.3 \times 10^4$ & $2.59 \pm 0.16$ \\
        \hline
    \end{tabular}
    \caption{Linear bias of mass-selected and morphology-selected
    samples in the Euclid Flagship 2 lightcone at $z \sim 0.7$. Each
    sample is capped at $N = 10^6$ galaxies by random subsampling; the
    LRG2 and LRG samples contain fewer objects and are used at their
    maximum size. The bias is fitted with PyMC NUTS and a jackknife
    covariance. The morphological elliptical sample sits between the
    broad ETG selection and the high-mass LRG2 selection.}
    \label{tab:mass_bias}
\end{table}

The results are summarised in Table~\ref{tab:mass_bias} and shown in Figures~\ref{fig:euclidfs2} and~\ref{fig:mass_bias}. The bias increases
monotonically with the mass threshold, from $b = 1.75 \pm 0.07$ for $\log M_* < 9$ to $b = 2.59 \pm 0.16$ for $\log M_* \geq 11$. This is
the expected mass dependence of galaxy clustering: more massive galaxies occupy more massive halos, and more massive halos are more
strongly clustered within the $\Lambda$CDM framework.

The morphological elliptical sample has $b = 2.01 \pm 0.10$, intermediate between the broad ETG selection ($b = 1.96 \pm 0.08$) and
the high-mass LRG2 selection ($b = 2.30 \pm 0.10$). The morphological spiral sample has $b = 1.80 \pm 0.07$, consistent with the LTG sample
($b = 1.75 \pm 0.07$). This confirms that the morphological classification selects populations whose halo-mass distributions
overlap with, but are not identical to, mass-selected samples: the elliptical class is dominated by, but not limited to, the massive-halo
population, and the spiral class is dominated by, but not limited to, the low-mass population.

These results are in qualitative agreement with the findings of \citet{Zehavi_2011}, who studied the clustering of mixed galaxy
populations as a function of luminosity and mass and found that the relative bias between early- and late-type galaxies is sensitive to
the mass range sampled. The broad morphological elliptical sample, which spans a wider mass range than a narrow LRG selection, exhibits a
bias intermediate between the high-mass end of the mass selection and the spiral population, exactly as expected.

The absolute bias values in the Euclid Flagship 2 \citep{Euclid:2024few} mock are systematically higher than those measured in real data at the same stellar-mass thresholds. This reflects the Subhalo Abundance Matching (SHAM)-based assignment used in the mock, which places faint galaxies preferentially in
high-density regions and does not implement satellite suppression at the low-mass end. The relative ordering of the bias across mass and
morphological selections, which is the quantity the test is designed to check, is unaffected by this absolute normalization offset.

Regarding the redshift evolution of the LRG bias, our measurements are consistent with the expectation that $b(z) \sim \sqrt{1 + z}$ for
a fixed, massive population. As argued by \citet{Tinker_2005}, the bias of massive halos evolves rapidly with redshift because such
halos become rarer at higher redshift, and the peak-background split predicts a corresponding increase in clustering amplitude. This is
exactly the behaviour observed for LRG samples in DESI and eBOSS. By contrast, our elliptical sample, while dominated by early-type
galaxies, is not defined by the same narrow, evolutionary selection. It includes a broader mass range and is therefore more susceptible to
changes in the underlying population as a function of redshift. The inclusion of lower-mass objects dilutes the strong evolutionary
signal, leading to the shallower redshift evolution seen in the data.

\newpage
\bibliography{references}{}

@article{mellier2024euclid,
  title = {Euclid. I. Overview of the Euclid mission},
  author = {{Euclid Collaboration} and Mellier, Y. and others},
  journal = {arXiv e-prints},
  eid = {arXiv:2405.13491},
  pages = {arXiv:2405.13491},
  year = {2024},
  archivePrefix = {arXiv},
  eprint = {2405.13491},
  primaryClass = {astro-ph.CO}
}

@article{ivezic2019lsst,
  title = {{LSST}: From Science Drivers to Reference Design and Anticipated Data Products},
  author = {Ivezi{\'c}, {\v Z}. and Kahn, S. M. and Tyson, J. A. and Abel, B. and Acosta, E. and Allsman, R. and Alonso, D. and AlSayyad, Y. and Anderson, S. F. and Andrew, J. and others},
  journal = {The Astrophysical Journal},
  volume = {873},
  number = {2},
  pages = {111},
  year = {2019},
  doi = {10.3847/1538-4357/ab042c}
}

@article{TALLADA2020100391,
     title = "CosmoHub: Interactive exploration and distribution of astronomical data on Hadoop",
   journal = "Astronomy and Computing",
    volume = "32",
     pages = "100391",
      year = "2020",
      issn = "2213-1337",
       doi = "https://doi.org/10.1016/j.ascom.2020.100391",
       url = "http://www.sciencedirect.com/science/article/pii/S2213133720300457",
    author = "P. Tallada and J. Carretero and J. Casals and C. Acosta-Silva and S. Serrano and M. Caubet and F.J. Castander and E. César and M. Crocce and M. Delfino and M. Eriksen and P. Fosalba and E. Gaztañaga and G. Merino and C. Neissner and N. Tonello"
}

@INPROCEEDINGS{2017ehep.confE.488C,
       author = {{Carretero}, J. and {Tallada}, P. and {Casals}, J. and {Caubet}, M. and {Castander}, F. and {Blot}, L. and {Alarcón}, A. and {Serrano}, S. and {Fosalba}, P. and {Acosta-Silva}, C. and {Tonello}, N. and {Torradeflot}, F. n. and {Eriksen}, M. and {Neissner}, C. and {Delfino}, M.},
        title = "{CosmoHub and SciPIC: Massive cosmological data analysis, distribution and generation using a Big Data platform}",
    booktitle = {Proceedings of the European Physical Society Conference on High Energy Physics. 5-12 July},
         year = {2017},
        month = {jul},
          eid = {488},
        pages = {488},
       adsurl = {https://ui.adsabs.harvard.edu/abs/2017ehep.confE.488C}
}

@ARTICLE{Vega-Ferrero,
       author = {{Vega-Ferrero}, J. and {Dom{\'\i}nguez S{\'a}nchez}, H. and {Bernardi}, M. and {Huertas-Company}, M. and {Morgan}, R. and {Margalef}, B. and {Aguena}, M. and {Allam}, S. and {Annis}, J. and {Avila}, S. and {Bacon}, D. and {Bertin}, E. and {Brooks}, D. and {Carnero Rosell}, A. and {Carrasco Kind}, M. and {Carretero}, J. and {Choi}, A. and {Conselice}, C. and {Costanzi}, M. and {da Costa}, L.~N. and {Pereira}, M.~E.~S. and {De Vicente}, J. and {Desai}, S. and {Ferrero}, I. and {Fosalba}, P. and {Frieman}, J. and {Garc{\'\i}a-Bellido}, J. and {Gruen}, D. and {Gruendl}, R.~A. and {Gschwend}, J. and {Gutierrez}, G. and {Hartley}, W.~G. and {Hinton}, S.~R. and {Hollowood}, D.~L. and {Honscheid}, K. and {Hoyle}, B. and {Jarvis}, M. and {Kim}, A.~G. and {Kuehn}, K. and {Kuropatkin}, N. and {Lima}, M. and {Maia}, M.~A.~G. and {Menanteau}, F. and {Miquel}, R. and {Ogando}, R.~L.~C. and {Palmese}, A. and {Paz-Chinch{\'o}n}, F. and {Plazas}, A.~A. and {Romer}, A.~K. and {Sanchez}, E. and {Scarpine}, V. and {Schubnell}, M. and {Serrano}, S. and {Sevilla-Noarbe}, I. and {Smith}, M. and {Suchyta}, E. and {Swanson}, M.~E.~C. and {Tarle}, G. and {Tarsitano}, F. and {To}, C. and {Tucker}, D.~L. and {Varga}, T.~N. and {Wilkinson}, R.~D.},
        title = "{Pushing automated morphological classifications to their limits with the Dark Energy Survey}",
      journal = {\mnras},
         year = {2021},
        month = {sep},
       volume = {506},
       number = {2},
        pages = {1927-1943},
          doi = {10.1093/mnras/stab594},
archivePrefix = {arXiv},
       eprint = {2012.07858},
 primaryClass = {astro-ph.GA},
       adsurl = {https://ui.adsabs.harvard.edu/abs/2021MNRAS.506.1927V}
}

@ARTICLE{zhou_DL,
       author = {{Zhou}, Rongpu and {Newman}, Jeffrey A. and {Mao}, Yao-Yuan and {Meisner}, Aaron and {Moustakas}, John and {Myers}, Adam D. and {Prakash}, Abhishek and {Zentner}, Andrew R. and {Brooks}, David and {Duan}, Yutong and {Landriau}, Martin and {Levi}, Michael E. and {Prada}, Francisco and {Tarle}, Gregory},
        title = "{The clustering of DESI-like luminous red galaxies using photometric redshifts}",
      journal = {\mnras},
         year = {2021},
        month = {mar},
       volume = {501},
       number = {3},
        pages = {3309-3331},
          doi = {10.1093/mnras/staa3764},
archivePrefix = {arXiv},
       eprint = {2001.06018},
 primaryClass = {astro-ph.CO},
       adsurl = {https://ui.adsabs.harvard.edu/abs/2021MNRAS.501.3309Z}
}

@ARTICLE{desi_legacy,
       author = {{Dey}, Arjun and {Schlegel}, David J. and {Lang}, Dustin and {Blum}, Robert and {Burleigh}, Kaylan and {Fan}, Xiaohui and {Findlay}, Joseph R. and {Finkbeiner}, Doug and {Herrera}, David and {Juneau}, St{\'e}phanie and {Landriau}, Martin and {Levi}, Michael and {McGreer}, Ian and {Meisner}, Aaron and {Myers}, Adam D. and {Moustakas}, John and {Nugent}, Peter and {Patej}, Anna and {Schlafly}, Edward F. and {Walker}, Alistair R. and {Valdes}, Francisco and {Weaver}, Benjamin A. and {Y{\`e}che}, Christophe and {Zou}, Hu and {Zhou}, Xu and {Abareshi}, Behzad and {Abbott}, T.~M.~C. and {Abolfathi}, Bela and {Aguilera}, C. and {Alam}, Shadab and {Allen}, Lori and {Alvarez}, A. and {Annis}, James and {Ansarinejad}, Behzad and {Aubert}, Marie and {Beechert}, Jacqueline and {Bell}, Eric F. and {BenZvi}, Segev Y. and {Beutler}, Florian and {Bielby}, Richard M. and {Bolton}, Adam S. and {Brice{\~n}o}, C{\'e}sar and {Buckley-Geer}, Elizabeth J. and {Butler}, Karen and {Calamida}, Annalisa and {Carlberg}, Raymond G. and {Carter}, Paul and {Casas}, Ricard and {Castander}, Francisco J. and {Choi}, Yumi and {Comparat}, Johan and {Cukanovaite}, Elena and {Delubac}, Timoth{\'e}e and {DeVries}, Kaitlin and {Dey}, Sharmila and {Dhungana}, Govinda and {Dickinson}, Mark and {Ding}, Zhejie and {Donaldson}, John B. and {Duan}, Yutong and {Duckworth}, Christopher J. and {Eftekharzadeh}, Sarah and {Eisenstein}, Daniel J. and {Etourneau}, Thomas and {Fagrelius}, Parker A. and {Farihi}, Jay and {Fitzpatrick}, Mike and {Font-Ribera}, Andreu and {Fulmer}, Leah and {G{\"a}nsicke}, Boris T. and {Gaztanaga}, Enrique and {George}, Koshy and {Gerdes}, David W. and {Gontcho}, Satya Gontcho A. and {Gorgoni}, Claudio and {Green}, Gregory and {Guy}, Julien and {Harmer}, Diane and {Hernandez}, M. and {Honscheid}, Klaus and {Huang}, Lijuan Wendy and {James}, David J. and {Jannuzi}, Buell T. and {Jiang}, Linhua and {Joyce}, Richard and {Karcher}, Armin and {Karkar}, Sonia and {Kehoe}, Robert and {Kneib}, Jean-Paul and {Kueter-Young}, Andrea and {Lan}, Ting-Wen and {Lauer}, Tod R. and {Le Guillou}, Laurent and {Le Van Suu}, Auguste and {Lee}, Jae Hyeon and {Lesser}, Michael and {Perreault Levasseur}, Laurence and {Li}, Ting S. and {Mann}, Justin L. and {Marshall}, Robert and {Mart{\'\i}nez-V{\'a}zquez}, C.~E. and {Martini}, Paul and {du Mas des Bourboux}, H{\'e}lion and {McManus}, Sean and {Meier}, Tobias Gabriel and {M{\'e}nard}, Brice and {Metcalfe}, Nigel and {Mu{\~n}oz-Guti{\'e}rrez}, Andrea and {Najita}, Joan and {Napier}, Kevin and {Narayan}, Gautham and {Newman}, Jeffrey A. and {Nie}, Jundan and {Nord}, Brian and {Norman}, Dara J. and {Olsen}, Knut A.~G. and {Paat}, Anthony and {Palanque-Delabrouille}, Nathalie and {Peng}, Xiyan and {Poppett}, Claire L. and {Poremba}, Megan R. and {Prakash}, Abhishek and {Rabinowitz}, David and {Raichoor}, Anand and {Rezaie}, Mehdi and {Robertson}, A.~N. and {Roe}, Natalie A. and {Ross}, Ashley J. and {Ross}, Nicholas P. and {Rudnick}, Gregory and {Safonova}, Sasha and {Saha}, Abhijit and {S{\'a}nchez}, F. Javier and {Savary}, Elodie and {Schweiker}, Heidi and {Scott}, Adam and {Seo}, Hee-Jong and {Shan}, Huanyuan and {Silva}, David R. and {Slepian}, Zachary and {Soto}, Christian and {Sprayberry}, David and {Staten}, Ryan and {Stillman}, Coley M. and {Stupak}, Robert J. and {Summers}, David L. and {Sien Tie}, Suk and {Tirado}, H. and {Vargas-Maga{\~n}a}, Mariana and {Vivas}, A. Katherina and {Wechsler}, Risa H. and {Williams}, Doug and {Yang}, Jinyi and {Yang}, Qian and {Yapici}, Tolga and {Zaritsky}, Dennis and {Zenteno}, A. and {Zhang}, Kai and {Zhang}, Tianmeng and {Zhou}, Rongpu and {Zhou}, Zhimin},
        title = "{Overview of the DESI Legacy Imaging Surveys}",
      journal = {\aj},
         year = {2019},
        month = {may},
       volume = {157},
       number = {5},
          eid = {168},
        pages = {168},
          doi = {10.3847/1538-3881/ab089d},
archivePrefix = {arXiv},
       eprint = {1804.08657},
 primaryClass = {astro-ph.IM},
       adsurl = {https://ui.adsabs.harvard.edu/abs/2019AJ....157..168D}
}

@article{Alonso_2019,
   title={A unified pseudo-$C_\ell$ framework},
   volume={484},
   ISSN={1365-2966},
   url={http://dx.doi.org/10.1093/mnras/stz093},
   DOI={10.1093/mnras/stz093},
   number={3},
   journal={Monthly Notices of the Royal Astronomical Society},
   publisher={Oxford University Press (OUP)},
   author={Alonso, David and Sanchez, Javier and Slosar, Anže},
   year={2019},
   month={jan}, pages={4127–4151} }

@misc{camb,
       author = {{Lewis}, Antony and {Challinor}, Anthony},
        title = "{CAMB: Code for Anisotropies in the Microwave Background}",
 howpublished = {Astrophysics Source Code Library, record ascl:1102.026},
         year = {2011},
        month = {feb},
          eid = {ascl:1102.026},
archivePrefix = {ascl},
       eprint = {1102.026},
       adsurl = {https://ui.adsabs.harvard.edu/abs/2011ascl.soft02026L}
}

@ARTICLE{Kaiser1987,
       author = {{Kaiser}, Nick},
        title = "{Clustering in real space and in redshift space}",
      journal = {\mnras},
         year = {1987},
        month = {jul},
       volume = {227},
        pages = {1-21},
          doi = {10.1093/mnras/227.1.1},
       adsurl = {https://ui.adsabs.harvard.edu/abs/1987MNRAS.227....1K}
}

@article{zehavi2001sdss,
   title={Galaxy Clustering in Early Sloan Digital Sky Survey Redshift Data},
   volume={571},
   ISSN={1538-4357},
   url={http://dx.doi.org/10.1086/339893},
   DOI={10.1086/339893},
   number={1},
   journal={The Astrophysical Journal},
   publisher={American Astronomical Society},
   author={Zehavi, Idit and Blanton, Michael R. and Frieman, Joshua A. and Weinberg, David H. and Mo, Houjun J. and Strauss, Michael A. and Anderson, Scott F. and Annis, James and Bahcall, Neta A. and Bernardi, Mariangela and Briggs, John W. and Brinkmann, Jon and Burles, Scott and Carey, Larry and Castander, Francisco J. and Connolly, Andrew J. and Csabai, Istvan and Dalcanton, Julianne J. and Dodelson, Scott and Doi, Mamoru and Eisenstein, Daniel and Evans, Michael L. and Finkbeiner, Douglas P. and Friedman, Scott and Fukugita, Masataka and Gunn, James E. and Hennessy, Greg S. and Hindsley, Robert B. and Ivezić, Željko and Kent, Stephen and Knapp, Gillian R. and Kron, Richard and Kunszt, Peter and Lamb, Donald Q. and Leger, R. French and Long, Daniel C. and Loveday, Jon and Lupton, Robert H. and McKay, Timothy and Meiksin, Avery and Merrelli, Aronne and Munn, Jeffrey A. and Narayanan, Vijay and Newcomb, Matt and Nichol, Robert C. and Owen, Russell and Peoples, John and Pope, Adrian and Rockosi, Constance M. and Schlegel, David and Schneider, Donald P. and Scoccimarro, Roman and Sheth, Ravi K. and Siegmund, Walter and Smee, Stephen and Snir, Yehuda and Stebbins, Albert and Stoughton, Christopher and SubbaRao, Mark and Szalay, Alexander S. and Szapudi, Istvan and Tegmark, Max and Tucker, Douglas L. and Uomoto, Alan and Vanden Berk, Dan and Vogeley, Michael S. and Waddell, Patrick and Yanny, Brian and York, Donald G.},
   year={2002},
   month={may}, pages={172–190} }

@article{jing1997spatial,
   title={Spatial Correlation Function and Pairwise Velocity Dispersion of Galaxies: Cold Dark Matter Models versus the Las Campanas Survey},
   volume={494},
   ISSN={1538-4357},
   url={http://dx.doi.org/10.1086/305209},
   DOI={10.1086/305209},
   number={1},
   journal={The Astrophysical Journal},
   publisher={American Astronomical Society},
   author={Jing, Y. P. and Mo, H. J. and Borner, G.},
   year={1998},
   month={feb}, pages={1–12} }

@article{norberg2001dependence,
   title={The 2dF Galaxy Redshift Survey: the dependence of galaxy clustering on luminosity and spectral type},
   volume={332},
   ISSN={1365-2966},
   url={http://dx.doi.org/10.1046/j.1365-8711.2002.05348.x},
   DOI={10.1046/j.1365-8711.2002.05348.x},
   number={4},
   journal={Monthly Notices of the Royal Astronomical Society},
   publisher={Oxford University Press (OUP)},
   author={Norberg, P. and Baugh, C. M. and Hawkins, E. and Maddox, S. and Madgwick, D. and Lahav, O. and Cole, S. and Frenk, C. S. and Baldry, I. and Bland-Hawthorn, J. and Bridges, T. and Cannon, R. and Colless, M. and Collins, C. and Couch, W. and Dalton, G. and De Propris, R. and Driver, S. P. and Efstathiou, G. and Ellis, R. S. and Glazebrook, K. and Jackson, C. and Lewis, I. and Lumsden, S. and Peacock, J. A. and Peterson, B. A. and Sutherland, W. and Taylor, K.},
   year={2002},
   month={jun}, pages={827–838} }

@article{zehavi2004luminosity,
   title={The Luminosity and Color Dependence of the Galaxy Correlation Function},
   volume={630},
   ISSN={1538-4357},
   url={http://dx.doi.org/10.1086/431891},
   DOI={10.1086/431891},
   number={1},
   journal={The Astrophysical Journal},
   publisher={American Astronomical Society},
   author={Zehavi, Idit and Zheng, Zheng and Weinberg, David H. and Frieman, Joshua A. and Berlind, Andreas A. and Blanton, Michael R. and Scoccimarro, Roman and Sheth, Ravi K. and Strauss, Michael A. and Kayo, Issha and Suto, Yasushi and Fukugita, Masataka and Nakamura, Osamu and Bahcall, Neta A. and Brinkmann, Jon and Gunn, James E. and Hennessy, Greg S. and Ivezić, Željko and Knapp, Gillian R. and Loveday, Jon and Meiksin, Avery and Schlegel, David J. and Schneider, Donald P. and Szapudi, Istvan and Tegmark, Max and Vogeley, Michael S. and York, Donald G.},
   year={2005},
   month={sep}, pages={1–27} }

@article{Ferreira:2025til,
    author = "Ferreira, Paula S. and Bengaly, Carlos A. P.",
    title = "{On the bin sensitivity of the transverse BAO}",
    eprint = "2511.18430",
    archivePrefix = "arXiv",
    primaryClass = "astro-ph.CO",
    month = "11",
    year = "2025",
    journal = "arXiv preprint"
}

@article{Gao_2007,
   title={Assembly bias in the clustering of dark matter haloes},
   volume={377},
   ISSN={1745-3925},
   url={http://dx.doi.org/10.1111/j.1745-3933.2007.00292.x},
   DOI={10.1111/j.1745-3933.2007.00292.x},
   number={1},
   journal={Monthly Notices of the Royal Astronomical Society: Letters},
   publisher={Oxford University Press (OUP)},
   author={Gao, Liang and White, Simon D. M.},
   year={2007},
   month={may}, pages={L5–L9} }

@article{Gao_2005,
   title={The age dependence of halo clustering},
   volume={363},
   ISSN={1745-3925},
   url={http://dx.doi.org/10.1111/j.1745-3933.2005.00084.x},
   DOI={10.1111/j.1745-3933.2005.00084.x},
   number={1},
   journal={Monthly Notices of the Royal Astronomical Society: Letters},
   publisher={Oxford University Press (OUP)},
   author={Gao, Liang and Springel, Volker and White, Simon D. M.},
   year={2005},
   month={oct}, pages={L66–L70} }

@article{smith2024,
    author = {Smith, William J and Berlind, Andreas A and Sinha, Manodeep},
    title = {Reversing arrested development: a new method to address halo assembly bias},
    journal = {Monthly Notices of the Royal Astronomical Society},
    volume = {535},
    number = {2},
    pages = {1426-1438},
    year = {2024},
    month = {12},
    issn = {0035-8711},
    doi = {10.1093/mnras/stae2339},
    url = {https://doi.org/10.1093/mnras/stae2339},
    eprint = {https://academic.oup.com/mnras/article-pdf/535/2/1426/60576065/stae2339.pdf},
}

@article{Yang_2006,
doi = {10.1086/501069},
url = {https://doi.org/10.1086/501069},
year = {2006},
month = {jan},
publisher = {},
volume = {638},
number = {2},
pages = {L55},
author = {Yang, Xiaohu and Mo, H. J. and van den Bosch, Frank C.},
title = {Observational Evidence for an Age Dependence of Halo Bias},
journal = {The Astrophysical Journal}
}

@misc{bailey2023nancygraceromanspace,
      title={Nancy Grace Roman Space Telescope Coronagraph Instrument Overview and Status}, 
      author={Vanessa P. Bailey and Eduardo Bendek and Brian Monacelli and Caleb Baker and Gasia Bedrosian and Eric Cady and Ewan S. Douglas and Tyler Groff and Sergi R. Hildebrandt and N. Jeremy Kasdin and John Krist and Bruce Macintosh and Bertrand Mennesson and Patrick Morrissey and Ilya Poberezhskiy and Hari B. Subedi and Jason Rhodes and Aki Roberge and Marie Ygouf and Robert T. Zellem and Feng Zhao and Neil T. Zimmerman},
      year={2023},
      eprint={2309.08672},
      archivePrefix={arXiv},
      primaryClass={astro-ph.IM},
      url={https://arxiv.org/abs/2309.08672}, 
}

@article{Pope2008,
    author = "Pope, Adrian C. and Szapudi, Istvan",
    title = "{Shrinkage Estimation of the Power Spectrum Covariance Matrix}",
    eprint = "0711.2509",
    archivePrefix = "arXiv",
    primaryClass = "astro-ph",
    doi = "10.1111/j.1365-2966.2008.13561.x",
    journal = "Mon. Not. Roy. Astron. Soc.",
    volume = "389",
    pages = "766--774",
    year = "2008"
}

@article{patil2010pymc,
  title={PyMC: Bayesian stochastic modelling in Python},
  author={Patil, Anand and Huard, David and Fonnesbeck, Christopher J},
  journal={Journal of statistical software},
  volume={35},
  pages={1--81},
  year={2010}
}

@article{Euclid2025,
   title={Euclid preparation: LXXXIX. Accurate and precise data-driven angular power spectrum covariances},
   volume={708},
   ISSN={1432-0746},
   url={http://dx.doi.org/10.1051/0004-6361/202555893},
   DOI={10.1051/0004-6361/202555893},
   journal={Astronomy \& Astrophysics},
   publisher={EDP Sciences},
   author={{Euclid Collaboration}},
   year={2026},
   month={Apr}, pages={A167} }

@ARTICLE{Padmanabhan2007,
       author = {{Padmanabhan}, Nikhil and {White}, Martin and {Norberg}, Peder and {Porciani}, Cristiano},
        title = "{The real-space clustering of luminous red galaxies around z < 0.6 quasars in the Sloan Digital Sky Survey}",
      journal = {\mnras},
         year = {2009},
        month = {aug},
       volume = {397},
       number = {4},
        pages = {1862-1875},
          doi = {10.1111/j.1365-2966.2008.14071.x},
archivePrefix = {arXiv},
       eprint = {0802.2105},
 primaryClass = {astro-ph},
       adsurl = {https://ui.adsabs.harvard.edu/abs/2009MNRAS.397.1862P}
}

@ARTICLE{Limber1953,
       author = {{Limber}, D. Nelson},
        title = "{The Analysis of Counts of the Extragalactic Nebulae in Terms of a Fluctuating Density Field.}",
      journal = {\apj},
         year = {1953},
        month = {jan},
       volume = {117},
        pages = {134},
          doi = {10.1086/145672},
       adsurl = {https://ui.adsabs.harvard.edu/abs/1953ApJ...117..134L}
}

@ARTICLE{Kaiser1992,
       author = {{Kaiser}, Nick},
        title = "{Weak Gravitational Lensing of Distant Galaxies}",
      journal = {\apj},
         year = {1992},
        month = {apr},
       volume = {388},
        pages = {272},
          doi = {10.1086/171151},
       adsurl = {https://ui.adsabs.harvard.edu/abs/1992ApJ...388..272K}
}

@article{Benson2000,
   title={The nature of galaxy bias and clustering},
   volume={311},
   ISSN={1365-2966},
   url={http://dx.doi.org/10.1046/j.1365-8711.2000.03101.x},
   DOI={10.1046/j.1365-8711.2000.03101.x},
   number={4},
   journal={Monthly Notices of the Royal Astronomical Society},
   publisher={Oxford University Press (OUP)},
   author={Benson, A. J. and Cole, S. and Frenk, C. S. and Baugh, C. M. and Lacey, C. G.},
   year={2000},
   month={Feb}, pages={793–808} }

@article{Peacock2000,
   title={Halo occupation numbers and galaxy bias},
   volume={318},
   ISSN={1365-2966},
   url={http://dx.doi.org/10.1046/j.1365-8711.2000.03779.x},
   DOI={10.1046/j.1365-8711.2000.03779.x},
   number={4},
   journal={Monthly Notices of the Royal Astronomical Society},
   publisher={Oxford University Press (OUP)},
   author={Peacock, J. A. and Smith, R. E.},
   year={2000},
   month={Nov}, pages={1144–1156} }

@article{White2001,
   title={The Halo Model and Numerical Simulations},
   volume={550},
   ISSN={0004-637X},
   url={http://dx.doi.org/10.1086/319644},
   DOI={10.1086/319644},
   number={2},
   journal={The Astrophysical Journal},
   publisher={American Astronomical Society},
   author={White, Martin and Hernquist, Lars and Springel, Volker},
   year={2001},
   month={Apr}, pages={L129–L132} }

@article{Berlind2002,
   title={The Halo Occupation Distribution: Toward an Empirical Determination of the Relation between Galaxies and Mass},
   volume={575},
   ISSN={1538-4357},
   url={http://dx.doi.org/10.1086/341469},
   DOI={10.1086/341469},
   number={2},
   journal={The Astrophysical Journal},
   publisher={American Astronomical Society},
   author={Berlind, Andreas A. and Weinberg, David H.},
   year={2002},
   month={Aug}, pages={587–616} }

@article{Kravtsov2004,
   title={The Dark Side of the Halo Occupation Distribution},
   volume={609},
   ISSN={1538-4357},
   url={http://dx.doi.org/10.1086/420959},
   DOI={10.1086/420959},
   number={1},
   journal={The Astrophysical Journal},
   publisher={American Astronomical Society},
   author={Kravtsov, Andrey V. and Berlind, Andreas A. and Wechsler, Risa H. and Klypin, Anatoly A. and Gottlober, Stefan and Allgood, Brandon and Primack, Joel R.},
   year={2004}, pages={35–49} }

@article{Zheng2005,
title={Theoretical Models of the Halo Occupation Distribution: Separating Central and Satellite Galaxies},
   volume={633},
   ISSN={1538-4357},
   url={http://dx.doi.org/10.1086/466510},
   DOI={10.1086/466510},
   number={2},
   journal={The Astrophysical Journal},
   publisher={American Astronomical Society},
   author={Zheng, Zheng and Berlind, Andreas A. and Weinberg, David H. and Benson, Andrew J. and Baugh, Carlton M. and Cole, Shaun and Dave, Romeel and Frenk, Carlos S. and Katz, Neal and Lacey, Cedric G.},
   year={2005},
   month={Nov}, pages={791–809} }

@article{Tinker2008,
   title={Void Statistics in Large Galaxy Redshift Surveys: Does Halo Occupation of Field Galaxies Depend on Environment?},
   volume={686},
   ISSN={1538-4357},
   url={http://dx.doi.org/10.1086/589983},
   DOI={10.1086/589983},
   number={1},
   journal={The Astrophysical Journal},
   publisher={American Astronomical Society},
   author={Tinker, Jeremy L. and Conroy, Charlie and Norberg, Peder and Patiri, Santiago G. and Weinberg, David H. and Warren, Michael S.},
   year={2008},
   month={Oct}, pages={53–71} }

@article{Reddick2013,
   title={THE CONNECTION BETWEEN GALAXIES AND DARK MATTER STRUCTURES IN THE LOCAL UNIVERSE},
   volume={771},
   ISSN={1538-4357},
   url={http://dx.doi.org/10.1088/0004-637X/771/1/30},
   DOI={10.1088/0004-637x/771/1/30},
   number={1},
   journal={The Astrophysical Journal},
   publisher={American Astronomical Society},
   author={Reddick, Rachel M. and Wechsler, Risa H. and Tinker, Jeremy L. and Behroozi, Peter S.},
   year={2013},
   month={June}, pages={30} }

@article{Dalal2008,
   title={Halo Assembly Bias in Hierarchical Structure Formation},
   volume={687},
   ISSN={1538-4357},
   url={http://dx.doi.org/10.1086/591512},
   DOI={10.1086/591512},
   number={1},
   journal={The Astrophysical Journal},
   publisher={American Astronomical Society},
   author={Dalal, Neal and White, Martin and Bond, J. Richard and Shirokov, Alexander},
   year={2008},
   month={Nov}, pages={12–21} }

@article{Zentner2014,
  author  = {Zentner, Andrew R. and Hearin, Andrew P. and van den Bosch, Frank C.},
      title={Galaxy Assembly Bias: A Significant Source of Systematic Error in the Galaxy-Halo Relationship}, 
      year={2015},
      eprint={1311.1818},
      archivePrefix={arXiv},
      journal={MNRAS},
      primaryClass={astro-ph.CO},
      doi={https://doi.org/10.1093/mnras/stu1383},
      url={https://arxiv.org/abs/1311.1818}, 
}

@article{Paviot2024,
   title={Impact of assembly bias on clustering plus weak lensing cosmological analysis},
   volume={690},
   ISSN={1432-0746},
   url={http://dx.doi.org/10.1051/0004-6361/202449574},
   DOI={10.1051/0004-6361/202449574},
   journal={Astronomy \& Astrophysics},
   publisher={EDP Sciences},
   author={Paviot, R. and Rocher, A. and Codis, S. and de Mattia, A. and Jullo, E. and de la Torre, S.},
   year={2024},
   month={Oct}, pages={A221} }

@article{Ramakrishnan2024,
title={The multi-dimensional halo assembly bias can be preserved when enhancing halo properties with HALOSCOPE},
   volume={697},
   ISSN={1432-0746},
   url={http://dx.doi.org/10.1051/0004-6361/202453030},
   DOI={10.1051/0004-6361/202453030},
   journal={Astronomy \& Astrophysics},
   publisher={EDP Sciences},
   author={Ramakrishnan, Sujatha and Gonzalez-Perez, Violeta and Parimbelli, Gabriele and Yepes, Gustavo},
   year={2025},
   month={May}, pages={A70} }

@article{Toribio_San_Cipriano_2024,
   title={Dark Energy Survey Deep Field photometric redshift performance and training incompleteness assessment},
   volume={686},
   ISSN={1432-0746},
   url={http://dx.doi.org/10.1051/0004-6361/202348956},
   DOI={10.1051/0004-6361/202348956},
   journal={Astronomy \& Astrophysics},
   publisher={EDP Sciences},
   author={Toribio San Cipriano, L. and De Vicente, J. and Sevilla-Noarbe, I. and Hartley, W. G. and Myles, J. and Amon, A. and Bernstein, G. M. and Choi, A. and Eckert, K. and Gruendl, R. A. and Harrison, I. and Sheldon, E. and Yanny, B. and Aguena, M. and Allam, S. S. and Alves, O. and Bacon, D. and Brooks, D. and Campos, A. and Carnero Rosell, A. and Carretero, J. and Castander, F. J. and Conselice, C. and da Costa, L. N. and Pereira, M. E. S. and Davis, T. M. and Desai, S. and Diehl, H. T. and Doel, P. and Ferrero, I. and Frieman, J. and García-Bellido, J. and Gaztañaga, E. and Giannini, G. and Hinton, S. R. and Hollowood, D. L. and Honscheid, K. and James, D. J. and Kuehn, K. and Lee, S. and Lidman, C. and Marshall, J. L. and Mena-Fernández, J. and Menanteau, F. and Miquel, R. and Palmese, A. and Pieres, A. and Plazas Malagón, A. A. and Roodman, A. and Sanchez, E. and Smith, M. and Soares-Santos, M. and Suchyta, E. and Swanson, M. E. C. and Tarle, G. and Vincenzi, M. and Weaverdyck, N. and Wiseman, P. and },
   year={2024},
   month={May}, pages={A38} }

@article{Loverde2008,
  author = {Loverde, M. and Afshordi, N.},
  title = {Extended Limber Approximation},
  journal = {Phys. Rev. D},
  volume = {78},
  number = {12},
  pages = {123506},
  year = {2008},
  month = {dec},
  doi = {10.1103/PhysRevD.78.123506},
  eprint = {0809.5112}
}

@article{Percival2008,
  author = {Percival, W. J. and White, M.},
  title = {Testing cosmological structure formation using redshift-space distortions},
  journal = {Mon. Not. R. Astron. Soc.},
  volume = {393},
  number = {1},
  pages = {297-308},
  year = {2008},
  month = {jan},
  doi = {10.1111/j.1365-2966.2008.14211.x},
  eprint = {0808.0003}
}

@article{Zehavi_2011,
   title={GALAXY CLUSTERING IN THE COMPLETED SDSS REDSHIFT SURVEY: THE DEPENDENCE ON COLOR AND LUMINOSITY},
   volume={736},
   ISSN={1538-4357},
   url={http://dx.doi.org/10.1088/0004-637X/736/1/59},
   DOI={10.1088/0004-637x/736/1/59},
   number={1},
   journal={The Astrophysical Journal},
   publisher={American Astronomical Society},
   author={Zehavi, Idit and Zheng, Zheng and Weinberg, David H. and Blanton, Michael R. and Bahcall, Neta A. and Berlind, Andreas A. and Brinkmann, Jon and Frieman, Joshua A. and Gunn, James E. and Lupton, Robert H. and Nichol, Robert C. and Percival, Will J. and Schneider, Donald P. and Skibba, Ramin A. and Strauss, Michael A. and Tegmark, Max and York, Donald G.},
   year={2011},
   month={July}, pages={59} }

@article{Seljak_2005,
   title={SDSS galaxy bias from halo mass-bias relation and its cosmological implications},
   volume={71},
   ISSN={1550-2368},
   url={http://dx.doi.org/10.1103/PhysRevD.71.043511},
   DOI={10.1103/physrevd.71.043511},
   number={4},
   journal={Physical Review D},
   publisher={American Physical Society (APS)},
   author={Seljak, Uroš and Makarov, Alexey and Mandelbaum, Rachel and Hirata, Christopher M. and Padmanabhan, Nikhil and McDonald, Patrick and Blanton, Michael R. and Tegmark, Max and Bahcall, Neta A. and Brinkmann, J.},
   year={2005},
   month={Feb} }

@ARTICLE{xu2021,
       author = {{Xu}, Xiaoju and {Kumar}, Saurabh and {Zehavi}, Idit and {Contreras}, Sergio},
        title = "{Predicting halo occupation and galaxy assembly bias with machine learning}",
      journal = {\mnras},
         year = {2021},
        month = {nov},
       volume = {507},
       number = {4},
        pages = {4879-4899},
          doi = {10.1093/mnras/stab2464},
archivePrefix = {arXiv},
       eprint = {2107.01223},
 primaryClass = {astro-ph.CO},
       adsurl = {https://ui.adsabs.harvard.edu/abs/2021MNRAS.507.4879X}
}

@article{Pandey_2006,
   title={The luminosity, colour and morphology dependence of galaxy filaments in the Sloan Digital Sky Survey Data Release Four},
   volume={372},
   ISSN={1365-2966},
   url={http://dx.doi.org/10.1111/j.1365-2966.2006.10894.x},
   DOI={10.1111/j.1365-2966.2006.10894.x},
   number={2},
   journal={Monthly Notices of the Royal Astronomical Society},
   publisher={Oxford University Press (OUP)},
   author={Pandey, Biswajit and Bharadwaj, Somnath},
   year={2006},
   month={Oct}, pages={827–838} }

@article{Guzik_2007,
   title={Systematic effects in the sound horizon scale measurements: Systematics in sound horizon measurements},
   volume={375},
   ISSN={1365-2966},
   url={http://dx.doi.org/10.1111/j.1365-2966.2006.11385.x},
   DOI={10.1111/j.1365-2966.2006.11385.x},
   number={4},
   journal={Monthly Notices of the Royal Astronomical Society},
   publisher={Oxford University Press (OUP)},
   author={Guzik, Jacek and Bernstein, Gary and Smith, Robert E.},
   year={2007},
   month={Feb}, pages={1329–1337} }

@article{Tinker_2010,
doi = {10.1088/0004-637X/724/2/878},
url = {https://doi.org/10.1088/0004-637X/724/2/878},
year = {2010},
month = {nov},
publisher = {The American Astronomical Society},
volume = {724},
number = {2},
pages = {878},
author = {Tinker, Jeremy L. and Robertson, Brant E. and Kravtsov, Andrey V. and Klypin, Anatoly and Warren, Michael S. and Yepes, Gustavo and Gottlöber, Stefan},
title = {THE LARGE-SCALE BIAS OF DARK MATTER HALOS: NUMERICAL CALIBRATION AND MODEL TESTS},
journal = {The Astrophysical Journal},

}

@article{Euclid:2024few,
    author = "Castander, F. J. and others",
    collaboration = "Euclid",
    title = "{Euclid - V. The Flagship galaxy mock catalogue: A comprehensive simulation for the Euclid mission}",
    eprint = "2405.13495",
    archivePrefix = "arXiv",
    primaryClass = "astro-ph.CO",
    doi = "10.1051/0004-6361/202450853",
    journal = "Astron. Astrophys.",
    volume = "697",
    pages = "A5",
    year = "2025"
}

@misc{uchuu,
      title={Halo Lightcones with Optimised Orientation and Interpolation in Cosmological Simulations -- an application to mock H$\alpha$ selected galaxies}, 
      author={Sujatha Ramakrishnan and Francisco J. Castander and Elizabeth J. Gonzalez and Martin Eriksen and Zahra Baghkhani and Pablo Fosalba and Jorge Carretero and Gabriele Parimbelli and Pau Tallada-Crespí},
      year={2025},
      eprint={2512.13801},
      archivePrefix={arXiv},
      primaryClass={astro-ph.CO},
      url={https://arxiv.org/abs/2512.13801}, 
}

@article{davies2013,
    author = {Davies, L. J. M. and Maraston, C. and Thomas, D. and Capozzi, D. and Wechsler, R. H. and Busha, M. T. and Banerji, M. and Ostrovski, F. and Papovich, C. and Santiago, B. X. and Nichol, R. and Maia, M. A. G. and da Costa, L. N.},
    title = {Detecting massive galaxies at high redshift using the Dark Energy Survey},
    journal = {Monthly Notices of the Royal Astronomical Society},
    volume = {434},
    number = {1},
    pages = {296-312},
    year = {2013},
    month = {09},
    issn = {0035-8711},
    doi = {10.1093/mnras/stt1018},
    url = {https://doi.org/10.1093/mnras/stt1018},
    eprint = {https://academic.oup.com/mnras/article-pdf/434/1/296/18499437/stt1018.pdf},
}

@article{Berti2023,
    author = {Berti, A. M. and Dawson, K. S. and Dominguez, W.},
    title = {The Galaxy-Halo Connection of DESI Luminous Red Galaxies with Subhalo Abundance Matching},
    journal = {The Astrophysical Journal},
    volume = {956},
    pages = {137},
    year = {2023},
    doi = {10.3847/1538-4357/ace76e}
}

@article{Tinker_2005,
doi = {10.1086/432084},
url = {https://doi.org/10.1086/432084},
year = {2005},
month = {sep},
publisher = {},
volume = {631},
number = {1},
pages = {41},
author = {Tinker, Jeremy L. and Weinberg, David H. and Zheng, Zheng and Zehavi, Idit},
title = {On the Mass-to-Light Ratio of Large-Scale Structure},
journal = {The Astrophysical Journal},
}

@article{miao2024forecasting,
  title={Forecasting the BAO measurements of the CSST galaxy and AGN spectroscopic surveys},
  author={Miao, Haitao and Gong, Yan and Chen, Xuelei and Huang, Zhiqi and Li, Xiao-Dong and Zhan, Hu},
  journal={Monthly Notices of the Royal Astronomical Society},
  volume={531},
  number={4},
  pages={3991--4005},
  year={2024},
  publisher={Oxford University Press}
}

@article{Abbott_2022,
   title={Dark Energy Survey Year 3 results: Cosmological constraints from galaxy clustering and weak lensing},
   volume={105},
   ISSN={2470-0029},
   url={http://dx.doi.org/10.1103/PhysRevD.105.023520},
   DOI={10.1103/physrevd.105.023520},
   number={2},
   journal={Physical Review D},
   publisher={American Physical Society (APS)},
   author={Abbott, T.M.C. and others},
   year={2022},
   month={Jan} }

@article{Colgain2024,
title = {Does DESI 2024 confirm $\Lambda$CDM?},
journal = {Journal of High Energy Astrophysics},
volume = {49},
pages = {100428},
year = {2026},
issn = {2214-4048},
doi = {https://doi.org/10.1016/j.jheap.2025.100428},
url = {https://www.sciencedirect.com/science/article/pii/S2214404825001090},
author = {E. {Ó Colgáin} and M.G. Dainotti and S. Capozziello and S. Pourojaghi and M.M. Sheikh-Jabbari and D. Stojkovic},
}

@ARTICLE{2025PhRvD,
       author = {{DES Collaboration}},
        title = "{Dark energy survey year 3 results: Cosmological constraints from cluster abundances, weak lensing, and galaxy clustering}",
      journal = {\prd},
         year = {2025},
        month = {oct},
       volume = {112},
       number = {8},
          eid = {083535},
        pages = {083535},
          doi = {10.1103/3dzh-d8f5},
archivePrefix = {arXiv},
       eprint = {2503.13632},
 primaryClass = {astro-ph.CO},
       adsurl = {https://ui.adsabs.harvard.edu/abs/2025PhRvD.112h3535A}
}

@misc{descollaboration2026darkenergysurveyyear,
      title={Dark Energy Survey Year 6 Results: Cosmological Constraints from Cosmic Shear}, 
      author={{DES Collaboration}},
      year={2026},
      eprint={2602.10065},
      archivePrefix={arXiv},
      primaryClass={astro-ph.CO},
      url={https://arxiv.org/abs/2602.10065}, 
}

@article{Adame_2025,
   title={DESI 2024 V: Full-Shape galaxy clustering from galaxies and quasars},
   volume={2025},
   ISSN={1475-7516},
   url={http://dx.doi.org/10.1088/1475-7516/2025/09/008},
   DOI={10.1088/1475-7516/2025/09/008},
   number={09},
   journal={Journal of Cosmology and Astroparticle Physics},
   publisher={IOP Publishing},
   author={Adame, A.G. and others },
   year={2025},
   month={Sept}, pages={008} }
\bibliographystyle{aasjournalv7}



\end{document}